\documentclass[twocolumn]{aastex631}

\usepackage{booktabs}
\usepackage{amsmath}
\usepackage{tabularx}   
\usepackage{multirow}
\usepackage{mathtools}
\usepackage{soul}
\usepackage[T1]{fontenc}
\usepackage{amssymb}
\newcolumntype{M}{m{1.5cm}}%

\newcommand\clearrow{\global\let\rowmac\relax}
\newcommand{\bdv}[1]{\mbox{\boldmath$#1$}}
\def\au{{\rm AU}}

\def\B{{\rm B}}
\def\Sc{{\rm S}}
\def\L{{\rm L}}
\def\E{{\rm E}}
\def\bpi{{\bdv\pi}}
\def\bmu{{\bdv\mu}}

\graphicspath{{./}{figures/}}
\begin{document}
\title{Gaia22dkvLb: A Microlensing Planet Potentially Accessible to Radial-Velocity Characterization}

\correspondingauthor{Subo Dong}
\email{dongsubo@pku.edu.cn}

\author[0009-0007-5754-6206]{Zexuan Wu}
\affil{Department of Astronomy, School of Physics, Peking University,
	Yiheyuan Rd. 5, Haidian District, Beijing, China, 100871 \\}
\affil{Kavli Institute of Astronomy and Astrophysics, Peking University,
	Yiheyuan Rd. 5, Haidian District, Beijing, China, 100871 \\}
\author[0000-0002-1027-0990]{Subo Dong}
\affil{Department of Astronomy, School of Physics, Peking University,
	Yiheyuan Rd. 5, Haidian District, Beijing, China, 100871 \\}
\affil{Kavli Institute of Astronomy and Astrophysics, Peking University,
	Yiheyuan Rd. 5, Haidian District, Beijing, China, 100871 \\}
\author{Tuan Yi}
\affil{Department of Astronomy, School of Physics, Peking University,
	Yiheyuan Rd. 5, Haidian District, Beijing, China, 100871 \\}
\affil{Kavli Institute of Astronomy and Astrophysics, Peking University,
	Yiheyuan Rd. 5, Haidian District, Beijing, China, 100871 \\}
\author{Zhuokai Liu}
\affil{Department of Astronomy, School of Physics, Peking University,
	Yiheyuan Rd. 5, Haidian District, Beijing, China, 100871 \\}
\affil{Kavli Institute of Astronomy and Astrophysics, Peking University,
	Yiheyuan Rd. 5, Haidian District, Beijing, China, 100871 \\}
\author{Kareem El-Badry}
\affil{Department of Astronomy, California Institute of Technology, Pasadena, CA 91125, USA\\}
\author{Andrew Gould}
\affil{Max-Planck Institute for Astronomy, K\"onigstuhl 17, D-69117 Heidelberg, Germany\\}
\affil{Department of Astronomy, Ohio State University, 140 W. 18th Ave., Columbus, OH 43210, USA\\}
\author[0000-0002-9658-6151]{{\L}. Wyrzykowski}
\affil{Astronomical Observatory, University of Warsaw, Al. Ujazdowskie 4, 00-478 Warszawa, Poland}
\author{K.A.~Rybicki}
\affil{Department of Particle Physics and Astrophysics, Weizmann Institute of Science, Rehovot 76100, Israel}
\affil{Astronomical Observatory, University of Warsaw, Al. Ujazdowskie 4, 00-478 Warszawa, Poland}
\author{Etienne Bachelet}
\affil{IPAC, Mail Code 100-22, Caltech, 1200 E. California Blvd., Pasadena, CA 91125, USA}
\author{Grant W. Christie}
\affil{Auckland Observatory, Auckland, New Zealand}
\author[0000-0001-8179-1147]{L. de Almeida}
\affil{Laborat\'orio Nacional de Astrof\'isica, Rua Estados Unidos 154, 37504-364 Itajub\'a---MG, Brazil}
\author{L.~A.~G. Monard}
\affil{Klein Karoo Observatory, Centre for Backyard Astrophysics, Calitzdorp, South Africa}
\author{J.~McCormick}
\affil{Farm Cove Observatory, Centre for Backyard Astrophysics, Pakuranga, Auckland, New Zealand}
\author{Tim Natusch}
\affil{Mathematical Sciences Department, Auckland University of Technology, Auckland, New Zealand }
\affil{Auckland Observatory, Auckland, New Zealand}
\author[0000-0001-6434-9429]{P. Zieli{\'n}ski}
\affil{Institute of Astronomy, Faculty of Physics, Astronomy and Informatics, Nicolaus Copernicus University in Toru{\'n}, Grudzi\k{a}dzka 5, 87-100 Toru{\'n}, Poland}
\author{Huiling Chen}
\affil{Department of Astronomy, School of Physics, Peking University,
	Yiheyuan Rd. 5, Haidian District, Beijing, China, 100871 \\}
\affil{Kavli Institute of Astronomy and Astrophysics, Peking University,
	Yiheyuan Rd. 5, Haidian District, Beijing, China, 100871 \\}
\author{Yang Huang}
\affiliation{University of Chinese Academy of Sciences, Beijing 100049,  China\\}
\affiliation{National Astronomical Observatories, Chinese Academy of Sciences, Beijing 100012,  China\\}
\author[0000-0002-7866-4531]{Chang Liu}
\affiliation{Department of Physics and Astronomy, Northwestern University, 2145 Sheridan Rd, Evanston, IL 60208, USA}
\affiliation{Center for Interdisciplinary Exploration and Research in Astrophysics (CIERA), Northwestern University, 1800 Sherman Ave, Evanston, IL 60201, USA}
\author[0000-0003-2125-0183]{A.~M\'erand}
\affiliation{European Southern Observatory, Karl-Schwarzschild-Str. 2, 85748 Garching, Germany}
\author[0000-0001-7016-1692]{Przemek Mr{\'o}z}
\affiliation{Astronomical Observatory, University of Warsaw, Al. Ujazdowskie 4, 00-478 Warszawa, Poland}
\author{Jinyi Shangguan}
\affiliation{Max Planck Institute for Extraterrestrial Physics, Giessenbachstr. 1, 85748, Garching, Germany}
\author[0000-0001-5207-5619]{Andrzej Udalski}
\affiliation{Astronomical Observatory, University of Warsaw, Al. Ujazdowskie 4, 00-478 Warszawa, Poland}
\author{J. Woillez}
\affiliation{European Southern Observatory, Karl-Schwarzschild-Str. 2, 85748 Garching, Germany}
\author{Huawei Zhang}
\affil{Department of Astronomy, School of Physics, Peking University,
	Yiheyuan Rd. 5, Haidian District, Beijing, China, 100871 \\}
\affil{Kavli Institute of Astronomy and Astrophysics, Peking University,
	Yiheyuan Rd. 5, Haidian District, Beijing, China, 100871 \\}
\author[0000-0003-0125-8700]{Franz-Josef Hambsch}
\affil{Vereniging Voor Sterrenkunde (VVS), Oostmeers 122 C, 8000 Brugge, Belgium\\}
\affil{AAVSO, 185 Alewife Brook Parkway, Suite 410, Cambridge, MA 02138, USA\\}
\affil{Groupe Europ\'{e}en d'Observations Stellaires (GEOS), 23 Parc de Levesville, 28300 Bailleau l'Ev\^{e}que, France}
\affil{Bundesdeutsche Arbeitsgemeinschaft f\"{u}r Ver\"{a}nderliche Sterne (BAV), Munsterdamm 90, 12169 Berlin, Germany\\}
\author[0000-0001-8916-8050]{P. J. Miko{\l}ajczyk}
\affil{Astronomical Observatory, University of Warsaw, Al. Ujazdowskie 4, 00-478 Warszawa, Poland}
\affil{Astronomical Institute, University of Wroc{\l}aw, ul. M. Kopernika 11, 51-622 Wroc{\l}aw, Poland}
\author{M. Gromadzki}
\affil{Astronomical Observatory, University of Warsaw, Al. Ujazdowskie 4, 00-478 Warszawa, Poland}
\author{M. Ratajczak}
\affil{Astronomical Observatory, University of Warsaw, Al. Ujazdowskie 4, 00-478 Warszawa, Poland}
\author{Katarzyna Kruszy{\'n}ska}
\affil{Astronomical Observatory, University of Warsaw, Al. Ujazdowskie 4, 00-478 Warszawa, Poland}
\affil{Las Cumbres Observatory, 6740 Cortona Drive, Goleta, CA 93117, USA}
\author{N. Ihanec}
\affil{Astronomical Observatory, University of Warsaw, Al. Ujazdowskie 4, 00-478 Warszawa, Poland}
\author{Uliana Pylypenko}
\affil{Astronomical Observatory, University of Warsaw, Al. Ujazdowskie 4, 00-478 Warszawa, Poland}
\author{M. Sitek}
\affil{Astronomical Observatory, University of Warsaw, Al. Ujazdowskie 4, 00-478 Warszawa, Poland}
\author{K. Howil}
\affil{Astronomical Observatory, University of Warsaw, Al. Ujazdowskie 4, 00-478 Warszawa, Poland}
\author{Staszek Zola}
\affil{Astronomical Observatory of the Jagiellonian University, ul. Orla 171, 30-244 Krakow, Poland\\}
\author{Olga Michniewicz}
\affil{Janusz Gil Institute of Astronomy, University of Zielona G\'ora, Poland\\}
\author{Michal Zejmo}
\affil{Janusz Gil Institute of Astronomy, University of Zielona G\'ora, Poland\\}
\author{Fraser Lewis}
\affil{Faulkes Telescope Project, UK\\}
\affil{The Schools Observatory, UK\\}
\author{Mateusz Bronikowski}
\affil{Center for Astrophysics and Cosmology, University of Nova Gorica, Vipavska 11c, 5270 Ajdov\v{s}\v{c}ina, Slovenia\\}
\author{Stephen Potter}
\affil{South African Astronomical Observatory, Observatory Road, Observatory, 7925, Cape Town, South Africa\\}
\affil{Department of Physics, University of Johannesburg, PO Box 524, Auckland Park 2006, South Africa}
\author{Jan Andrzejewski}
\affil{Janusz Gil Institute of Astronomy, University of Zielona G\'ora, Poland\\}
\author{Jaroslav Merc}
\affil{Astronomical Institute, Faculty of Mathematics and Physics, Charles University, V Hole\v{s}ovi\v{c}k\'{a}ch 2, 180 00 Prague, Czech Republic\\}
\author[0000-0001-6279-0552]{Rachel Street}
\affil{Las Cumbres Observatory, 6740 Cortona Drive, Suite 102, Goleta, CA 93117, USA}
\author{Akihiko Fukui}
\affil{Komaba Institute for Science, The University of Tokyo, 3-8-1 Komaba, Meguro, Tokyo 153-8902, Japan}
\affil{Instituto de Astrof\'isica de Canarias, V\'ia L\'actea s/n, E-38205 La Laguna, Tenerife, Spain}
\author{R. Figuera Jaimes}
\affil{Millennium Institute of Astrophysics MAS, Nuncio Monsenor Sotero Sanz 100, Of. 104, Providencia, Santiago, Chile}
\affil{Instituto de Astrof\'isica, Facultad de F\'isica, Pontificia Universidad Cat\'olica de Chile, Av. Vicu\~na Mackenna 4860, 7820436 Macul, Santiago, Chile}
\author{V. Bozza}
\affil{Dipartimento di Fisica ``E.R. Caianiello'', Universit\`a di Salerno, Via Giovanni Paolo II 132, Fisciano, 84084, Italy.}
\affil{Istituto Nazionale di Fisica Nucleare, Sezione di Napoli, Via Cintia, 80126, Napoli, Italy.}
\author{P. Rota}
\affil{Dipartimento di Fisica ``E.R. Caianiello'', Universit\`a di Salerno, Via Giovanni Paolo II 132, Fisciano, 84084, Italy.}
\author{A. Cassan}
\affil{Institut d'Astrophysique de Paris, Sorbonne Universit\'e, CNRS, UMR 7095, 98 bis bd Arago, F-75014 Paris, France}
\author{M. Dominik}
\affil{University of St Andrews, Centre for Exoplanet Science, SUPA School of Physics \& Astronomy, North Haugh, St Andrews, KY16 9SS, United Kingdom}
\author[0000-0001-8411-351X]{Y. Tsapras}
\affil{Zentrum f{\"u}r Astronomie der Universit{\"a}t Heidelberg, Astronomisches Rechen-Institut, M{\"o}nchhofstr. 12-14, 69120 Heidelberg, Germany}
\author{M. Hundertmark}
\affil{Zentrum f{\"u}r Astronomie der Universit{\"a}t Heidelberg, Astronomisches Rechen-Institut, M{\"o}nchhofstr. 12-14, 69120 Heidelberg, Germany}
\author{J. Wambsganss}
\affil{Zentrum f{\"u}r Astronomie der Universit{\"a}t Heidelberg, Astronomisches Rechen-Institut, M{\"o}nchhofstr. 12-14, 69120 Heidelberg, Germany}
\author{K. B\k{a}kowska}
\affil{Institute of Astronomy, Faculty of Physics, Astronomy and Informatics, Nicolaus Copernicus University in Toru{\'n}, Grudzi\k{a}dzka 5, 87-100 Toru{\'n}, Poland}
\author{A. S{\l}owikowska}
\affil{Joint Institute for VLBI ERIC (JIVE), Oude Hoogeveensedijk 4, 7991 PD Dwingeloo, The Netherlands}
\affil{Institute of Astronomy, Faculty of Physics, Astronomy and Informatics, Nicolaus Copernicus University in Toru{\'n}, Grudzi\k{a}dzka 5, 87-100 Toru{\'n}, Poland}

\begin{abstract}
	We report discovering an exoplanet from following up a microlensing event alerted by Gaia. The event Gaia22dkv is toward a { disk source} rather than the traditional bulge microlensing fields. { Our primary analysis yields a Jovian planet with $M_{\rm p} = 0.59 ^{+0.15}_{-0.05}\,M_{\rm J}$ at a projected orbital separation $r_{\perp} = 1.4^{+0.8}_{-0.3}$\,AU, and the host is a $\sim 1.1\,M_\odot$ turnoff star at $\sim 1.3\,$kpc. At $r'\approx14$, the host} is far brighter than any previously discovered microlensing planet host, opening up the opportunity of testing the microlensing model with radial velocity (RV) observations.
	RV data can be used to measure the planet's orbital period and eccentricity, and they also enable searching for inner planets of the microlensing cold Jupiter, as expected from the ``inner-outer correlation'' inferred from {\it Kepler} and RV discoveries.
	Furthermore, we show that Gaia astrometric microlensing will not only allow precise measurements of its angular Einstein radius $\theta_{\rm E}$, but also directly measure the microlens parallax vector and unambiguously break a geometric light-curve degeneracy, leading to definitive characterization of the lens system.
\end{abstract}
\keywords{gravitational lensing;}

\section{Introduction}\label{sec:intro}
Almost 200 exoplanets have been detected with gravitational microlensing. In a microlensing event, the light from a background star (the ``source'') is bent in the gravitational field of the foreground mass (the ``lens''), and a planet in the lens system can be detected from a short-lasting signal in the microlensing light curve. Most of the exoplanets discovered with microlensing are near or beyond the snow line, and the majority of their host stars are significantly more distant than $\sim$\,1\,kpc. In contrast, other detection methods overwhelmingly probe nearby hosts, and the two most prolific methods at present, transit and radial velocity (RV), predominantly find planets orbiting within the snow line. On the one hand, these distinguishing properties of the microlensing planetary systems provide a uniquely powerful probe of the Galactic exoplanet demography (see \citealt{ZhuDong2021} for a recent review). On the other hand, microlensing planetary systems are generally inaccessible to follow-up observations by other methods (except for a handful of potentially possible cases near the sensitivity limit of present RV capabilities, such as OGLE-2018-BLG-0740Lb at $V\sim18$ by \citealt{Han2019}), posing a limitation to further characterizing the detected planets and probing the existence of other planets in those systems.

The recent surge of time-domain surveys covering a considerable portion of the sky are opening up fresh possibilities of searching for microlensing planets in regions other than the traditional Galactic bulge fields.
The first such discovery was a Neptune-mass planet inside the snow line \citep{Nucita2018, Fukui2019} found in a near-field (source at $\approx 660$\,pc) microlensing event Kojima-1 (a.k.a., TCP J05074264+2447555), whose microlensing nature was recognized with the survey data of All-Sky Automatic Survey for Supernovae (ASAS-SN; \citealt{Shappee2014, 2017PASP..129j4502K}). The interferometric observations \citep{Dong2019} carried out by the GRAVITY instrument \citep{grav17} mounted on the Very Large Telescope Interferometer (VLTI) enabled measuring the physical parameters of Kojima-1, revealing that the planet has mass $M_{\rm planet} = 19.0\pm3.0\,M_\oplus$ at projected planet-host separation $r_\perp = 0.78\pm0.05\,{\rm AU}$ orbiting the host star with mass $M=0.495\pm0.063\,M_\odot$ at $429\pm21\,{\rm pc}$ \citep{Zang2020}.

In this paper, we report the discovery of an extrasolar planet from our follow-up observations of a near-field microlensing event Gaia22dkv detected by the {\it Gaia} mission, which has found hundreds of microlensing events over the entire sky (see, e.g., \citealt{Wyrzykowski2023}). Our primary analysis suggests that Gaia22dkvLb is a Jovian planet inside the snow line, and it is potentially accessible to high-precision radial velocity (RV) facilities.

\section{Observations and Data Reductions} \label{sec:obs_dre}

\begin{figure*}[ht!]
	\epsscale{1.15}
	\plotone{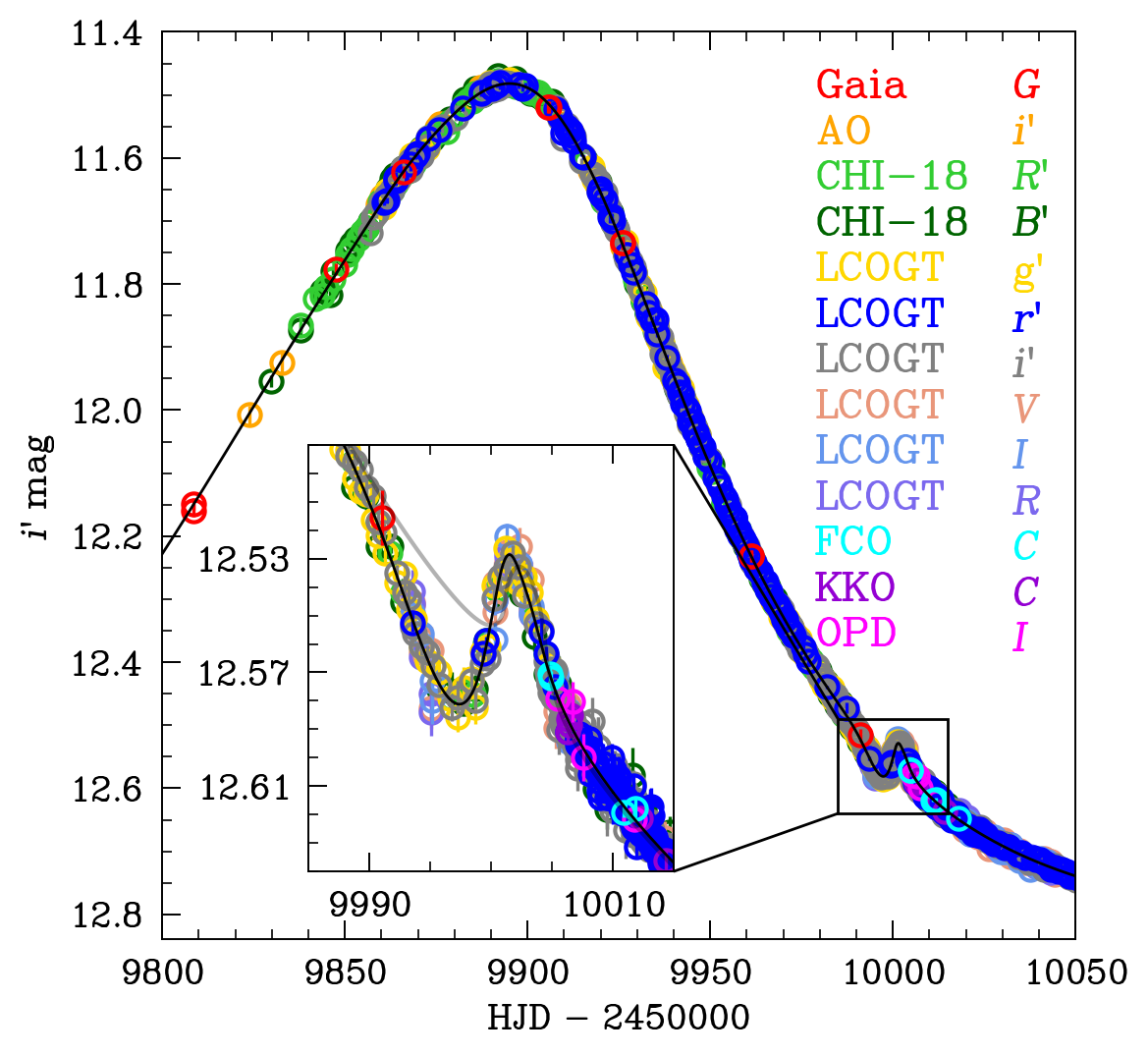}
	\caption{Gaia22dkv's light curve and the best-fit planetary microlensing model (black solid line) considering microlens parallax and orbital motion effects. Data points observed by various sites are shown as open circles in different colors respectively. The magnitudes are transformed into $i'$ band using the magnifications from the best-fit model. The inset panel presents the zoomed-in view of the region in the neighborhood of the planetary anomaly. The grey line represents the best-fit binary-source (1L2S) model, which fails to match the anomaly.
		\label{fig:best_fit}}
\end{figure*}

Gaia22dkv was discovered by Gaia Science Alerts
	{ \citep{GSC}} on 2022 August 16 UT and announced\footnote{\url{http://gsaweb.ast.cam.ac.uk/alerts/alert/Gaia22dkv/}} as a microlensing candidate  on 2022 August 19 UT.
The event's equatorial coordinates are
$\left(\alpha,\,\delta\right)_\mathrm{J2000} = \left(10^{\rm h}07^{\rm m}04.56^{\rm s},-66\arcdeg10\arcmin51.20\arcsec\right)$
, corresponding to Galactic coordinates
$\left(l,\,b\right)_\mathrm{J2000} = \left(287.36783 ,\, -8.41019\right)$.

We followed up Gaia22dkv as part of an ongoing program to identify  and characterize bright and long timescale microlensing events to be observed by VLTI-GRAVITY, and the main scientific goal of the program is to search for isolated dark stellar remnants (i.e., neutron stars and black holes).  Our program started in 2018, and we have systematically processed microlensing alerts and performed photometric follow-up observations using the 1-m telescopes of Las Cumbres Observatory global network (LCOGT, \citealt{2013PASP..125.1031B}) and several small telescopes.

At the time of the Gaia alert, Gaia22dkv was not accessible to LCOGT due to its high airmass. We began to take images in the $Blue$ and $Red$ bands (denoted as $B'$ and $R'$, respectively, throughout the paper) of the Astrodon LRGB filters on 2022 August 20 UT ({$\mathrm{HJD}' \coloneqq \mathrm{HJD} - 2450000\approx 9811.89$}) using the 0.18-m Newtonian telescope (CHI-18) located at the El Sauce Observatory in Chile on every clear night. We also performed imaging observations with the 40-cm telescope at the Auckland Observatory (AO) in SDSS $i'$ band since 2022 August 22 UT. Our LCOGT observations started on October 06 2022 with a daily cadence in SDSS $i'$ and $r'$ bands.

VLTI-GRAVITY observations with the Auxiliary Telescopes (ATs) under the Target-of-Opportunity (ToO) program 108.220D.007 were carried out on 2022 November 3, 2022 November 30, and 2022 December 15. No fringe was found during the first epoch, and for the latter two epochs, fringes were present only for two baselines, inadequate for inferring the closure phase. Therefore,  we obtain no useful VLTI data.

As the brightness of Gaia22dkv continued to decline and approached the baseline, the prospect of obtaining useful VLTI data with ATs diminished. On 2023 February 01, we thus decided to change the LCOGT observations to a weekly cadence, while the cadence of CHI-18 observations stayed unchanged. We perform real-time, automatic photometry on the LCOGT images. From an inspection of the LCOGT photometry on UT 2023 March 01 ({$\mathrm{HJD}' \approx 10004.5$}), we noticed a deviation from the single-lens microlensing model. We subsequently confirmed the anomaly using the CHI-18 data after processing the images. On the same day, we alerted to Microlensing Follow-Up Network ($\mu$FUN) requesting high-cadence follow-up observations. Three sites associated with $\mu$FUN observed this event, including the 0.36-m telescope at the Farm Cove Observatory (FCO) with clear ($C$) filter, the 0.36-m telescope at the Klein Karoo Observatory (KKO) with clear ($C$) filter and the 0.6-m and 1.6-m telescopes in Pico dos Dias Observatory (OPD) with Johnson-Cousin $I$-band filter.
	{ This event was also independently followed up by a number of sites with the coordination via a web-based system named Black Hole Target and Observation Manager (BHTOM) based on Las Cumbres Observatory's Target and Observation Manager (TOM) Toolkit \citep{TOM}.
		BHTOM is a web platform for coordinating time-domain observations using a heterogeneous network of telescopes from around the world. The system collects the calibrated images from the telescopes and performs an automated PSF photometry as well as standardisation to Gaia Synthetic Photometry filters.
		For Gaia22dkv, the following telescopes collected photometric data: Las Cumbres Observatory (1-m, SAAO South Africa, CTIO Chile, SSO Australia), PROMPT5 (0.4-m, CTIO, Chile), PROMPT6 (0.4-m, CTIO, Chile), PROMPT-MO-1 (0.4-m, Meckering Observatory, Australia), ROAD Observatory (0.4-m, Chile), UZPW (0.5-m, Chile), Go-Chile-GoT1 (0.4-m, Chile), Lesedi (1-m, SAAO, South Africa), Danish (1.54-m, La Silla, Chile).
		In total, BHTOM collected and processed 3189 data points for Gaia22dkv spanning 220 days between $\mathrm{HJD'} = 9814.42$ and $10034.04$. More details on the technical setup of the telescopes are available on the BHTOM webpage\footnote{\url{http://bhtom.space}}.}

	{  We first check the quality of all acquired images and reject those taken under poor seeing ($>10\arcsec$) conditions or with very low atmospheric transmission, and as a result, 175 out of 6003 CHI-18 images and 63 out of 1616 LCOGT images are excluded from further analysis. Using the \texttt{PmPyeasy} photometric pipeline \citep{Chen2022}, we conduct aperture photometry on LCOGT (including those collected via BHTOM, which were partly obtained by the OMEGA collaboration), CHI-18, FCO, and OPD data and point-spread-function (PSF) photometry on the AO data using \texttt{DoPHOT} \citep{1993PASP..105.1342S}. The BHTOM images from sites other than LCOGT were processed with PSF photometry using CCDPhot and standardized with the Cambridge Photometric Calibration Server (CPCS) as described in \citet{CPCS}. These latter sets of BHTOM data show qualitative agreement with other data sets, including the region around the planetary anomaly, but they exhibit significantly larger scatters than the others and thus are not used in the subsequent analysis.} The Gaia $G$ photometric data downloaded from the Gaia Science Alerts website do not have photometric uncertainties, and we adopt $\sigma_G = 0.02$ as suggested by \citet{GSC} for a bright source ($G \lesssim 13$). When modeling the light curves, we employ the commonly-adopted procedure of rescaling the photometric errors \citep[see, e.g.,][]{2012ApJ...755..102Y} to reach $\chi^2 / \mathrm{dof} \approx 1$ for each data set.
	{ On 2022 December 2 UT, we took a high-resolution spectrum with the 10-m Southern African Large Telescope \citep[SALT;][]{2006SPIE.6267E..0ZB}.\footnote{SALT project ID: 2022-2-SCI-016, PI: P. Zieli{\'n}ski} The observation was conduted close to the peak, at magnification $A\sim 4.3$. The High-Resolution Spectrograph \citep[HRS;][]{2014SPIE.9147E..6TC} was used in order to obtain the {\'e}chelle spectrum registered in red and blue channels covering 3700 - 8800\,\AA. The Low Resolution mode (LR) of HRS instrument ($R\sim15000$) was applied. We took a 60\,s exposure with a $1.673$\arcsec\, slit width, and slow readout speed.
		The raw spectroscopic data were reduced in a standard way with the SALT MIDAS-based pipeline \citep{2016MNRAS.459.3068K,2017ASPC..510..480K}, i.e., bias-subtracted, flat-field normalized and wavelength calibrated by using ThAr comparison lamp. The obtained spectrum is extracted to a 1-D version, merged by orders, sky emission subtracted and corrected for the heliocentric velocity. Then, we cleaned the spectrum by removing some artificial features and cosmic ray spikes. In addition, due to the poor quality of the spectrum in the blue edge of the wavelength range, we removed some parts with $\lambda<4300$\,\AA  \,\,as well as between 7590 - 7670\,\AA, which was problematic in a continuum fitting. In this way, the resulting spectrum used in further analysis covers the wavelength range 4300 - 8800\,\AA\,\,with a signal-to-noise ratio SNR~$\sim33$, on average.}

On 2023 April 23 UT, we took spectroscopic follow-up observation using the Magellan Inamori Kyocera Echelle \citep[MIKE;][]{MIKE2003} spectrograph
mounted on the Magellan 6.5-m telescope. We used a 0.5\arcsec\, slit and took a 900\,s exposure. We employ the CarPy \citep{Kelson2000,Kelson2003} package to reduce the raw spectroscopic data, following the standard procedures of bias subtraction, flat-field correction, automatic cosmic-ray removal, 1-D spectral orders extraction, wavelength calibration, order merge, and applying heliocentric RV correction. Then we manually identify and remove un-cleaned cosmic-rays and detector defects. The resulting spectrum covers 3400\,\AA\, - 9000\,\AA, with a spectral resolution $R\sim40000$ and median signal-to-noise ratio ${\rm SNR}\sim100$ on the red side and $R\sim55000$ with median ${\rm SNR}\sim38$ on the blue side.
Because the SNR is relatively low at blue wavelengths $\lambda\lesssim3800$\,\AA\,and has serious telluric contaminations on the red side with $\lambda\gtrsim6800$\,\AA, we use the spectrum between 3800\,\AA\, -- 6800\,\AA\, for subsequent analysis.

\begin{figure}[htbp]
	\epsscale{1.15}
	\plotone{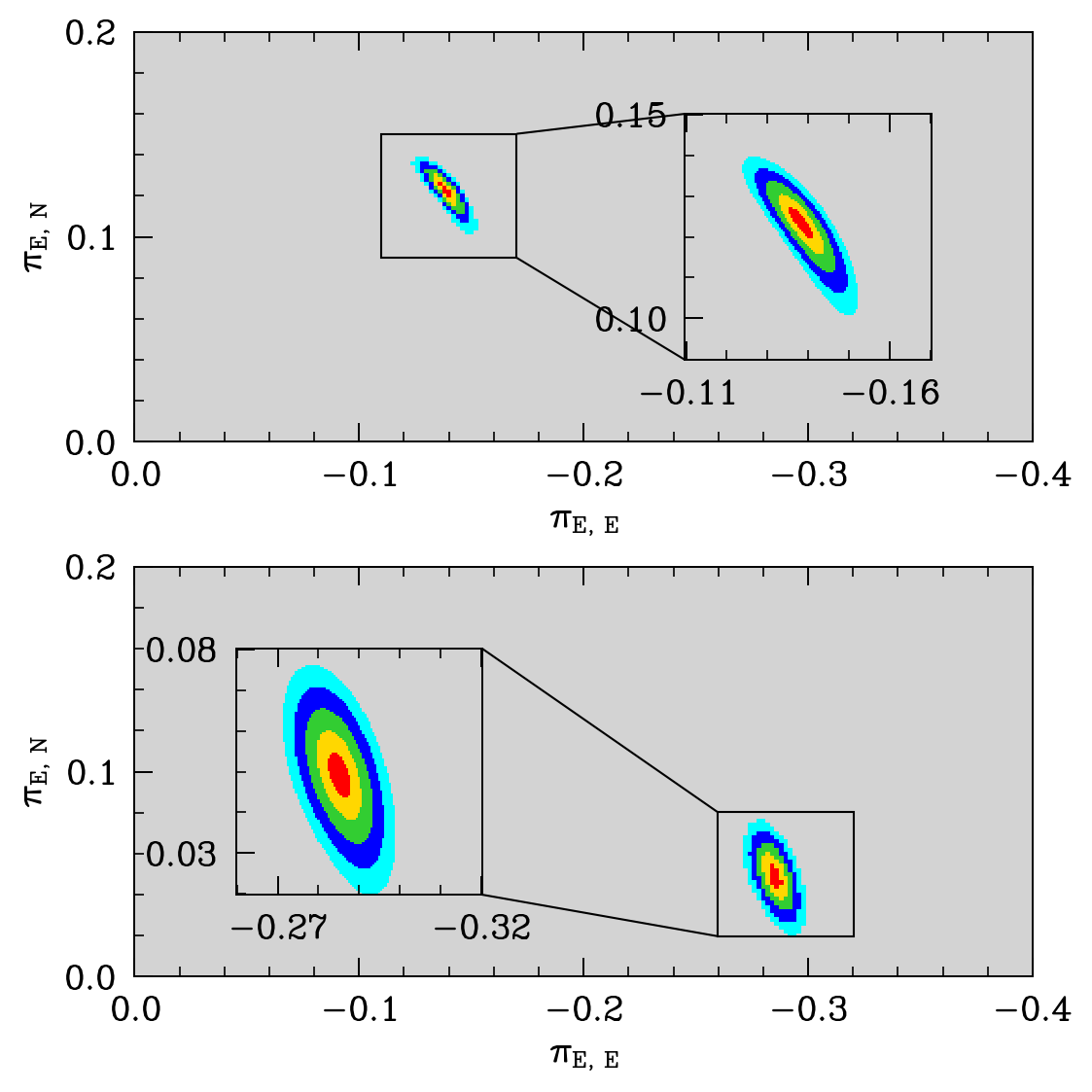}
	\caption{Two degenerate microlens parallax solutions from PSPL analysis of the light curve with the region in the neighborhood of the planet anomaly removed. The upper and the lower panels display the $u_0+$ and $u_0-$ grid-search results, respectively. { The solutions with $\Delta \chi^2 < 1,\, 4,\, 9,\, 16, \,25$ and $>25$ of the best fit are color-coded with red, yellow, green, blue, cyan and grey.} The insets display zoomed-in views.
		\label{fig:parallax_grid}}
\end{figure}

\begin{figure}[ht!]
	\epsscale{1.15}
	\plotone{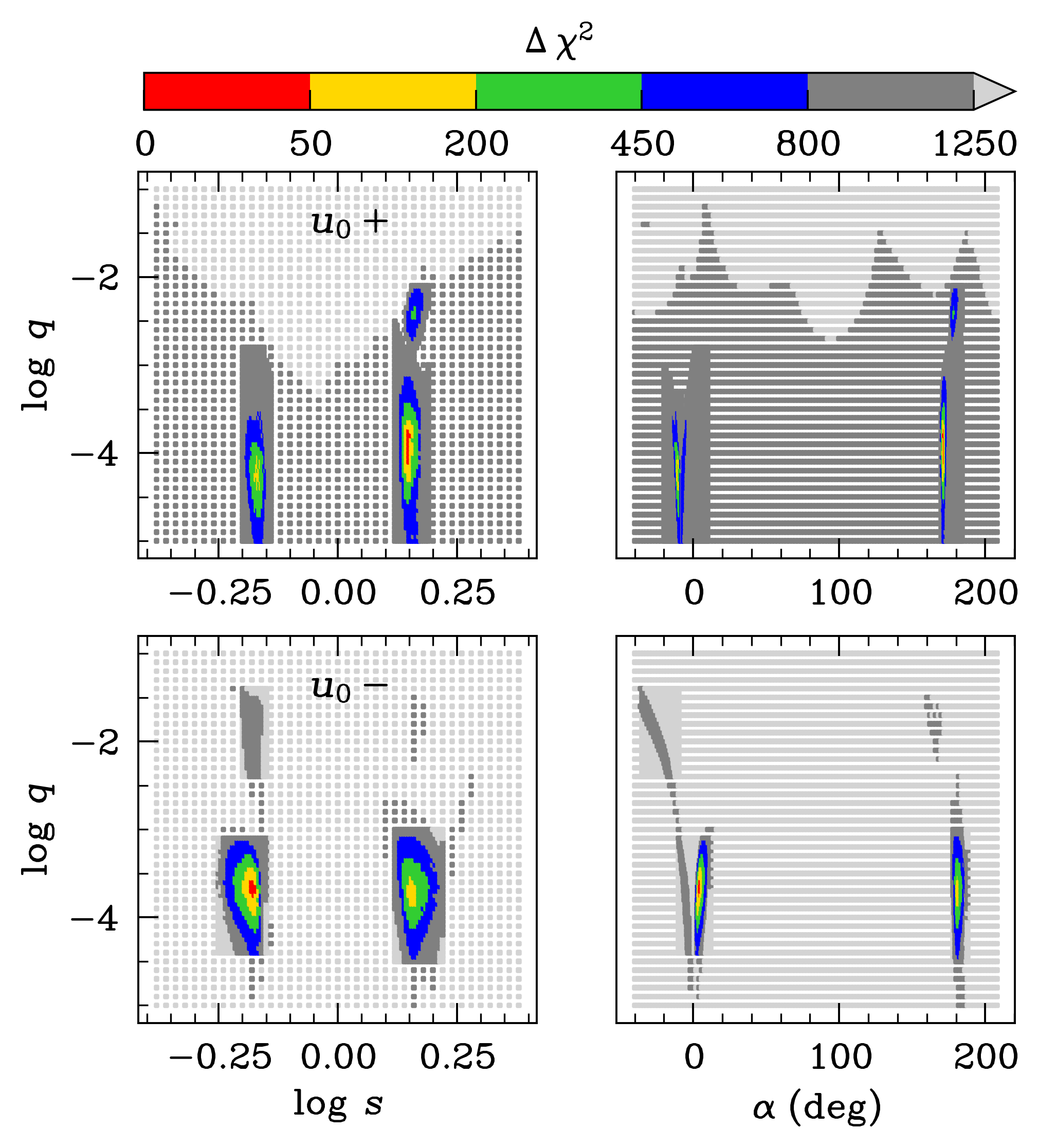}
	\caption{The $\Delta \chi^2$ map of the binary-lens point-source grid-search results on the $\left(\log s,\, \log q\right)$ and $\left(\alpha,\, \log q\right)$ planes.
	The upper and lower panels show $u_0+$ and $u_0-$ grid searches, respectively.
	The solutions with $\Delta \chi^2 < 50,\, 200,\, 450,\, 800,\, 1250$ and $>1250$ compared with the best fits ($\chi^2_{\rm min}\approx868$ for $u_0+$ and  $\chi^2_{\rm min}\approx690$ for $u_0-$) are color-coded with red, yellow, green, blue, grey, and light grey, respectively.
	\label{fig:grid_search}}
\end{figure}

\section{Microlens Modeling} \label{sec:analysis}
The overall light curve of Gaia22dkv (see Figure~\ref{fig:best_fit}) has a striking asymmetry, with the post-peak declining at a significantly steeper rate than the pre-peak rising. As discussed later in this section, this long-term deviation from the standard, symmetric point-source point-lens (PSPL) model \citep{1986ApJ...304....1P} is well explained by the annual microlens parallax effect due to the Earth's orbital motion \citep{Gould1992}.  There is a short-duration anomaly on the light curve, and in the following, we show that it is the type of planetary microlensing signature as discussed in \citet{1992ApJ...396..104G}.

Before performing the more complicated modeling necessary to interpret the anomaly, we fit the light curve by excluding the region in the neighborhood of the anomaly with $\mathrm{HJD}'=[9980, 10020]$
to PSPL models, including microlens parallax. The PSPL model includes 5 parameters $\Theta_{\rm PSPL} = \left(t_0,\, u_0,\, t_\E,\, \pi_{\mathrm{E,N}}, \pi_{\mathrm{E,E}}\right)$ to compute the magnification $A$ of the background source star as a function of time $t$. $\left(t_0,\,u_0,\,t_\E\right)$ denote the time of the closest source-lens approach, the impact parameter normalized by the Einstein radius, and the timescale to cross the Einstein radius, respectively. The microlens parallax vector $\bpi_\E=\left(\pi_\mathrm{E,N}, \pi_{\mathrm{E,E}}\right)$ is expressed in terms of its north and east equatorial coordinates, as defined in \citet{Gould2004}. The observed flux $f_i\left(t\right)$ for each dataset $i$ is modeled as $f_i \left(t\right) = f_{\Sc,i} A \left(t\right)  + f_{\B,i}$, where the source flux $ f_{\Sc,i} $ and the blended (i.e., un-lensed) flux $ f_{\B,i}$ within the PSF are the two flux parameters. The blended flux is referred to as the ``blend'' throughout the text. The microlens parallax solutions are subject to a 2-fold discrete ``constant acceleration'' degeneracy \citep{Smith03}, and the two corresponding sets of solutions ($u_0+$ and $u_0-$) from grid searches are shown in the upper and lower panels of Figure~\ref{fig:parallax_grid}, respectively.

In \S~\ref{subsec:2L1S}, we analyze the light curve using the binary-lens\footnote{As a common practice in the microlensing literature, a binary lens is a generic term referring to a lens system with two point masses, including a planetary lens for which the mass ratio is low.} (2L1S) model, and we consider the effects of microlens parallax and planetary orbital motion. We also explore alternative models, including the so-called xallarap effect that can mimic microlens parallax \citep{xallarap, Smith2002} in \S~\ref{sec:xallarap}, and the single-lens binary-source (1L2S) model that can sometimes produce a planet-like signal \citep{1998ApJ...506..533G, HanGould1997, 2013ApJ...778...55H} in \S~\ref{subsec:1L2S}, respectively.
\begin{table*}
	\renewcommand{\arraystretch}{1.1}
	\caption{2L1S model parameters and $1\sigma$ uncertainties for $u_0+$. \label{tab:2L1S_u0p_orb_parameters}}
	\begin{tabular*}{\textwidth}{@{} M @{\extracolsep{\fill}} CCCCCCCCCCCCC@{} }
		\hline
		Model & t_0 & u_0 & t_\E &\log\rho & s & q & \alpha & \pi_\mathrm{E,\,N} & \pi_\mathrm{E,\,E} & w_1 & w_2 & w_3 & P_\perp\\
		$\chi^2$ & \left(\mathrm{HJD'}\right) &  & \left(\mathrm{day}\right)  &  &  & \times 10^3 & \left(\mathrm{deg}\right) &  & &\mathrm{yr}^{-1} &\mathrm{yr}^{-1}&\mathrm{yr}^{-1} & \mathrm{yr}
		\\\hline
		\bf UPS$-$3.3 & 9894.96 & 0.208 & 177.0 & -2.12 & 0.638 & 0.510 & -14.1 & 0.119 & -0.143 & 0.57 & -3.14 & 0.50 & 1.58 \\
		\bf 1918.0 & ^{+0.04}_{-0.03} & ^{+0.004}_{-0.004} & ^{+2.6}_{-2.8} & <-1.69 & ^{+0.003}_{-0.004} & ^{+0.031}_{-0.031} & ^{+0.5}_{-0.5} & ^{+0.002}_{-0.003} & ^{+0.002}_{-0.002} & ^{+0.08}_{-0.11} & ^{+0.12}_{-0.10} & ^{+0.07}_{-0.06} & ^{+0.12}_{-0.09} \\\hline
		UPS$-$3.8 & 9894.88 & 0.197 & 185.1 & -2.28 & 0.651 & 0.143 & -10.8 & 0.128 & -0.138 & -0.47 & -1.44 & -0.06 & 2.02 \\
		1930.2 & ^{+0.03}_{-0.03} & ^{+0.004}_{-0.004} & ^{+2.7}_{-2.8} & <-1.96 & ^{+0.003}_{-0.003} & ^{+0.032}_{-0.032} & ^{+0.4}_{-0.4} & ^{+0.002}_{-0.002} & ^{+0.002}_{-0.002} & ^{+0.20}_{-0.19} & ^{+0.26}_{-0.20} & ^{+0.33}_{-0.24} & ^{+0.91}_{-1.01} \\\hline
		LPS$-$2.4 & 9895.28 & 0.216 & 173.1 & -2.58 & 0.619 & 4.150 & 1.1 & 0.115 & -0.142 & 4.80 & 10.93 & -1.95 & 0.30 \\
		1928.0 & ^{+0.13}_{-0.12} & ^{+0.004}_{-0.004} & ^{+2.8}_{-3.1} & <-1.47 & ^{+0.003}_{-0.003} & ^{+0.305}_{-0.268} & ^{+0.6}_{-0.7} & ^{+0.002}_{-0.002} & ^{+0.001}_{-0.001} & ^{+0.75}_{-1.45} & ^{+0.99}_{-0.52} & ^{+1.24}_{-0.83} & ^{+0.08}_{-0.15} \\\hline
		LFS$-$2.6 & 9895.27 & 0.194 & 189.9 & -1.56 & 0.639 & 2.776 & 1.1 & 0.127 & -0.140 & 3.63 & 9.07 & 4.37 & 0.52 \\
		1927.7 & ^{+0.11}_{-0.10} & ^{+0.006}_{-0.004} & ^{+4.6}_{-4.6} & ^{+0.09}_{-0.10} & ^{+0.003}_{-0.003} & ^{+0.486}_{-0.444} & ^{+1.2}_{-1.4} & ^{+0.003}_{-0.004} & ^{+0.002}_{-0.002} & ^{+0.57}_{-0.56} & ^{+0.62}_{-0.69} & ^{+0.44}_{-0.40} & ^{+0.06}_{-0.04} \\ \hline
		WPS$-$4.0 & 9894.91 & 0.193 & 188.2 & -3.29 & 1.487 & 0.099 & 172.2 & 0.129 & -0.136 & 0.20 & -0.01 & 9.68 & 0.65 \\
		1972.0 & ^{+0.03}_{-0.03} & ^{+0.003}_{-0.003} & ^{+2.6}_{-2.5} & <-2.61 & ^{+0.006}_{-0.006} & ^{+0.003}_{-0.003} & ^{+0.3}_{-0.3} & ^{+0.002}_{-0.002} & ^{+0.001}_{-0.001} & ^{+0.02}_{-0.02} & ^{+0.01}_{-0.01} & ^{+0.02}_{-0.01} & ^{+0.002}_{-0.002} \\\hline
	\end{tabular*}
	\tablecomments{
	The ``FS'' or ``PS'' contained in labels refer to solutions with significant finite source effect or solutions consistent with points source. For PS solutions, we report the 3 $\sigma$ upper limits in $\log \rho$ column.
	The wide solutions ($s>1$) are labeled with the prefix ``W''.
	The close solutions ($s<1$) are labeled with the prefix ``U''(upper planetary caustic) and ``L''(lower planetary caustic) based on which planetary caustic the source trajectory crosses/approaches. {The label and $\chi^2$ of the globally best-fit solution UPS$-$3.3 are highlighted in boldface.} }
\end{table*}

\begin{table*}
	\renewcommand{\arraystretch}{1.1}
	\caption{2L1S model parameters and $1\sigma$ uncertainties for $u_0-$. \label{tab:2L1S_u0m_orb_parameters}}
	\begin{tabular*}{\textwidth}{@{} M @{\extracolsep{\fill}} CCCCCCCCCCCCC@{} }
		\hline
		Model & t_0 & u_0 & t_\E &\log\rho & s & q & \alpha & \pi_\mathrm{E,\,N} & \pi_\mathrm{E,\,E} & w_1 & w_2 & w_3 & P_\perp\\
		$\chi^2$ & \left(\mathrm{HJD'}\right) &  & \left(\mathrm{day}\right)  &  &  & \times 10^3 & \left(\mathrm{deg}\right) &  & &\mathrm{yr}^{-1} &\mathrm{yr}^{-1}&\mathrm{yr}^{-1} & \mathrm{yr} \\ \hline
		LPS$-$3.0 & 9895.16 & -0.216 & 216.2 & -1.71 & 0.630 & 1.047 & 9.9 & 0.045 & -0.285 & 1.52 & 3.08 & 0.66 & 1.13 \\
		1918.5 & ^{+0.08}_{-0.08} & ^{+0.005}_{-0.006} & ^{+7.5}_{-7.2} & <-1.52 & ^{+0.004}_{-0.005} & ^{+0.041}_{-0.048} & ^{+0.3}_{-0.3} & ^{+0.004}_{-0.005} & ^{+0.001}_{-0.001} & ^{+0.08}_{-0.10} & ^{+0.08}_{-0.11} & ^{+0.15}_{-0.16} & ^{+0.13}_{-0.16} \\\hline
		LPS$-$3.9 & 9894.85 & -0.200 & 242.3 & -2.27 & 0.651 & 0.115 & 5.7 & 0.054 & -0.289 & -0.61 & -0.57 & 0.31 & 4.47 \\
		1928.8 & ^{+0.03}_{-0.02} & ^{+0.002}_{-0.002} & ^{+2.8}_{-3.1} & <-2.02 & ^{+0.002}_{-0.002} & ^{+0.014}_{-0.026} & ^{+0.2}_{-0.3} & ^{+0.001}_{-0.002} & ^{+0.001}_{-0.001} & ^{+0.09}_{-0.14} & ^{+0.11}_{-0.30} & ^{+0.24}_{-0.42} & ^{+0.76}_{-2.24} \\\hline
		UFS$-$2.6 & 9894.87 & -0.191 & 260.4 & -1.62 & 0.646 & 2.180 & -6.2 & 0.050 & -0.291 & 3.30 & -10.00 & 4.00 & 0.49 \\
		1923.5 & ^{+0.14}_{-0.13} & ^{+0.005}_{-0.005} & ^{+11.0}_{-9.0} & ^{+0.07}_{-0.13} & ^{+0.003}_{-0.004} & ^{+0.344}_{-0.365} & ^{+1.0}_{-0.8} & ^{+0.003}_{-0.003} & ^{+0.002}_{-0.001} & ^{+0.51}_{-0.51} & ^{+0.65}_{-0.53} & ^{+0.33}_{-0.42} & ^{+0.05}_{-0.04} \\\hline
		UPS$-$2.6 & 9894.76 & -0.209 & 227.1 & -2.77 & 0.629 & 2.871 & -5.5 & 0.052 & -0.280 & 4.84 & -10.56 & 1.40 & 0.28 \\
		1918.9 & ^{+0.30}_{-0.21} & ^{+0.006}_{-0.006} & ^{+8.2}_{-7.3} & <-1.62 & ^{+0.005}_{-0.004} & ^{+0.245}_{-0.208} & ^{+0.6}_{-0.5} & ^{+0.003}_{-0.004} & ^{+0.001}_{-0.001} & ^{+0.31}_{-0.28} & ^{+0.28}_{-0.33} & ^{+0.11}_{-0.18} & ^{+0.02}_{-0.03} \\\hline
		UPS$-$3.1 & 9895.32 & -0.198 & 244.6 & -1.86 & 0.648 & 0.738 & -2.0 & 0.047 & -0.287 & 1.19 & -6.94 & 1.73 & 0.79 \\
		1924.8 & ^{+0.25}_{-0.19} & ^{+0.004}_{-0.004} & ^{+6.7}_{-5.3} & <-1.63 & ^{+0.004}_{-0.003} & ^{+0.029}_{-0.038} & ^{+0.5}_{-0.5} & ^{+0.003}_{-0.004} & ^{+0.001}_{-0.001} & ^{+0.08}_{-0.11} & ^{+0.13}_{-0.12} & ^{+0.06}_{-0.07} & ^{+0.03}_{-0.02} \\\hline
		WPS$-$4.8 & 9894.94 & -0.193 & 253.4 & -3.37 & 1.496 & 0.023 & -176.7 & 0.055 & -0.287 & 0.96 & -2.01 & 6.56 & 0.91 \\
		1983.6 & ^{+0.05}_{-0.05} & ^{+0.003}_{-0.003} & ^{+5.3}_{-5.4} & <-2.74 & ^{+0.007}_{-0.007} & ^{+0.020}_{-0.008} & ^{+0.2}_{-0.1} & ^{+0.002}_{-0.002} & ^{+0.001}_{-0.001} & ^{+0.15}_{-0.25} & ^{+0.10}_{-0.06} & ^{+1.06}_{-0.79} & ^{+0.10}_{-0.10} \\\hline
	\end{tabular*}
\end{table*}

\begin{figure*}[htbp]
	\epsscale{1.15}
	\plotone{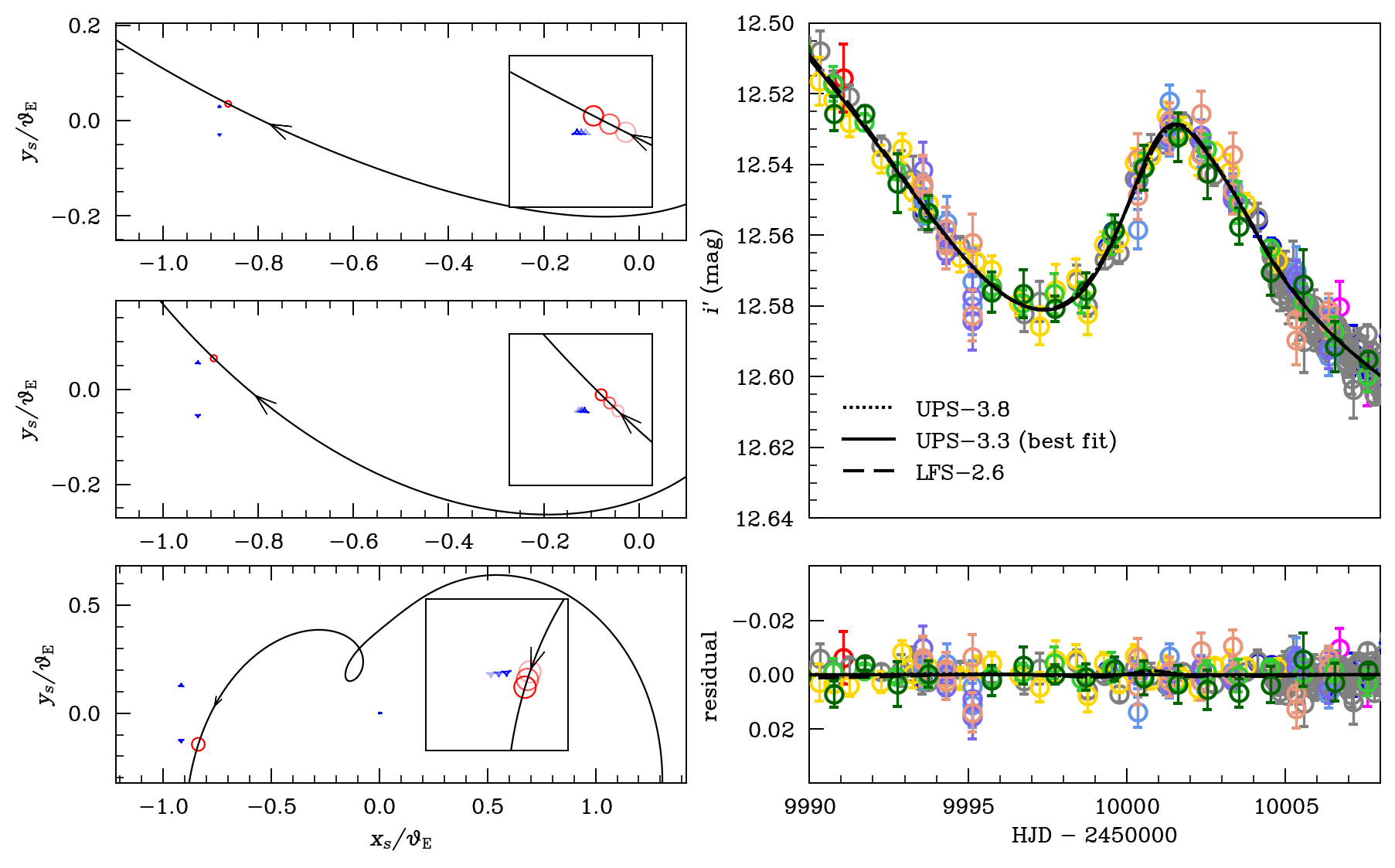}
	\caption{(For $u_0+$ cases) The source trajectories and light curves in the neighborhood of the planet anomaly for several $u_0+$ solutions. Left:
		The upper and middle panels show the caustics and source trajectories for solutions UPS$-3.8$, UPS$-3.3$ (the global best-fit solution) and LFS$-2.6$, respectively.
		The trajectories with arrows indicating the direction of motion are plotted with respect to the binary-lens axis. The insets of each sub-panel on the left display the caustics (blue) and source position (red) at $t_{0,\mathrm{kep}} - 1\,{\rm d}$, $t_{0,\mathrm{kep}}$ and $t_{0,\mathrm{kep}} + 1\,{\rm d}$ with increasing opacities. The radii of the red circles indicate the best-fit source sizes (upper limits) for the FS (PS) solutions.
		Right: The light curves and best-fit models of the four solutions (UPS$-3.8$: dotted line, UPS$-3.3$: solid lines, LFS$-2.6$: dashed lines)
		are shown in the upper panel. The residuals are displayed in the lower panel.
		\label{fig:traju0+}}
\end{figure*}
\begin{figure*}[htbp]
	\epsscale{1.15}
	\plotone{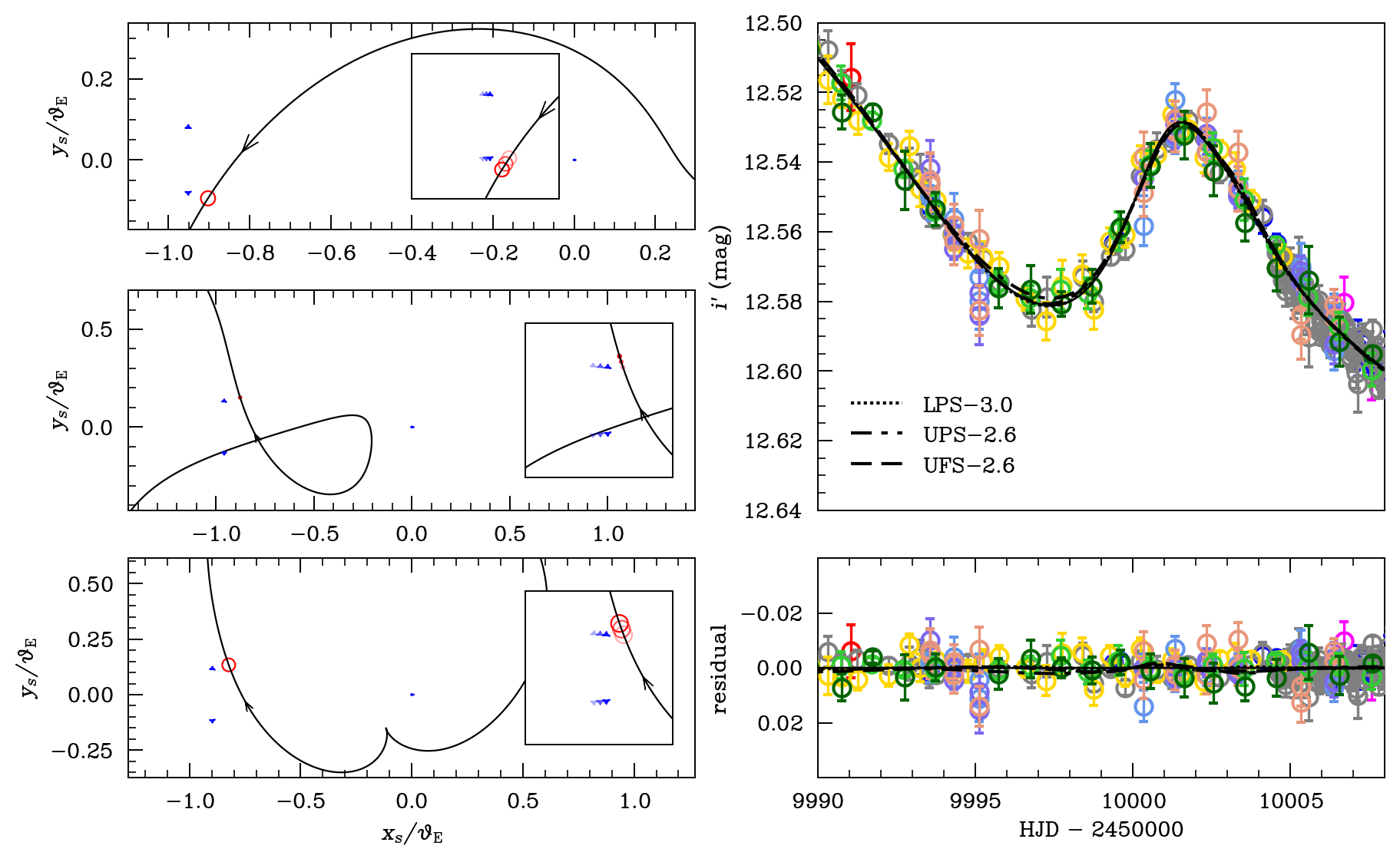}
	\caption{(For $u_0-$ cases) The source trajectories and light curves for three $u_0-$ solutions. Left: The source trajectories and caustics for the LPS$-3.0$ (upper), UPS$-2.6$ (middle) and UFS$-2.6$ (lower) solutions. Right: The light curves in the neighborhood of the planet anomaly and the best-fit models (upper) and corresponding residuals (lower) for the LPS$-3.0$ (dotted lines), UPS$-2.6$ (dotted-dashed lines) and UFS$-2.6$ (dashed lines) solutions.
		\label{fig:traju0-}}
\end{figure*}

\subsection{2L1S Modeling} \label{subsec:2L1S}
We start the 2L1S modeling analysis by employing the static binary lens model, which includes three parameters $(s,\, q,\, \alpha)$ to describe the binary systems: $s$ is the projected separation between the binary components normalized to the angular Einstein radius $\theta_\E$, $q$ is the binary mass ratio,
and $\alpha$ is the angle between the source trajectory and the binary-lens axis. An additional parameter relative to the PSPL model is $\rho = \theta_{*} / \theta_\E$, which is the angular source radius $\theta_{*}$ normalized by $\theta_\E$, and it is used to model the finite-source effects when the light curve deviates from the point-source approximation ($\rho=0$).

We use the \texttt{VBBinaryLensing} package
\citep{2018MNRAS.479.5157B}
to compute the binary-lens magnification $ A \left( t;\, \Theta_{\rm 2L1S}\right)$ at a given set of model parameters $\Theta_{\rm 2L1S}$ and time $t$. \texttt{VBBinaryLensing} uses the optimized root solver by \citet{SkowronGould12} to perform the point source (PS) calculations, and the advanced contour integration algorithm \citep{2010MNRAS.408.2188B} for calculating the finite source (FS) effects.

We probe the parameter space of 2L1S models on a fixed
$\left(\log s,\,\log q ,\, \alpha,\, \log \rho\right)$ grid of $-0.8 \le \log s \le 0.8 ,\, -5 \le \log q \le 0 $ and $ 0 \le \alpha \le 2\pi$ with $81$, $51$ and $181$ equally spaced values, respectively. We set $\left(t_0 ,\,  u_0 ,\,  t_\mathrm{eff} \equiv u_0 t_\E\right)$ free, which are fitted using Markov Chain Monte Carlo (MCMC) with the \texttt{EMCEE} ensemble sampler
\citep{emcee}, and their initial values are seeded using their PSPL model estimates. The flux parameters $(f_{\Sc,i}, f_{\B,i})$ for site $i$ are determined by linear fits \citep[see, e.g.,][]{Gould03}. The parallax parameters $\left(\pi_\mathrm{E,N} ,\, \pi _\mathrm{E,E}\right)$ are held fixed using the PSPL values.  Due to the two-fold degeneracy for microlens parallax, we perform two sets of grid searches for $u_0+$ and $u_0-$ solutions separately.

We conduct the 2L1S grid search for PS models and FS models fixed at $\log\rho=\{-2.5,\, -1.5\}$, respectively. Figure~\ref{fig:grid_search} shows the PS grid search results with multiple minima in the planetary regime ($\log q\lesssim -2$). The FS grid searches yield local minima in similar $(\log s,\,\log q ,\, \alpha)$ regions.
The $u_0+$ and $u_0-$ grid-search search results, as separately shown in the upper and lower panels of Figure~\ref{fig:grid_search}, exhibit different patterns. The light-curve profile of the planetary anomaly critically depends on the geometry of the source-lens trajectory when crossing/passing by the caustics, which differs for the $u_0+$ and $u_0-$ cases due to the presence of strong microlens parallax, giving rise to different perturbations.

We next analyze the binary orbital motion, which can also introduce significant distortions in the source-lens trajectory. We assume a circular orbit and use the parameterization adopted in \texttt{VBBinaryLensing}, which  includes three parameters $\left(w_1, \, w_2,\, w_3\right)$ describing the first derivatives of $s$, $\alpha$ and the relative radial velocity at a reference time  $t_{0, \mathrm{kep}}$, respectively:
\begin{equation*}
	w_1  = \frac{1}{s}\frac{\mathrm{d} s}{\mathrm{d}t }, \quad
	w_2  = \frac{\mathrm{d} \alpha}{\mathrm{d}t }, \quad
	w_3  = \frac{1}{s}\frac{\mathrm{d} s_z}{\mathrm{d}t }.
\end{equation*}
See \citet{Skowron2011} and \citet{VBBLastrometry} for relevant discussions on the parameterization. We choose the reference time ($t_{0, \mathrm{kep}} = \mathrm{HJD}' = 10001.8$) to be around the peak of the planetary bump.
We seed all local minima from the static-binary grid searches to investigate the lens-orbital motion effect using MCMC. All the local minima are listed in Table~\ref{tab:2L1S_u0p_orb_parameters} and \ref{tab:2L1S_u0m_orb_parameters}, respectively.

We classify these solutions according to two types of degeneracies and label them accordingly in the first columns of Table~\ref{tab:2L1S_u0p_orb_parameters} and \ref{tab:2L1S_u0m_orb_parameters}.
The first type is a discrete degeneracy regarding the source size. For most solutions, the planetary bump is consistent with a cusp approach by a point source (i.e., $\rho=0$ at $<3\,\sigma$ level), and they have ``PS'' in the labels. In the table, we report the best-fit $\log\rho$ values and their $3\,\sigma$ upper limits. For other solutions, the source is resolved via caustic structures, corresponding to those with ``FS'' in the labels. The second type of degeneracy is related to the caustic topology.
There is a well-known degeneracy between close ($s<1$) and wide ($s>1$) binaries \citep{GriestSafizadeh1998, Dominik1999, An2005}. The wide-binary solutions are labeled with the prefix ``W''.
For a close binary, there are a pair of planetary caustics, and depending on whether the source trajectory crosses/approaches the upper or lower planetary caustics, we label the solution with the prefix ``U'' and ``L'', respectively.
We also distinguish the solutions by putting their $\log q$ values with one significant digit as the suffix. { There are 9 close-binary solutions, including 4 for $u_0+$ and 5 for $u_0-$, which are listed in Table~\ref{tab:2L1S_u0p_orb_parameters} and \ref{tab:2L1S_u0m_orb_parameters}, respectively. We also list the best-fit wide-binary solutions for $u_0+$ and $u_0-$ in the tables, respectively. Since all the wide-binary solutions are disfavored with $\Delta \chi^2 > 50$, we do not include them in the subsequent analysis.}

The $\chi^2$ value for each solution is given below its label in the first columns of Table~\ref{tab:2L1S_u0p_orb_parameters} and \ref{tab:2L1S_u0m_orb_parameters}. The globally best-fit solution UPS$-$3.3($u_0+$) is a point-source ($\rho<0.02$ at the 3\,$\sigma$ level), close-binary model with $q\approx 5\times10^{-4}$. { There are 5 other solutions with $\Delta \chi^2 < 9$ relative to UPS$-$3.3($u_0+$). Among them, two point-source close-binary solutions have $\Delta \chi^2 < 1$, namely, LPS$-$3.0($u_0-$, $\Delta \chi ^2=0.5$) and UPS$-$2.6($u_0-$, $\Delta \chi ^2=0.9$), and the best-fit finite-source close-binary solution UFS$-$2.6($u_0-$) has $\Delta \chi ^2=4.5$}. All 9 solutions have $\Delta\chi^2<12$, with mass ratios covering factor of $\sim 30$ $(1.5\times10^{-4}\lesssim q \lesssim 4\times10^{-3})$. We display several models and their corresponding source trajectories with their respective caustics for the $u_0+$ and $u_0-$ cases in Figure~\ref{fig:traju0+} and \ref{fig:traju0-}, respectively.

The planetary anomaly occurred follows a long interval $(\sim 100\, \mathrm{d})$ after the event peak, which is a substantial fraction of the orbital period (e.g., circular orbital period $P_{\rm circ}=2.96\pm 0.20$\,yr for the solution UPS-3.3($u_0+$)). This confirms the necessity to use the full orbital parametrization with three parameters rather than the linear approximation with two parameters. The orbital motion parameters may sometimes be strongly correlated with binary-lens parameters; for example, $q$ differs significantly between static models and those considering binary orbital motions in the analysis by \citet{Han2022}.
This raises a possible concern that our preceding grid search based on static binaries may miss local minima by not including orbital motion effects. The long-term deviations from rectilinear motion for Gaia22dkv are dominated by the strong microlens parallax, which is taken into account in the static binary search, so the orbital motion effects are unlikely to bias the solution search. Nevertheless, we address this possible concern by performing new grid searches on a unidimensional grid of $\log q$ with the lens-orbital motion effects included. The $\left(\log s,\, \alpha\right)$ are seeded based on the caustic geometry and the U/L degeneracies. We also attempt various initial $\log \rho$ values corresponding to the PS/FS degeneracy. However, no new local minima are found from our extensive searches.
\begin{figure}[htbp]
	\epsscale{1.15}
	\plotone{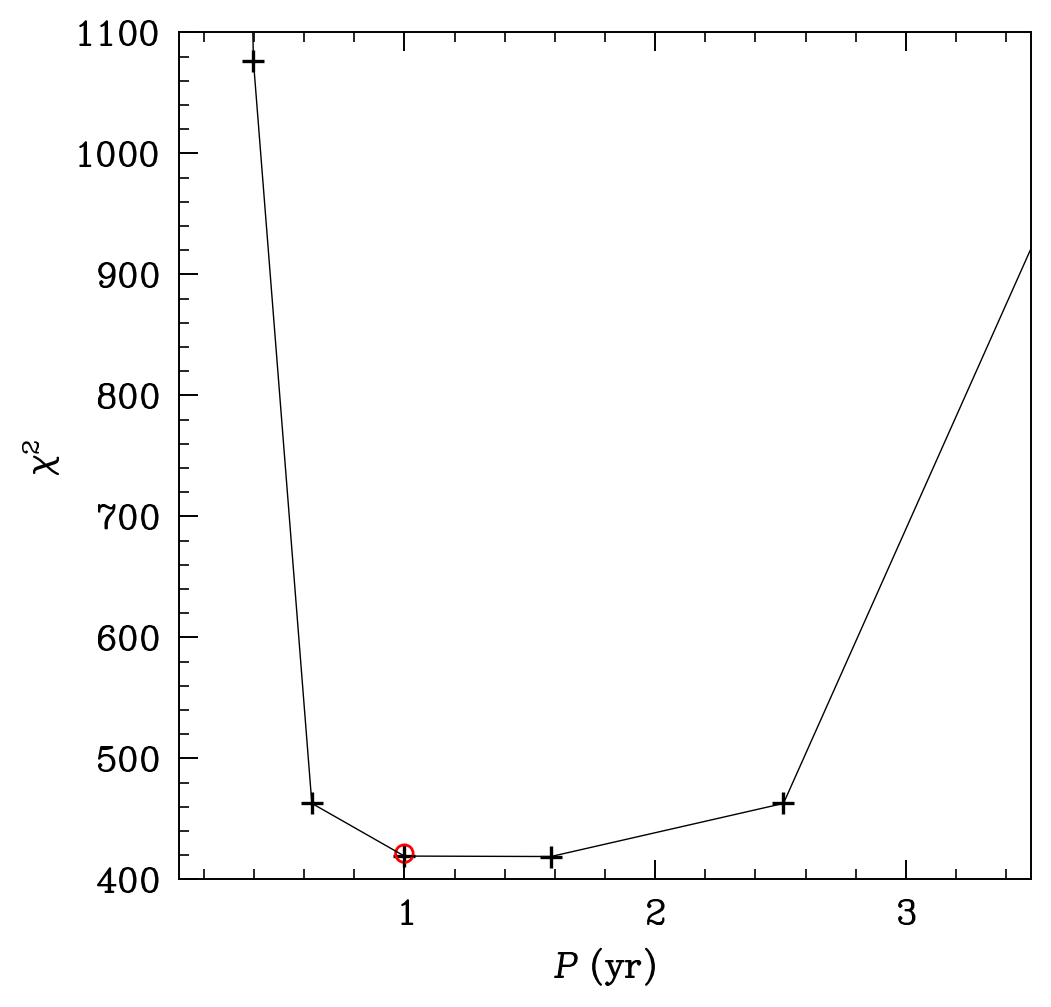}
	\caption{The black crosses connected by solid lines show the $\chi^2$ distributions for best-fit xallarap solutions at fixed binary-source orbital periods $P$. The red open circle at $P=1$\,yr shows the $\chi^2$ for the best-fit microlens parallax solution.
		\label{fig:P_chi2}}
\end{figure}
\begin{figure}[htbp]
	\epsscale{1.15}
	\plotone{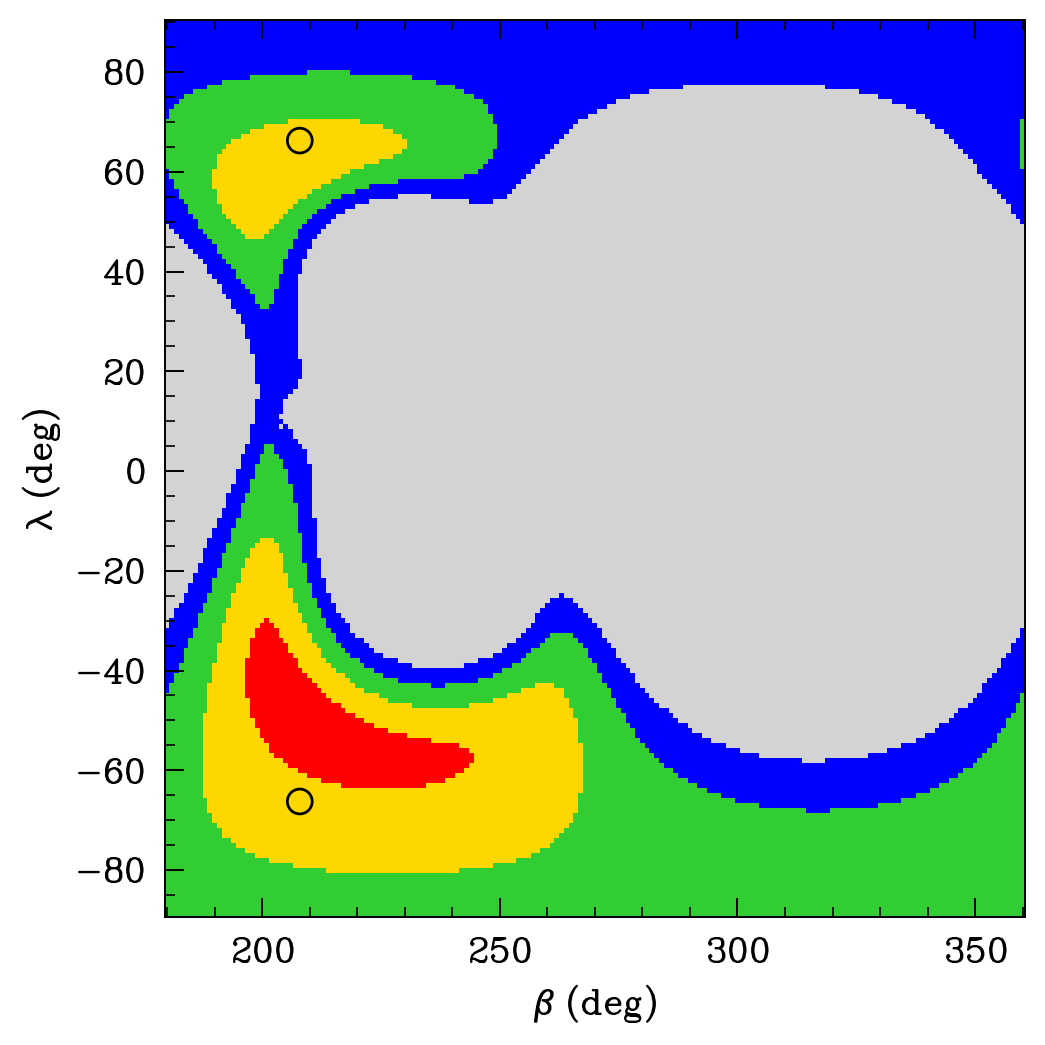}
	\caption{Results of xallarap fits by fixing $\beta$ and $\lambda$ at period $P=1\,\mathrm{yr}$ and setting $u_0 > 0$. The colors marked as red, yellow, green, blue and grey represent solutions with $\Delta \chi^2 < 1,\, 4,\, 9,\, 16$ and $25$ of the best fit.
		The ecliptic coordinates of Gaia22dkv, which represents the Earth parameter, are indicated by the black circles, with the lower and upper circles corresponding to the $u_0>0$ and $u_0<0$ cases due to perfect symmetry, respectively.
		\label{fig:xallarap_grid}}
\end{figure}
\subsection{Xallarap effects} \label{sec:xallarap}
If the source is in a binary system, its orbital motion can induce distortion of the microlensing light curve \citep{xallarap, HanGould1997}. This so-called xallarap effect can perfectly mimic parallax signals when mirroring Earth's orbital parameters \citep[see, e.g.,][]{Smith2002}. However, such finely-tuned parameters are {\it a priori} unlikely when randomly drawn from broadly distributed binary star orbital distributions. We follow the procedures of the test in \citet{Dong2009ob071} to assess whether the xallarap interpretation is preferable: the xallarap interpretation would be favored if the xallarap solution differed significantly from the Earth and having substantial $\chi^2$ improvements over microlens parallax. Otherwise parallax would be preferred.

We assume that the source is on a circular orbit, which requires 5 additional parameters to describe, namely
the period $P$ of the binary motion, the ecliptic longitude and latitude of the binary source orbit $\left(\lambda,\, \beta\right)$ and the xallarap vector's components $\left(\xi_\mathrm{E,N},\, \xi_\mathrm{E,E}\right)$.

The effects of binary-lens orbital motion can be entangled with parallax/xallarap effects, and to avoid such complications, we fit the light curve to PSPL models.
Like the PSPL+parallax modeling, we mask the light-curve region in the neighborhood of the anomaly region and perform PSPL+xallarap fitting.
We take advantage of the exact degeneracy
$\beta ' = \beta + 180^\circ,\, {\bdv\xi}_\E = -{\bdv\xi}_\E$
\citep{2005ApJ...633..914P}
and restrict our search to solutions with $180^\circ \le \beta < 360^\circ$, $-90^\circ<\lambda \le 90^\circ$.
We fix $\left(\lambda,\,\beta,\,\log P\right)$ on a dense grid and allow $\left(\xi_\mathrm{E,N},\, \xi_\mathrm{E,E}\right)$ together with PSPL parameters to vary freely in MCMC.
Compared to parallax, the best-fit xallarap solution improves the fit by $\Delta \chi^2 \sim 12$. We find that most of the improvement comes from dataset CHI-18 $R'$ ($\Delta \chi^2 \sim 9$). In contrast, there is only a minor improvement ($\Delta \chi^2 \sim 2$) in CHI-18 $B'$ despite their essentially identical coverages and similar photometric precision. This indicates low-level systematics commonly seen in parallax/xallarap modelings \citep{2005ApJ...633..914P}. To make a further check, we redo the modeling but with only Gaia and LCOGT/CTIO datasets. The best-fit xallarap solution improves the fit by a mere $\Delta \chi^2 = 2.5$ with extra 3 degrees of freedom (dof). Figure~\ref{fig:P_chi2} shows the $\chi^2$ values of best-fit xallarap solutions as a function of period $P$, and the $\chi^2$ minima are in the range of $P \in \left[1,\, 1.6\right]\,\mathrm{yr}$, which is consistent with Earth's orbital period of 1\,yr.
Figure~\ref{fig:xallarap_grid} shows the $\left(\lambda,\,\beta \right)$ $\chi^2$ map for the xallarap solutions at 1\,yr period, and the Earth's parameters (black open circle) have $\Delta \chi^2 < 4$.
Given the minor $\chi^2$ improvement and the similarity between xallarap parameters with those of the Earth,
we favor the microlens parallax interpretation over xallarap. Nevertheless, our test does not rule out xallarap or a combination of xallarap and parallax.

\subsection{1L2S Modeling} \label{subsec:1L2S}
In this section, we analyze the data using a 1L2S model, involving a stationary, luminous binary companion of the source and a single lens, to assess whether it can explain the short-duration anomaly. The magnification $A$ in the 1L2S model is the flux-weighted mean of the respective PSPL magnifications of the two sources,
\begin{equation}
	A_j=\frac{A_1+q_{F,\,j} A_2}{1+q_{F,\,j}},
\end{equation}
where $q_{F,\, j}$ is the flux ratio in the passband $j$ between the two sources. Note that the 1L2S magnification is wavelength-dependent unless the two sources have strictly the same color.

We first perform a linear regression between $R'$ and $B'$ band fluxes and find no evidence for deviation from a constant color, indicating that the two sources, if they existed, would have insignificant color differences.
Therefore, we use the same flux ratio $q_F$ for all bands. Since
the light curve is well-characterized by 2L1S models with no obvious residual; if the light curve in one band, say $B'$-band, is well-fitted by a 1L2S model, the other bands should achieve a similar goodness-of-fit with a flux ratio identical to $q_{F,\, B'}$.
For a 1L2S model to re-produces a short-duration planet-like bump, a high peak magnification of the fainter source is often needed, so we include the finite-source effect, which could be necessary for the high magnification.
The best-fit 1L2S model is worse than the best-fit planetary model {UPS$-$3.3}($u_0+$) by $\Delta \chi^2=3269$ and is shown (grey curve) in Figure~\ref{fig:best_fit}. In particular, the 1L2S model fails to explain the region preceding the rising wing of the anomaly, which is the characteristic negative excess of a planetary anomaly as discussed in  \cite{1998ApJ...506..533G}. We conclude that the 1L2S interpretation is ruled out.
\begin{table*}[htbp]
	\centering
	\renewcommand{\arraystretch}{1.1}
	\caption{Physical parameters derived with finite-source effects for FS solutions.\label{tab:fs_params}}

	\begin{tabular*}{\textwidth}{@{} c| @{\extracolsep{\fill}}M CCCCCCCCC@{} }
		\hline
		&{Solution}& M_{\rm h} & M_{p} & r_\perp & \mu_\mathrm{N,\,hel} & \mu_\mathrm{E,\,hel} & D_\L  & D_\Sc & \theta_\E & P_\perp\\
		& {} & M_\mathrm{\odot} & M_\mathrm{J} & \mathrm{AU} & \mathrm{mas\,yr}^{-1} & \mathrm{mas\,yr}^{-1} &  \mathrm{kpc} & \mathrm{kpc} & \mathrm{mas} & \mathrm{yr} \\
		\hline
		$u_0+$ & LFS$-$2.6 &
		0.303^{+0.076}_{-0.045} & 0.886^{+0.276}_{-0.166} & 0.48^{+0.11}_{-0.07} & 1.15^{+0.29}_{-0.17} & -0.76^{+0.12}_{-0.18} & 1.60^{+0.02}_{-0.03} & 1.86^{+0.04}_{-0.02} & 0.47^{+0.12}_{-0.07} & 0.60^{+0.13}_{-0.08} \\
		\hline
		$u_0-$ & UFS$-$2.6 &
		0.219^{+0.056}_{-0.039} & 0.507^{+0.128}_{-0.101} & 0.52^{+0.11}_{-0.08} & 1.09^{+0.28}_{-0.20} & -0.89^{+0.16}_{-0.23} & 1.51^{+0.04}_{-0.05} & 1.98^{+0.08}_{-0.05} & 0.53^{+0.13}_{-0.09} & 0.79^{+0.15}_{-0.12} \\
		\hline
	\end{tabular*}
\end{table*}

\begin{figure}[htbp]
	\epsscale{1.15}
	\plotone{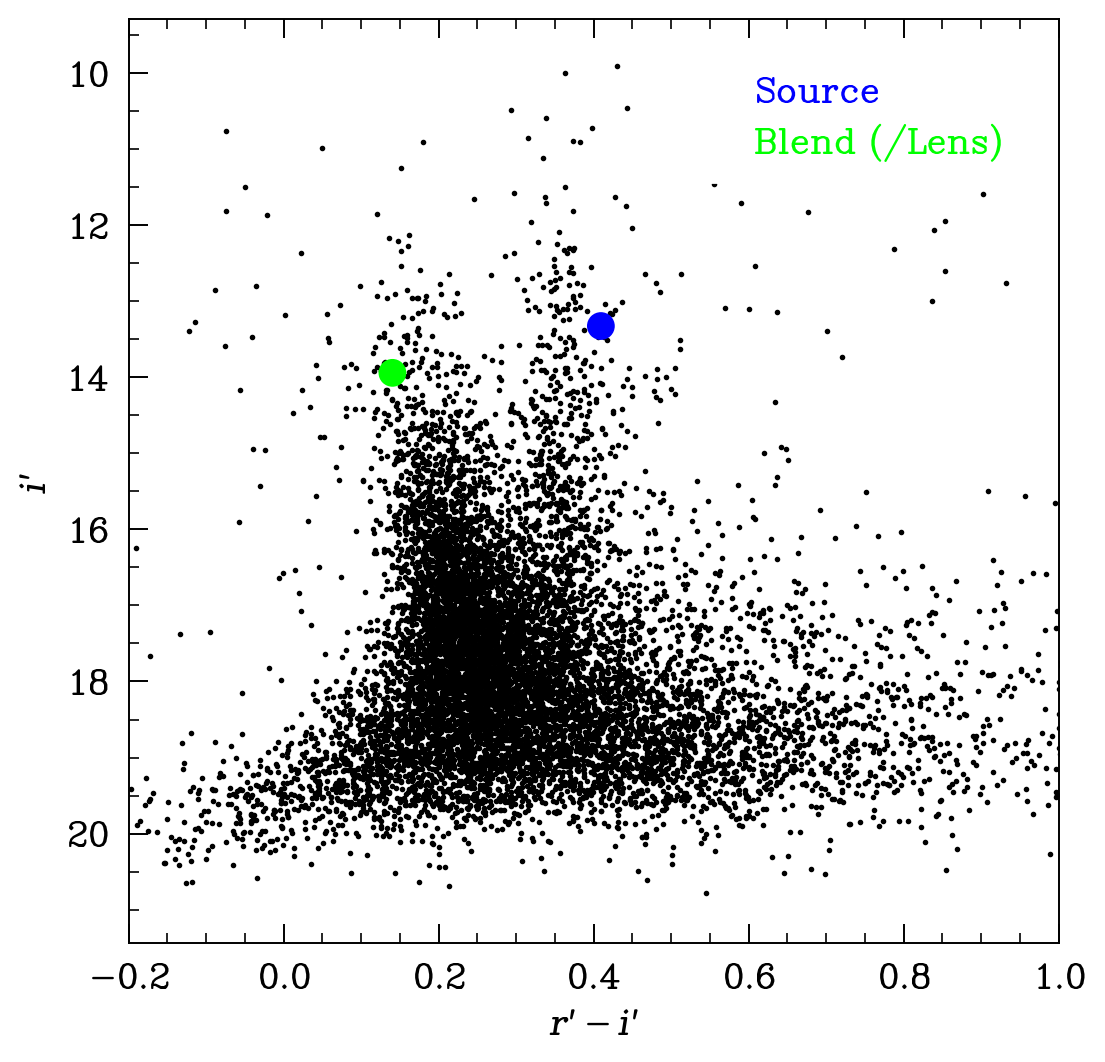}
	\caption{ The $r'-i'$ vs. $i'$ color-magnitude diagram (CMD) using stars in the LCOGT template images ($26.6\arcmin\times26.6\arcmin$), calibrated with the ATLAS All-Sky Stellar Reference Catalog (Refcat2; \citealt{Refcat2}) to the Pan-STARRS1 photometric system \citep{PS1}. The blue and green filled circles represent the source and the blend (i.e., the un-lensed light within the PSF), respectively. Our primary interpretation assumes that the blend is the lens. \label{fig:CMD}}
\end{figure}

\section{Physical Parameters} \label{sec:phy_param}
\subsection{Source Radius and the Finite-Source Effects} \label{subsec:theta_star}
The angular source size can be used as a ``ruler'' to measure or constrain the angular Einstein radius via the finite-source effects, $\theta_{\E} = {\theta_*}/{\rho}$. { We follow the standard procedure by using the source's de-reddened  color and magnitude to estimate $\theta_*$ \citep{Yoo2004}.
	We employ the three-dimensional extinction map by \citet{guo2021} to estimate the extinction corrections. The target is in a low-extinction region, and given the range of probable $D_S$ of $\sim2-6$\,kpc (see Tables~\ref{tab:fs_params}, \ref{tab:phy_params}, \ref{tab:phy_params_sp} and \ref{tab:phy_params_null} for estimates), the extinction map yields $E(B-V) = 0.132 ^{+0.025}_{-0.014}$.	Adopting the empirical selective-to-total extinction coefficients $\left(R_{i'},\,R_{r'}\right) = \left(1.791\,, 2.363\right)$ according to \citet{2023ApJS..264...14Z}  for our field, we estimated $A_{i'} = 0.236 ^{+0.045}_{-0.025}$ and $E(r'-i') = 0.075^{+0.014}_{-0.008}$.
		For the best-fit solution, we derive $\left(r'-i',\, i'\right)_\mathrm{s,\,0} = \left( 0.335^{+0.008}_{-0.014},\,  13.089 ^{+0.025}_{-0.045}\right)$. By using the surface brightness-color relation calibrated in the $i'$ and $r'$ bands \citep{2013ApJ...771...40B,2022AJ....164...75B},

		$\log _{10}\left(2 \theta_*\right)=(-0.298 \pm 0.044)(r'-i')_0^2 +(0.919 \pm 0.058)(r'-i')_0-0.2 i'_0+(0.767 \pm 0.010)$, we find that the source angular radius $\theta_* = 13.4 \pm 1.2 \,\mathrm{\mu as}$.
	}
For the FS solutions (i.e., in which $\rho$ is detected at high significance), the physical parameters can be unambiguously determined with
\begin{equation}
	\quad M_{\L}=\frac{\theta_{\E}}{\kappa \pi_{\E}}=\frac{\theta_*}{\kappa \pi_{\E}\rho}, \quad D_{\L}=\frac{\au}{\pi_{\L}}=\frac{\au}{\pi_{\E} \theta_{\E}+\pi_{\Sc}}.
\end{equation}
where $M_{\L}$ is the lens mass, $\pi_{\L}$ and $\pi_{\Sc}$ are the trigonometric parallaxes of the lens at distance $D_{\L}$ and source at distance $D_{\Sc}$, respectively, and $\kappa = 4\pi G/(c^2 \au ) = 8.14 \,{\rm mas}/M_\odot$ is a constant.
We list the derived physical parameters for these solutions in Table~\ref{tab:fs_params}. They all correspond to low-mass lenses with $\sim 0.2 - 0.3\,M_\odot$ at $\sim 1.5-1.6~\mathrm{kpc}$.
\subsection{Spectroscopic Source Distance} \label{subsec:src_dist}

\begin{figure}
	\centering
	\epsscale{1.15}
	\plotone{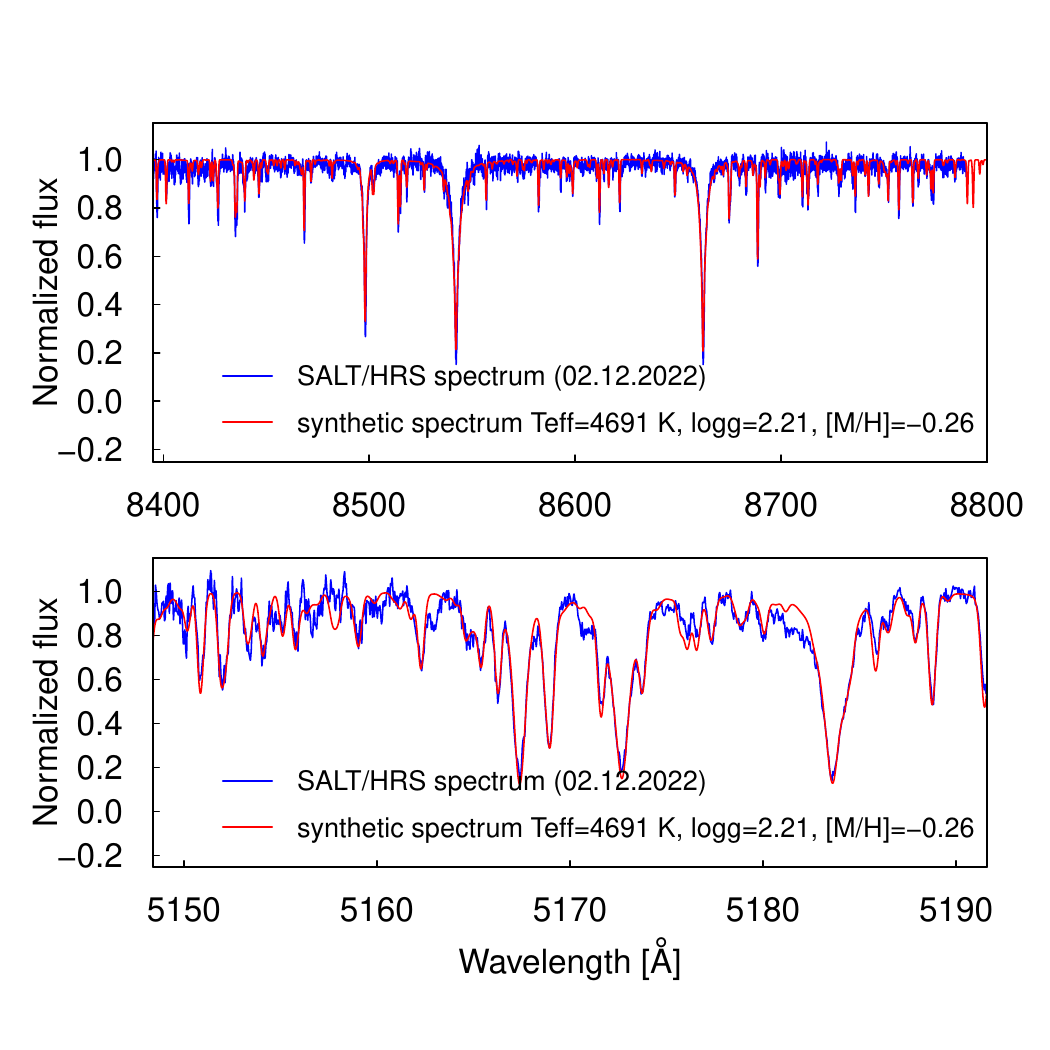}
	\caption{Upper: The spectrum from SALT/HRS ({\it blue}) and the best-fit synthetic spectrum ({\it red}). Lower: A zoomed region including Ca~II ({\it top}) and Mg~I ({\it bottom}).}
	\label{fig:spectra}
\end{figure}

{ We estimate the source distance $D_{\rm S}$ by using the SALT/HRS spectrum and the source fluxes.
The spectrum was taken at $A\sim 4.3$ and dominated by the source, allowing us to determine the atmospheric parameters of the source with minimal contamination.
We performed synthetic spectral fitting using the \texttt{iSpec}\footnote{\url{https://www.blancocuaresma.com/s/iSpec}} package \citep{BlancoCuaresma2014, BlancoCuaresma2019}, which integrates several well-known radiative transfer codes. We chose the SPECTRUM\footnote{\url{http://www.appstate.edu/~grayro/spectrum/spectrum.html}} code and generated a set of synthetic spectra based on a grid of MARCS atmospheric models \citep{Gustafsson2008} and solar abundances taken from \citet{Grevesse2007}. The synthetic spectra were fitted to the observed spectrum for $>300$ carefully selected lines (e.g., H, Ca, Mg, Fe, Na, Ti, etc.). The best-fit parameters are: $T_{\rm eff} = (4691 \pm 139)$~K, $\log g = (2.21 \pm 0.32)$, $\mathrm{[M/H]} = (-0.26 \pm 0.11)$~dex and $v_{\rm t} = (1.16 \pm 0.22)$~km~s$^{-1}$.
The spectrum and the fitted atmospheric parameters of the source are presented in Figure~\ref{fig:spectra}.
We then estimated the source luminosity by feeding these parameters to the theoretical stellar isochrones.
We employ the \texttt{MIST} package \cite[][]{2016ApJ...823..102C} and compute the atmospheric parameters for stars with masses between $0.1$ and $2\,\rm M_\odot$, stellar ages between $1$ and $15\,\rm Gyr$ and metallicities $\rm [Fe/H]$ between $-1.0$ and $ 0.5$. The stars matching with the measured spectroscopic  parameters span a wide range of luminosities, with $M_{r'}=-0.34^{+1.39}_{-0.44}$ and $M_{i'}=-0.60^{+1.38}_{-0.45}$, respectively. The source's color $M_{r'}-M_{i'}$ is well constrained from the isochrone analysis  as $0.26 \pm 0.02$, which is  bluer than the de-reddened color $(r'-i')_{s,\,0} = 0.335 ^{+0.008}_{-0.014}$ from our CMD analysis in \S~\ref{subsec:theta_star} at the $3 \sigma$ level, suggesting a possible bias in the derived spectroscopic parameters. A possible origin for the bias is the contamination from the blended light in spite of its low-level contribution ($\lesssim 20\%$), and we plan to carefully investigate it in a future work. Given this issue, we do not use color constraint directly when estimating the source distance. Combining with the source fluxes and extinction coefficients adopted in \S~\ref{subsec:theta_star}, we estimate the source distance  $D_S = 5.56^{+1.28}_{-2.36} \,\rm kpc$ for $i'$-band and $D_S = 5.71^{+1.28}_{-2.70}\,\rm kpc$ for $r'$ band, respectively.
}

\subsection{The Blended Light} \label{subsec:blend_lens}
From the light-curve analysis, we detect significant blending (the green filled circle in Figure~\ref{fig:CMD}), which is nearly equally bright in Gaia $G$ ($\sim 14$) as the source. The blended light can be due to the lens, a binary companion to the lens or the source, a random interloper within the PSF, or some combination of these possibilities.

From the Gaia Data Release 3 (DR3) catalog \citep{Gaia2016, GaiaDR3}, the number density is $9.8\times 10^{-5}\,\mathrm{arcsec}^{-2}$ for nearby field stars with $G < 14$. Thus the probability of a random field star falling in within $1$\arcsec\,of the source is $\approx 3 \times 10^{-4}$, suggesting that the blend is unlikely due to a random interloper.

In the following section, we assume that the blend is the lens and subsequently estimate the lens mass. We briefly discuss the possibilities of the blend as a companion of the source or the lens in \S~\ref{sec:summary}.

\subsection{Blend as the Lens}\label{subsec:blendaslens}
As discussed in \S~\ref{subsec:theta_star}, the FS solutions yield low-mass lenses at $\sim 1.5-1.6~\mathrm{kpc}$, which are far too faint compared to the blended flux. In this section, we consider the scenario in which blend is the lens for the PS solutions.

\begin{table*}[htbp]
	\centering
	\renewcommand{\arraystretch}{1.1}
	\caption{Physical parameters derived with blend flux and Gaia parallax.\label{tab:phy_params}}

	\begin{tabular*}{\textwidth}{@{} c| @{\extracolsep{\fill}}M CCCCCCCCC@{} }
		\hline
		&{Solution}& M_{\rm h} & M_{p} & r_\perp & \mu_\mathrm{N,\,hel} & \mu_\mathrm{E,\,hel} & D_\L  & D_\Sc & \theta_\E & P_\perp\\
		& {} & M_\mathrm{\odot} & M_\mathrm{J} & \mathrm{AU} & \mathrm{mas\,yr}^{-1} & \mathrm{mas\,yr}^{-1} &  \mathrm{kpc} & \mathrm{kpc} & \mathrm{mas} & \mathrm{yr} \\
		\hline
		\multirow{3}{*}{$u_0+$}
		& UPS$-3.3$ & 1.23^{+0.03}_{-0.09} & 0.67^{+0.05}_{-0.08} & 1.53^{+0.13}_{-0.13} & 4.61^{+0.28}_{-0.47} & -3.33^{+0.34}_{-0.20} & 1.29^{+0.08}_{-0.06} & 2.32^{+0.30}_{-0.26} & 1.86^{+0.07}_{-0.16} & 1.70^{+0.20}_{-0.19} \\
		& UPS$-3.8$ & 1.16^{+0.03}_{-0.04} & 0.17^{+0.06}_{-0.05} & 1.53^{+0.13}_{-0.14} & 4.45^{+0.23}_{-0.24} & -2.97^{+0.16}_{-0.15} & 1.32^{+0.06}_{-0.07} & 2.37^{+0.32}_{-0.29} & 1.78^{+0.07}_{-0.08} & 1.75^{+0.21}_{-0.21} \\
		& LPS$-2.4$ & 1.23^{+0.03}_{-0.04} & 5.36^{+0.34}_{-0.36} & 1.48^{+0.10}_{-0.13} & 4.53^{+0.18}_{-0.20} & -3.36^{+0.15}_{-0.13} & 1.30^{+0.06}_{-0.07} & 2.31^{+0.25}_{-0.26} & 1.84^{+0.06}_{-0.07} & 1.63^{+0.16}_{-0.18} \\

		\hline

		\multirow{4}{*}{$u_0-$}
		& LPS$-3.0$ & 1.13^{+0.03}_{-0.05} & 1.23^{+0.07}_{-0.08} & 1.60^{+0.10}_{-0.08} & 5.46^{+0.29}_{-0.32} & -5.20^{+0.32}_{-0.28} & 0.95^{+0.04}_{-0.04} & 3.54^{+0.83}_{-0.53} & 2.66^{+0.10}_{-0.12} & 1.90^{+0.15}_{-0.13} \\
		& UPS$-3.1$ & 1.04^{+0.04}_{-0.03} & 0.80^{+0.05}_{-0.05} & 1.63^{+0.09}_{-0.05} & 5.06^{+0.25}_{-0.22} & -4.34^{+0.18}_{-0.21} & 1.02^{+0.04}_{-0.04} & 3.80^{+0.79}_{-0.35} & 2.47^{+0.10}_{-0.08} & 2.05^{+0.15}_{-0.08} \\
		& UPS$-2.6$ & 1.10^{+0.05}_{-0.05} & 3.27^{+0.38}_{-0.34} & 1.59^{+0.08}_{-0.10} & 5.24^{+0.37}_{-0.34} & -4.82^{+0.35}_{-0.38} & 0.99^{+0.04}_{-0.05} & 3.55^{+0.61}_{-0.64} & 2.55^{+0.14}_{-0.14} & 1.91^{+0.13}_{-0.17} \\
		& LPS$-3.9$ & 1.05^{+0.05}_{-0.03} & 0.15^{+0.01}_{-0.01} & 1.62^{+0.10}_{-0.04} & 5.25^{+0.26}_{-0.17} & -4.58^{+0.15}_{-0.23} & 0.99^{+0.05}_{-0.03} & 3.76^{+0.93}_{-0.32} & 2.52^{+0.13}_{-0.08} & 2.01^{+0.17}_{-0.07} \\

		\hline
	\end{tabular*}
\end{table*}

\begin{table*}[htbp]
	\centering
	\renewcommand{\arraystretch}{1.1}
	\caption{Physical parameters derived with blend flux and \label{tab:phy_params_sp} source spectroscopic distance.}

	\begin{tabular*}{\textwidth}{@{} c| @{\extracolsep{\fill}}M CCCCCCCCC@{} }
		\hline
		&{Solution}& M_{\rm h} & M_{p} & r_\perp & \mu_\mathrm{N,\,hel} & \mu_\mathrm{E,\,hel} & D_\L  & D_\Sc & \theta_\E & P_\perp\\
		& {} & M_\mathrm{\odot} & M_\mathrm{J} & \mathrm{AU} & \mathrm{mas\,yr}^{-1} & \mathrm{mas\,yr}^{-1} &  \mathrm{kpc} & \mathrm{kpc} & \mathrm{mas} & \mathrm{yr} \\
		\hline
		\multirow{3}{*}{$u_0+$}
		& UPS$-3.3$ & 1.29^{+0.03}_{-0.05} & 0.68^{+0.04}_{-0.04} & 2.08^{+0.16}_{-0.18} & 4.83^{+0.14}_{-0.17} & -3.44^{+0.12}_{-0.12} & 1.66^{+0.09}_{-0.09} & 4.24^{+0.94}_{-0.63} & 1.95^{+0.05}_{-0.07} & 2.64^{+0.27}_{-0.29} \\
		& UPS$-3.8$ & 1.30^{+0.03}_{-0.05} & 0.20^{+0.04}_{-0.04} & 2.04^{+0.12}_{-0.20} & 4.97^{+0.11}_{-0.20} & -3.25^{+0.15}_{-0.10} & 1.57^{+0.07}_{-0.10} & 3.83^{+0.59}_{-0.63} & 1.99^{+0.04}_{-0.08} & 2.55^{+0.20}_{-0.32} \\
		& LPS$-2.4$ & 1.29^{+0.04}_{-0.06} & 5.67^{+0.39}_{-0.38} & 2.01^{+0.16}_{-0.22} & 4.73^{+0.15}_{-0.22} & -3.47^{+0.17}_{-0.12} & 1.68^{+0.10}_{-0.12} & 4.17^{+0.92}_{-0.74} & 1.92^{+0.06}_{-0.09} & 2.52^{+0.28}_{-0.35} \\

		\hline
		\multirow{4}{*}{$u_0-$}
		& UPS$-2.6$ & 1.04^{+0.03}_{-0.02} & 3.17^{+0.21}_{-0.22} & 1.66^{+0.10}_{-0.12} & 4.95^{+0.15}_{-0.12} & -4.49^{+0.16}_{-0.20} & 1.09^{+0.05}_{-0.06} & 4.35^{+1.35}_{-0.80} & 2.42^{+0.07}_{-0.05} & 2.09^{+0.17}_{-0.20} \\
		& UPS$-3.1$ & 1.03^{+0.02}_{-0.02} & 0.80^{+0.03}_{-0.04} & 1.70^{+0.09}_{-0.11} & 4.99^{+0.09}_{-0.11} & -4.34^{+0.11}_{-0.12} & 1.07^{+0.04}_{-0.05} & 4.53^{+1.19}_{-0.83} & 2.45^{+0.04}_{-0.05} & 2.17^{+0.15}_{-0.19} \\
		& LPS$-3.0$ & 1.06^{+0.05}_{-0.03} & 1.15^{+0.09}_{-0.07} & 1.64^{+0.09}_{-0.12} & 5.13^{+0.23}_{-0.17} & -4.87^{+0.23}_{-0.34} & 1.03^{+0.05}_{-0.07} & 4.19^{+1.17}_{-0.82} & 2.51^{+0.11}_{-0.08} & 2.04^{+0.16}_{-0.20} \\
		& LPS$-3.9$ & 1.04^{+0.01}_{-0.01} & 0.15^{+0.00}_{-0.00} & 1.68^{+0.02}_{-0.02} & 5.17^{+0.06}_{-0.05} & -4.45^{+0.05}_{-0.05} & 1.04^{+0.01}_{-0.01} & 4.29^{+0.27}_{-0.17} & 2.49^{+0.02}_{-0.02} & 2.14^{+0.04}_{-0.03} \\
		\hline
	\end{tabular*}
\end{table*}

\begin{table*}[htbp]
	\centering
	\renewcommand{\arraystretch}{1.1}
	\caption{Physical parameters derived with blend flux only.\label{tab:phy_params_null}}
	\begin{tabular*}{\textwidth}{@{} c| @{\extracolsep{\fill}}M CCCCCCCCC@{} }
		\hline
		&{Solution}& M_{\rm h} & M_{p} & r_\perp & \mu_\mathrm{N,\,hel} & \mu_\mathrm{E,\,hel} & D_\L  & D_\Sc & \theta_\E & P_\perp\\
		& {} & M_\mathrm{\odot} & M_\mathrm{J} & \mathrm{AU} & \mathrm{mas\,yr}^{-1} & \mathrm{mas\,yr}^{-1} &  \mathrm{kpc} & \mathrm{kpc} & \mathrm{mas} & \mathrm{yr} \\

		\hline
		\multirow{3}{*}{$u_0+$}
		& UPS$-3.3$ & 1.145^{+0.155}_{-0.078} & 0.591^{+0.152}_{-0.052} & 1.40^{+0.77}_{-0.34} & 4.23^{+0.79}_{-0.35} & -3.09^{+0.29}_{-0.52} & 1.27^{+0.43}_{-0.25} & 2.12^{+2.59}_{-0.65} & 1.71^{+0.28}_{-0.12} & 1.54^{+1.26}_{-0.50} \\
		& UPS$-3.8$ & 1.288^{+0.046}_{-0.272} & 0.250^{+0.040}_{-0.139} & 2.04^{+0.23}_{-1.11} & 4.92^{+0.35}_{-1.14} & -3.18^{+0.70}_{-0.27} & 1.58^{+0.10}_{-0.65} & 3.82^{+1.08}_{-2.56} & 1.97^{+0.10}_{-0.43} & 2.56^{+0.39}_{-1.67} \\
		& LPS$-2.4$ & 1.273^{+0.050}_{-0.200} & 5.657^{+0.540}_{-1.340} & 1.96^{+0.21}_{-1.11} & 4.70^{+0.30}_{-0.87} & -3.54^{+0.68}_{-0.17} & 1.66^{+0.11}_{-0.81} & 3.95^{+1.10}_{-2.80} & 1.90^{+0.10}_{-0.32} & 2.44^{+0.37}_{-1.68} \\

		\hline
		\multirow{4}{*}{$u_0-$}
		& UPS$-2.6$ & 1.077^{+0.080}_{-0.053} & 3.403^{+0.321}_{-0.571} & 1.40^{+0.42}_{-0.23} & 5.09^{+0.55}_{-0.31} & -4.62^{+0.26}_{-0.59} & 0.89^{+0.23}_{-0.12} & 2.43^{+4.05}_{-0.80} & 2.50^{+0.20}_{-0.14} & 1.60^{+0.75}_{-0.36} \\
		& UPS$-3.1$ & 1.046^{+0.036}_{-0.092} & 0.822^{+0.052}_{-0.105} & 1.68^{+0.14}_{-0.56} & 5.09^{+0.24}_{-0.57} & -4.45^{+0.50}_{-0.19} & 1.04^{+0.07}_{-0.28} & 4.21^{+1.83}_{-2.67} & 2.49^{+0.09}_{-0.24} & 2.13^{+0.26}_{-0.92} \\
		& LPS$-3.0$ & 1.157^{+0.016}_{-0.074} & 1.298^{+0.047}_{-0.151} & 1.69^{+0.13}_{-0.31} & 5.54^{+0.26}_{-0.46} & -5.36^{+0.48}_{-0.26} & 0.98^{+0.09}_{-0.12} & 4.27^{+2.16}_{-1.95} & 2.72^{+0.05}_{-0.19} & 2.04^{+0.26}_{-0.49} \\
		& LPS$-3.9$ & 1.050^{+0.045}_{-0.067} & 0.153^{+0.007}_{-0.012} & 1.61^{+0.16}_{-0.44} & 5.26^{+0.27}_{-0.36} & -4.50^{+0.33}_{-0.21} & 0.98^{+0.06}_{-0.26} & 3.61^{+1.60}_{-2.08} & 2.51^{+0.11}_{-0.16} & 1.99^{+0.27}_{-0.76} \\
		\hline
	\end{tabular*}
\end{table*}

For the PS solutions (i.e., in which $\rho$ is consistent with zero at $<3\,\sigma$), we can constrain the lens mass using the measured blended fluxes and $\pi_\E$.
The $\pi_\E$ measurement constrains $M_\L$ and $\pi_{\rm rel}$:
\begin{equation}
	\label{eq:microlensplx}
	\pi_\E= \sqrt{\frac{\pi_\mathrm{rel}}{\kappa M_\L}} = \sqrt{\left(\frac{1}{D_\L} - \frac{1}{D_\Sc}\right) \frac{\au}{\kappa M_\L}},
\end{equation}
where $\pi_{\rm rel}$ is the relative lens-source parallax. The trigonometric parallax of the baseline object measured by Gaia DR3 is $\pi_\text{base} = 0.58 \pm 0.05 \,\mathrm{mas}$,
which is the flux-weighted mean parallax by the lens and source \citep{Dong2019},
\begin{equation}
	\label{eq:pi_base}
	\pi_\text{base} = \eta_G \pi_\text{L} + (1-\eta_G)\pi_\text{S},
\end{equation}
where $\eta_G = f_{\B,G}/(f_{\Sc,G} + f_{\B,G})$ is the fraction of blended flux in Gaia $G$. Gaia DR3 reports that it has a high renormalized unit weight error ${\rm RUWE} = 4.38 \gg 1$,  indicating significant departures from the astrometric fitting \citep{Gaiaastrometry}. The amplitude of astrometric microlensing as discussed in \S~\ref{sec:astrometric} is too small to be responsible for the large RUWE. In particular, the Gaia DR3 measurements are based on data collected between 25 July 2014 and 28 May 2017, during which, the microlensing astrometric signals were negligible.

Without the single-epoch observations, we cannot assess how the Gaia DR3 parallax is biased.
	{ In the following, we first carry out our analysis by adopting the Gaia DR3 parallax measurement. Then we also discard the Gaia DR3 parallax and adopt the SALT/HRS spectroscopic source distance.}

\begin{figure}[htbp]
	\epsscale{1.15}
	\plotone{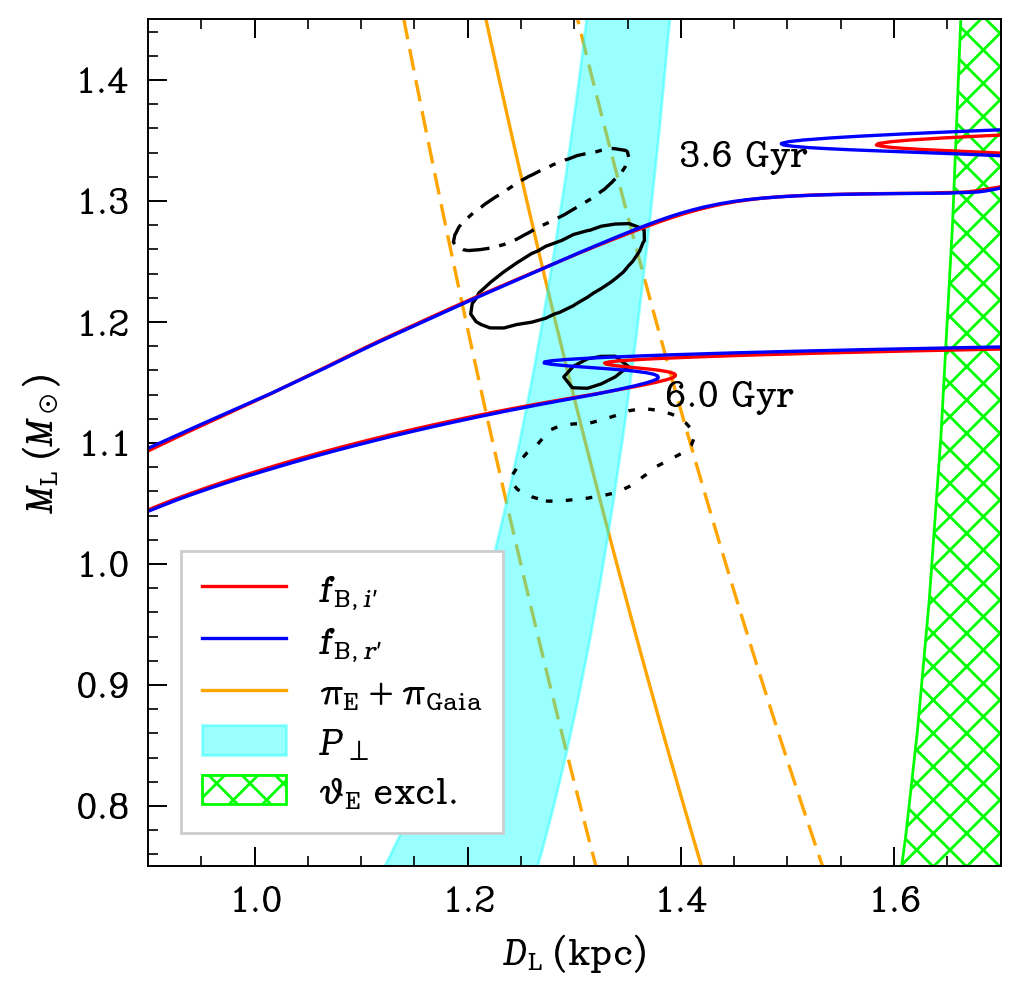}
	\caption{Physical constraints on $M_\L - D_\L$ for the best-fit solution. The contours are $1\,\sigma$ constraints by jointly fitting lens fluxes using isochrones, microlens parallax $\pi_{\rm E}$ and Gaia parallax $\pi_{\rm base}$ . The dotted, solid and dashed-dotted contours are constrained by isochrones at [Fe/H] = $-0.2$, $0$ and $0.2$, respectively. The red and blue lines show the isochrones for constraining fluxes in $i'$ and $r'$ bands, respectively. We display isochrones with stellar ages at 3.6 Gyr and 6 Gyr, which are the $1\,\sigma$ lower and upper limit of the posterior, respectively. The results are consistent with those from circular orbital motion using $P_\perp$ (blue). The green shaded region shows the $3\,\sigma$ exclusion region using the upper limit on $\theta_\E$ by fitting the finite-source effects, and it is also in consistency with the lens' physical constraints.
		\label{fig:best_constraints}}
\end{figure}

The measured blend fluxes place another constraint on $D_\L$ and $M_\L$. We employ the theoretical stellar isochrones in \texttt{MIST} package to compute the expected lens luminosity for a given lens mass $M_\L$ at a certain stellar age and metallicity [Fe/H]. We sample the MCMC posteriors of blended fluxes in $i'$ and $r'$ bands from the light-curve fitting and add the $\chi^2$ compensations using the measured blended fluxes in both $r'$ and $i'$ bands and expected lens fluxes on a grid of stellar ages and metallicities $\left(\mathrm{Age},\,\mathrm{[Fe/H]}\right)$.
We adopt two error terms $\left(e_\mathrm{cal},\,e_\mathrm{iso}\right)= \left(0.01,\, 0.01 \right)\,\mathrm{mag}$ to take into account of the photometric calibration errors and the theoretical uncertainties of the isochrones. We use the three-dimensional extinction map by \citet{guo2021} to estimate the extinction corrections at a given lens distance.
We impose flat priors in metallicities in the range of $\mathrm{[Fe/H]}\in \left\{ -0.2,\, 0,\, 0.2\right\}$ and in stellar age between 1 and 15 $\mathrm{Gyr}$.
{We also impose the constraint on $D_\Sc$ with measured source fluxes in $r'$ and $i'$.}
The $u_0+$ solutions generally yield lens masses $M_\L \sim 1.2\,M_\odot$ and distances $D_\L\sim 1.3\,{\rm kpc}$, whereas the $u_0-$ solutions favor $M_\L\sim 1.1\,M_\odot$  at $D_\L\sim 1
	.0\,{\rm kpc}$. Figure~\ref{fig:best_constraints} illustrates the constraints on $M_\L - D_\L$ from the flux and the $\pi_{\rm E}$ constraints for the globally best-fit solution, and the best-fit $M_\L$ and $D_\L$ are compatible with the lower limit on $\theta_{\rm E}$  from the $\rho$ upper limit. The metallicities and stellar ages are loosely constrained by the photometric data alone. The best-fit physical parameters derived using Gaia parallax are given in Table~\ref{tab:phy_params}.

{
We perform an alternative analysis by replacing the constraint of Equation~\ref{eq:pi_base} with the source spectroscopic distance estimated in \S~\ref{subsec:src_dist}. The corresponding physical parameters are given in Table~\ref{tab:phy_params_sp}. The estimated lens mass ($M_\L \sim 1.3\, M_\odot$ for $u_0+$ solutions) is broadly consistent with that derived from Gaia parallax, while the lens distance ($D_\L \sim 1.7\,\rm kpc$ for $u_0+$ solutions) is substantially larger and in $\sim 3 \,\sigma$ tension with the estimate adopting Gaia parallax.
The tension may be due to possible biases in both the Gaia DR3 parallax measurement (as indicated from the large RUWE) and the spectroscopic parameters (as suggested from its discrepancy with the source color), and we are not able to make firm conclusions. Due to these caveats, we proceed with a conservative approach by not adopting neither constraints. The results are given in Table~\ref{tab:phy_params_null}. For all point-source solutions, $M_\L$ is broadly consistent with ranging over $\sim 1.0-1.3\,M_\odot$ at $\sim 1.0-1.7\,\rm kpc$ with the implied $\pi_{\rm base}$ being in the range of $\sim 0.4-0.8$\,mas.
}

\subsection{Physical Constraints from Circular Orbital Motion}
\label{sec:phy_orbit}
The binary-lens orbital motion can place constraints on physical parameters of the lens system \citep[see, e.g.,][]{An2002, Dong2009ob071}. We carry out our analysis assuming circular orbits and evaluate the physical constraints under this assumption. We define the ``projected'' orbital period parameter $P_\perp$, which is directly related to observables and physical parameters,
\begin{equation}{\label{eqn:pperp_raw}}
	P_\perp = (M_\L / M_\odot) ^{-1/2} (r_\perp / \mathrm{AU}) ^{3/2}\,\mathrm{yr},
\end{equation}
where $r_\perp/\au = s \theta_\E/\pi _\L = s\kappa M_\L \pi_\E / \pi_\L $.
$P_\perp$ values derived using the above equation are given in the last columns of Table~\ref{tab:fs_params} and Table~\ref{tab:phy_params} for the FS and PS solutions, respectively.
The orbital period of a circular orbit and projected separation  ${r_\perp}$ can be expressed in terms of $\left(w_1, \, w_2,\, w_3\right)$ as follows,
\begin{equation}{\label{eqn:orbital_formula}}
	P_{\rm circ}=\frac{2\pi \sqrt{w_1^2+w_3^2}}{w_3 \sqrt{w_1^2+w_2^2+w_3^2}}, \quad \frac{r_\perp}{r}   = \frac{w_3}{\sqrt{w_1^2+w_3^2}}.
\end{equation}
Given that
\begin{equation*}
	P_\perp = P_{\rm circ} \left(\frac{r_\perp}{r}\right)^{3/2},
\end{equation*}
$P_\perp$ is thus related to the orbital parameters $\left(w_1, \, w_2,\, w_3\right)$:
\begin{equation}{\label{eqn:pperp2}}
	P_{\perp}=\frac{2\pi }{ \sqrt{w_1^2+w_2^2+w_3^2}}{\left(\frac{w_3}{\sqrt{w_1^2+w_3^2}}\right)}^{1/2}.
\end{equation}
The best-fit $P_\perp$ with uncertainties from the light-curve analysis are given in Table~\ref{tab:2L1S_u0p_orb_parameters} and Table~\ref{tab:2L1S_u0m_orb_parameters}.
{ These estimates can be compared with the corresponding constraints via Equation~\ref{eqn:pperp_raw} using the physical parameters derived in \S~\ref{subsec:blendaslens}. For the globally best-fit solution {\rm UPS}$-3.3 (u_0+)$, the $P_\perp$ value is in good agreement with those based on the blended flux (Table~\ref{tab:phy_params_null}) and Gaia parallax constraint (Table~\ref{tab:phy_params}).
We show that the $P_\perp$ constraint (blue band) from Equation~\ref{eqn:pperp2} on $\left(M_\L,\, D_\L\right)$ in Figure~\ref{fig:best_constraints}. We note that it is in $3\,\sigma $ disagreement with that based on the source spectroscopic distance constraint (Table~\ref{tab:phy_params_sp}). For some solutions, such as LPS$-$2.4($u_0+$) and UPS$-$2.6($u_0-$), $P_\perp$ values are in disagreements with all physical constraints, suggesting that they are unphysical under the assumption of circular orbits.
}
\section{Summary and Future Prospects} \label{sec:summary}

Our main interpretation of Gaia22dkv is under the assumption that the blend is the lens star itself and the microlens parallax effects rather than xallarap induce the light-curve distortions. Assuming circular planetary orbital motion, there are multiple degenerate planetary solutions with $\Delta{\chi^2}\lesssim 12$  from the light-curve analysis. For the best-fit solution, the lens is an $M_\L = 1.15^{+0.16}_{-0.08}\,M_\odot$ star at $D_\L = 1.27^{+0.43}_{-0.25}\,{\rm kpc}$ orbited by a planet with $M_{\rm p} = 0.59^{+0.15}_{-0.05}\,M_{\rm J}$ with a projected orbital separation of $r_{\perp} = 1.41^{+0.76}_{-0.36}$\,AU and a circular orbital period $P_{\rm circ}=2.96\pm 0.20$\,yr.

However, one or more of these assumptions may break down. The blend can be the binary companion of either the lens or the source, and the lens could be a low-mass star that is fainter than the blend. The lens could be a M dwarf lens as suggested by the FS solutions, and low-mass lens could also be compatible with the lower limits on $\theta_{\rm E}$ for the PS solutions. Additionally, the large RUWE from Gaia DR3 might hint at the existence of the binary orbital motion of a companion. Besides the common follow-up observations by ground-based Adaptive Optics (AOs) observations (see \citealt{masada} for comprehensive discussions), there are extraordinary opportunities to robustly test our primary interpretation in the near future.

As discussed in \S~\ref{sec:rv}, the exceptionally bright blend allows unprecedented opportunity for high-precision (on the order of $\mathrm{m\,s^{-1}}$) RV follow-ups of the planetary signal. And more epochs of spectra with RV precision on the order of $\mathrm{km\,s^{-1}}$ can make stringent tests on the lens/source companion scenarios. The astrometric issues such as the high RUWE will be clarified with the release of single-epoch Gaia astrometric observations in the future. As shown in \S~\ref{sec:astrometric}, Gaia astrometric data will not only be able to measure the $\theta_{\rm E}$ precisely, but also break the $u_0+/-$ degeneracy and constrain $\bpi_\E$ to mhigh precision, so it has the potential of definitively measure the physical properties of the lens.

\begin{figure}[htbp]
	\epsscale{1.15}
	\plotone{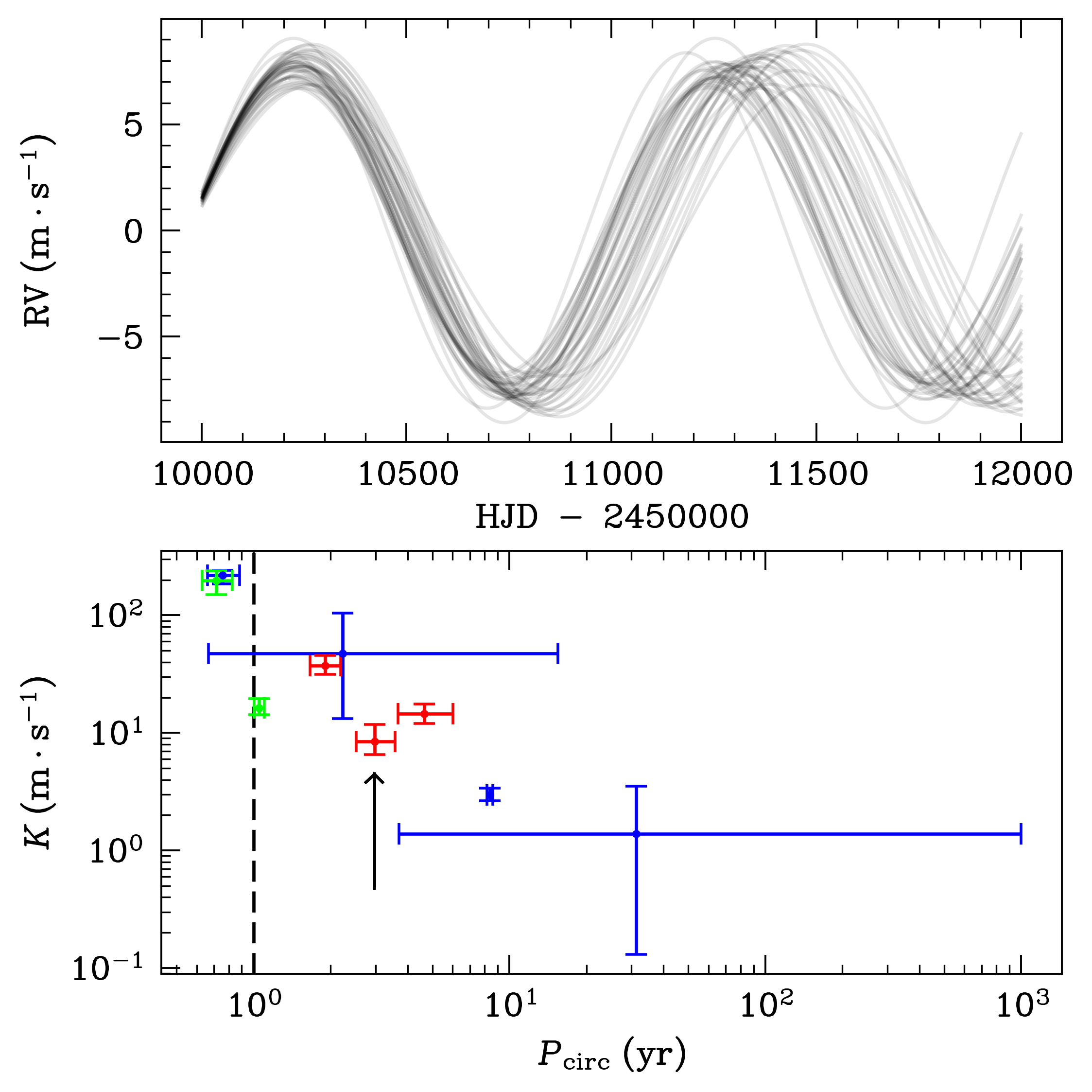}
	\caption{The upper panel displays the predicted radial velocity curves based on the orbital parameters of the best fit. The black lines represent the MCMC sample with $\Delta \chi^2 < 1$. The lower panel displays the RV semi amplitude $K$ and circular orbital period $P_{\rm circ}$ for all solutions. Each solution are shown as the color-coded filled dots with errorbars in red, yellow, green, blue, representing $\Delta\chi^2 < 1,\,4,\,9,\,16$ compared with best fit, respectively. One year period is indicated as dashed line. The best-fit solution UPS$-3.3$($u_0+$) is marked out with a black arrow.
		\label{fig:RV_curve}}
\end{figure}

\begin{figure*}[t]
	\includegraphics[width=\textwidth]{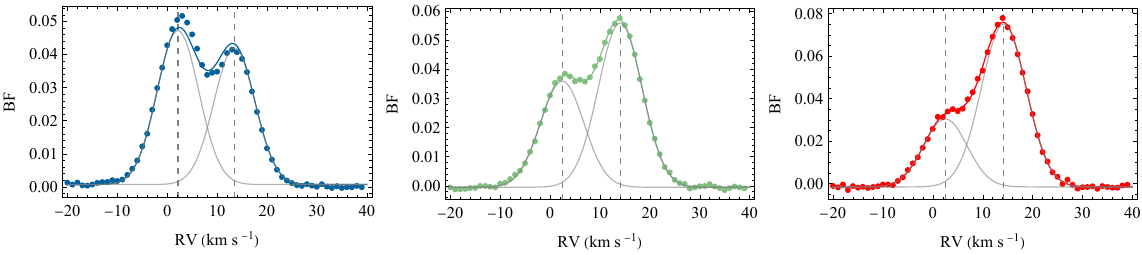}
	\caption{
	The spectral broadening function (BF) for the blue (left), green (middle), and red (right) regions of Gaia22dkv's MIKE spectrum.
	We identify two distinct peaks on the BFs, which correspond to the source (larger RV) and the lens (smaller RV) respectively.
	In each panel, the dots are the computed BF values to which we fit a function (solid color curve) using the sum of two Gaussians (grey curves). The RVs values (i.e., the centers of the Gaussians) are marked with dashed lines. Note that the BFs include the contributions from the instrumental spectral resolution $\sigma_{\rm instr}\sim 5 \, \mathrm{km\,s^{-1}}$, and thus the stellar line widths are smaller than the Gaussians.
	\label{fig:bf}}
\end{figure*}

\begin{figure}[t]
	\centering
	\includegraphics[width=1.0\columnwidth]{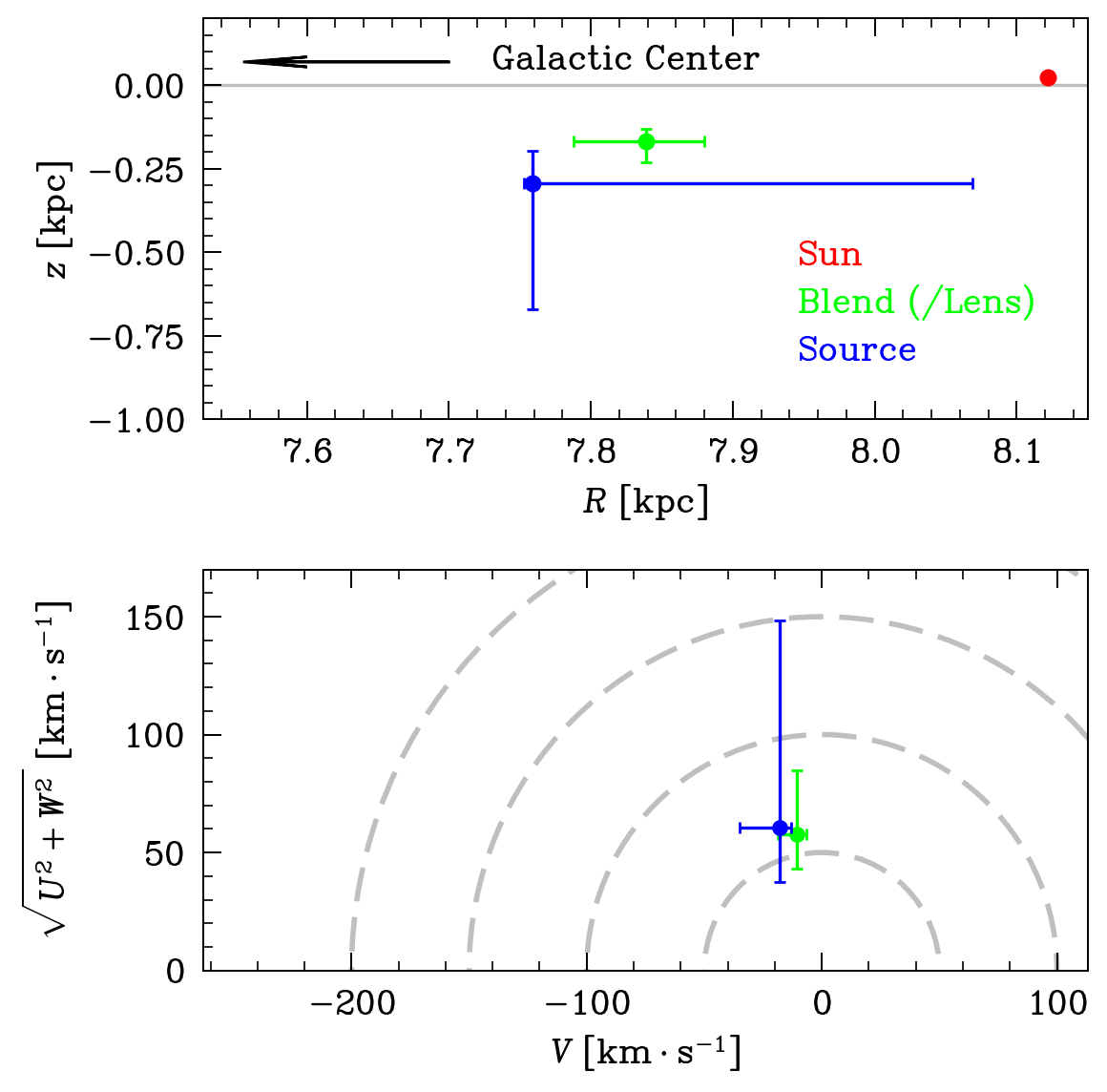}
	\caption{Galactic kinematics of the lens (green dot with error bar) and the source (blue dot with error bar) based on the best-fit solution.
	Upper: the Galactocentric distance $R$ and height $z$ with respect to the Galactic plane positions (the red dot represents the Sun at $z_{\odot}$ $=$ 25 pc; \citealt{Juric2008}).
	Lower:  the Toomre diagram. Dashed curves indicate constant peculiar velocities = 50, 100, 150, and 200 $\mathrm{km~s^{-1}}$.
	\label{fig:gala}}
\end{figure}

\subsection{RV Expectations and Galactic Kinematics}
\label{sec:rv}

There have been no microlensing planets characterized by RV observations. The primary difficulty is the faintness of the host (lens) star. One of the brightest microlensing planet hosts was OGLE-2018-BLG-0740 reached $V\sim18$ \citep{Han2019}, which is near the limit of achieving 10\,${\rm m\,s^{-1}}$ by the Echelle SPectrograph for Rocky Exoplanet and Stable Spectroscopic Observations (ESPRESSO) of VLT \citep{espresso}, the most sensitive high-precision RV instrument. By contrast, the blend of Gaia22dkv has $r \sim 14$ and $V \sim 14$, and if it is the lens, it is the brightest host for a microlensing planet to date. Furthermore, the best-fit planetary solution of { Gaia22dkv} yields a massive planet $\approx0.60\,M_{\rm J}$ on a relatively close-in orbit $(\approx 1.4\,{\rm AU})$, making it favorable for RV detections. For the globally best-fit solution (UPS$-3.3(u_0+)$), we derive that the expected RV signals will have a semi-amplitude of $\approx 8\, \mathrm{m\,s^{-1}}$ with orbital period of $P_{\rm circ}=2.96\pm 0.20$\,yr (see the upper panel of Figure~\ref{fig:RV_curve} for the RV forecast of this solution); the estimated RV amplitudes and periods for all solutions are shown in the lower panel of Figure~\ref{fig:RV_curve}. Note that a significant fraction of cold Jovian planets are on nearly circular orbits \citep{Shen2008, Zakamska2011}, but they have a broad eccentricity distribution \citep{WinnFabrycky2014} with $\bar{e}\sim0.3$. RV follow-up observations will be able to test the microlensing predictions, and in addition, RV data will enable measuring eccentricity and orbital period.

One possible complication for making precision RV observations of the lens is spectroscopically disentangling it from the source within the seeing disk. For Gaia22dkv, the source and blend are nearly equally bright, and thus, the feasibility of carrying out RV observations relies upon their separation in the velocity space.

We plan to derive the { accurate} stellar parameters, including the chemical abundances, by analyzing the MIKE spectrum in detail in a follow-up work. In this work, we present a preliminary analysis of the MIKE spectrum to primarily estimate the approximate RVs of the blend and the source.

We first make rough velocity estimates by using the spectral broadening function (BF) technique through linear inversion \citep{Rucinski1999}, which convolves a sharp-line spectral template with BFs to fit the observed spectrum.  We follow similar procedures in \citet{Yi2022} and use a grid of BT-Settl atmospheric models \citep{Allard2012} ($T_{\mathrm{eff}}:$ 4000K -- 7000 K, $\log{g}$: 1.0 dex -- 5.0 dex, $\mathrm{[Fe/H]}$:  $-$1.0 dex -- 0.5 dex) as our sharp-line templates. We analyze the MIKE spectrum in three wavelength regions separately, 3800\,\AA--4800\,\AA\,(``blue''), 4800\,\AA--5800\,\AA\,(``green''), and 5800\,\AA--6800\,\AA\,(``red''), where the flux ratios between the source and blend vary due to their different temperatures.
We calculate the BFs for all  templates and visually inspect each one, and we consistently find two distinct peaks in most of the BFs for the blue, green, and red regions.
The best-match templates have ($T_{\mathrm{eff}} = 5800$\,K, $\log{g} = 3.0$, $\mathrm{[Fe/H]} = -0.5$), ($T_{\mathrm{eff}} = 5200$\,K, $\log{g} = 2.5$, $\mathrm{[Fe/H]} = -0.5$), ($T_{\mathrm{eff}} = 4900$\,K, $\log{g} = 2.5$, $\mathrm{[Fe/H]} = -0.5$) for blue, green and red regions, respectively. Figure~\ref{fig:bf} shows the resulting BFs from the best-match templates. The two peaks correspond to the blend and the source: the peak with smaller radial velocity has smaller relative amplitudes in redder regions compared to the other peak, so the former peak is from the blend, which is bluer than the source. We fit the BFs by the sum of two Gaussian functions and obtain the parameter uncertainties with the bootstrap method. The results from three regions are consistent with each other, and by taking their weighted means, we get ${\rm RV}_\B = 2.4_{-0.1}^{+0.1}\,\mathrm{km~s^{-1}}$ and ${\rm RV}_\Sc = 14.1_{-0.1}^{+0.1}\,\mathrm{km~s^{-1}}$ for the blend and the source, respectively. The standard deviations of the best-fit Gaussian functions for the blend and the source are $\sigma_\B = 6.2_{-0.1}^{+0.1}\,\mathrm{km~s^{-1}}$ and $\sigma_\Sc = 6.3_{-0.1}^{+0.1}\,\mathrm{km~s^{-1}}$, respectively. Note that the instrument spectral resolution ($\sigma_{\rm instr}\sim 5\, \mathrm{km\,s^{-1}}$) significantly contributes to the measured line widths, and thus the stellar line widths should be notably smaller.

Next, we perform MCMC fitting of the MIKE spectrum with a combination of two stellar templates. We adopt the RVs from the BFs results as priors, and for simplicity, we set $\mathrm{[Fe/H]}$ close to zero  (allowing for small variations). The procedures for generating a model template are as follows: (1) For a given set of stellar parameters ($T_{\mathrm{eff}}$, $\log{g}$, $\mathrm{[Fe/H]}$), a  stellar template is derived from linear interpolation on the grid of synthetic spectra.  (2) The source and blend stellar templates are generated independently, shifted by respective RVs, and broadened by Gaussian kernels with $\sigma_{\mathrm{S/B}}$. (3) The two stellar templates are co-added after multiplying the source template by a flux ratio term. (4) The co-added template is re-binned onto the same wavelength grid as the MIKE spectrum and re-normalized by a fitted fourth-order polynomial as the pseudo continuum. The best-fit stellar parameters ($T_{\mathrm{eff, S}} = 4648~\mathrm{K}$, $\log{g}_{\mathrm{S}} = 3.0~\mathrm{dex}$; $T_{\mathrm{eff, B}} = 6032~\mathrm{K}$, $\log{g}_{\mathrm{B}} = 4.2~\mathrm{dex}$) are broadly consistent with the photometry-based analysis in \S~\ref{sec:phy_param}.  We obtain the best-fit RVs for the blend and source as ${\rm RV_{B}} = 2.9 \pm 0.1~\mathrm{km~s^{-1}}$ and ${\rm RV_{S}} = 14.3 \pm 0.1~\mathrm{km~s^{-1}}$, respectively. The best-fit standard deviations of the Gaussian profiles are  $\sigma_{\mathrm{B}} = 8.2 \pm 0.2~\mathrm{km~s^{-1}}$ and $\sigma_{\mathrm{S}} = 6.2 \pm 0.1~\mathrm{km~s^{-1}}$, respectively. Again, these line widths are considerably larger than the stellar line widths due to significant contributions by the instrumental spectral resolution ($\sigma_{\rm instr}\sim 5\, \mathrm{km\,s^{-1}}$). Therefore, the blend appears to be distinctly separated from the source by more than $\sim 11\, \mathrm{km\,s^{-1}}$ in the velocity space, and with a high-resolution spectrograph, we may obtain high-precision RV data using the source as the velocity reference. We expect to improve the RV analysis and further assess the feasibility of high-precision RV characterizations in the follow-up work with detailed spectral modeling.

With the RVs available, we derive the Galactic kinematics of the source and the lens for the best-fit solution (blend=lens). The best-fit lens and source parallaxes are $\pi_{\mathrm L} = 0.79_{-0.20}^{+0.19}$ mas, $\pi_{\mathrm S} = 0.47_{-0.26}^{+0.21}$ mas, respectively. Thus, they are at heights $z_{\mathrm L} = -166_{-63}^{+37}$ pc and  $z_{\mathrm S}=-293_{-378}^{+98}$ pc with respective to the Galactic plane (see the upper panel of Figure~\ref{fig:gala}). Their velocities relative to the local standard rest (LSR)
are estimated using the \texttt{pyasl} package in \texttt{PyAstronomy} adopting the peculiar velocity of the Sun as
$(U_{\mathrm{LSR}, \odot}, V_{\mathrm{LSR}, \odot}, W_{\mathrm{LSR}, \odot}) = (11.1, 12.24, 7.25)~\mathrm{km~s^{-1}}$ \citep{Schonrich2010}.
The lens and source have best-fit proper motions
$(\mu_{\mathrm{L, N}}, \, \mu_{\mathrm{L, E}}) = (5.72_{-0.20}^{+0.47}, \, -10.67_{-0.32}^{+0.16})\,\mathrm{mas~yr^{-1}}$ and $(\mu_{\mathrm{S, N}}, \, \mu_{\mathrm{S, E}}) = (1.49_{-0.32}^{+0.14}, \, -7.61_{-0.11}^{+0.22})\,\mathrm{mas~yr^{-1}}$, respectively.
{Adopting $\mathrm{RV}_{\mathrm{L}}=2.9\pm1.0\,\mathrm{km~s^{-1}}$ and $\mathrm{RV}_{\mathrm{S}}=14.3\pm1.0\,\mathrm{km~s^{-1}}$ derived from the MIKE spectrum\footnote{The RV uncertainty ($1.0\,\mathrm{km~s^{-1}}$) is dominated by the drift in the wavelength solution over the course of the night.},
we obtain
$(U_{\mathrm{LSR, L}}, V_{\mathrm{LSR, L}}, W_{\mathrm{LSR, L}}) = 
(-57.57^{+14.58}_{-27.32}, -10.67^{+4.30}_{-8.25}, -3.35^{+2.04}_{-2.51})
~\mathrm{km~s^{-1}}$ for the lens and
$(U_{\mathrm{LSR, S}}, V_{\mathrm{LSR, S}}, W_{\mathrm{LSR, S}}) = 
(-53.96^{+20.94}_{-76.47}, -17.81^{+4.92}_{-17.34}, -27.91^{+10.57}_{-42.76})
~\mathrm{km~s^{-1}}$ for the source, respectively.}
The lower panel of Figure~\ref{fig:gala} shows the locations of the lens and the source in the so-called Toomre diagram. They are near the transition between the thin and thick disk populations with some preference for the former.

\begin{figure}[htbp]
	\epsscale{1.15}
	\plotone{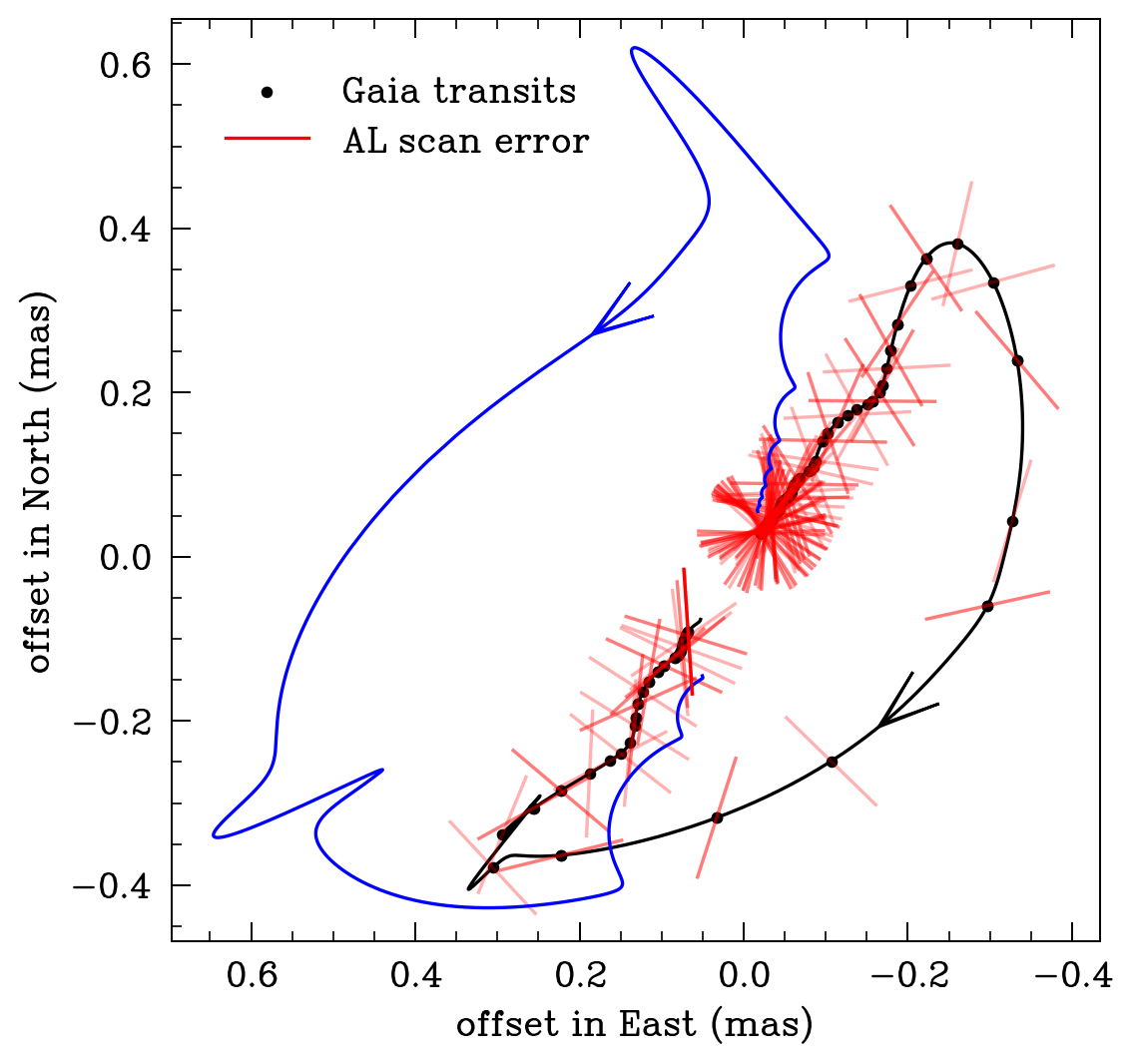}
	\caption{The simulated Gaia astrometric data and fitted models of the microlens relative to that of the baseline object (i.e., subtracting the baseline coordinates and the effects of its trigonometric parallax and proper motion).
			{ The black dots and red lines display the simulated Gaia observations and along-scan error bars. The error bars in the across-scan direction are two orders of magnitude larger than those in the along-scan and are not shown.} The black solid line represents the best-fit $u_0+$ model.
		The blue line shows the best-fit $u_0-$, which has $\Delta{\chi^2} \approx 115$ higher than $u_0+$.
		\label{fig:astro_tracks}}
\end{figure}

\begin{figure*}[htbp]
	\epsscale{1.15}
	\plotone{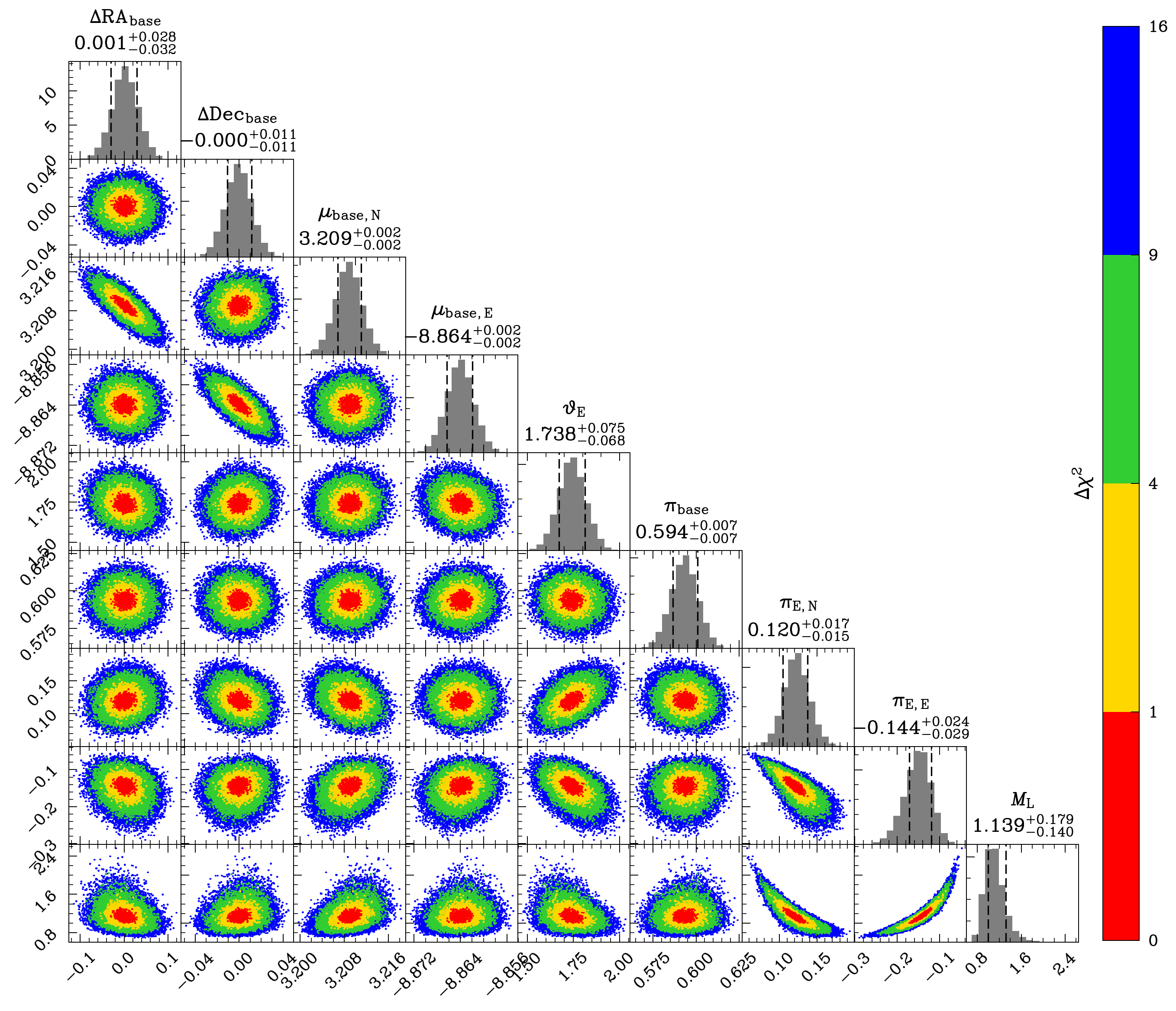}
	\caption{The posteriors of astrometric lensing model parameters from MCMC sampling.
		The red, yellow, green, blue area represent $\Delta{\chi^2}$ within $1, 4, 9, 16$ of the best fit. The model parameters with light-curve priors, i.e. $(t_0,\,u_0,\,t_\E,\,f_\B/f_\Sc)$, are not shown.
		\label{fig:astro_chi2}}
\end{figure*}

\begin{figure}[htbp]
	\epsscale{1.0}
	\plotone{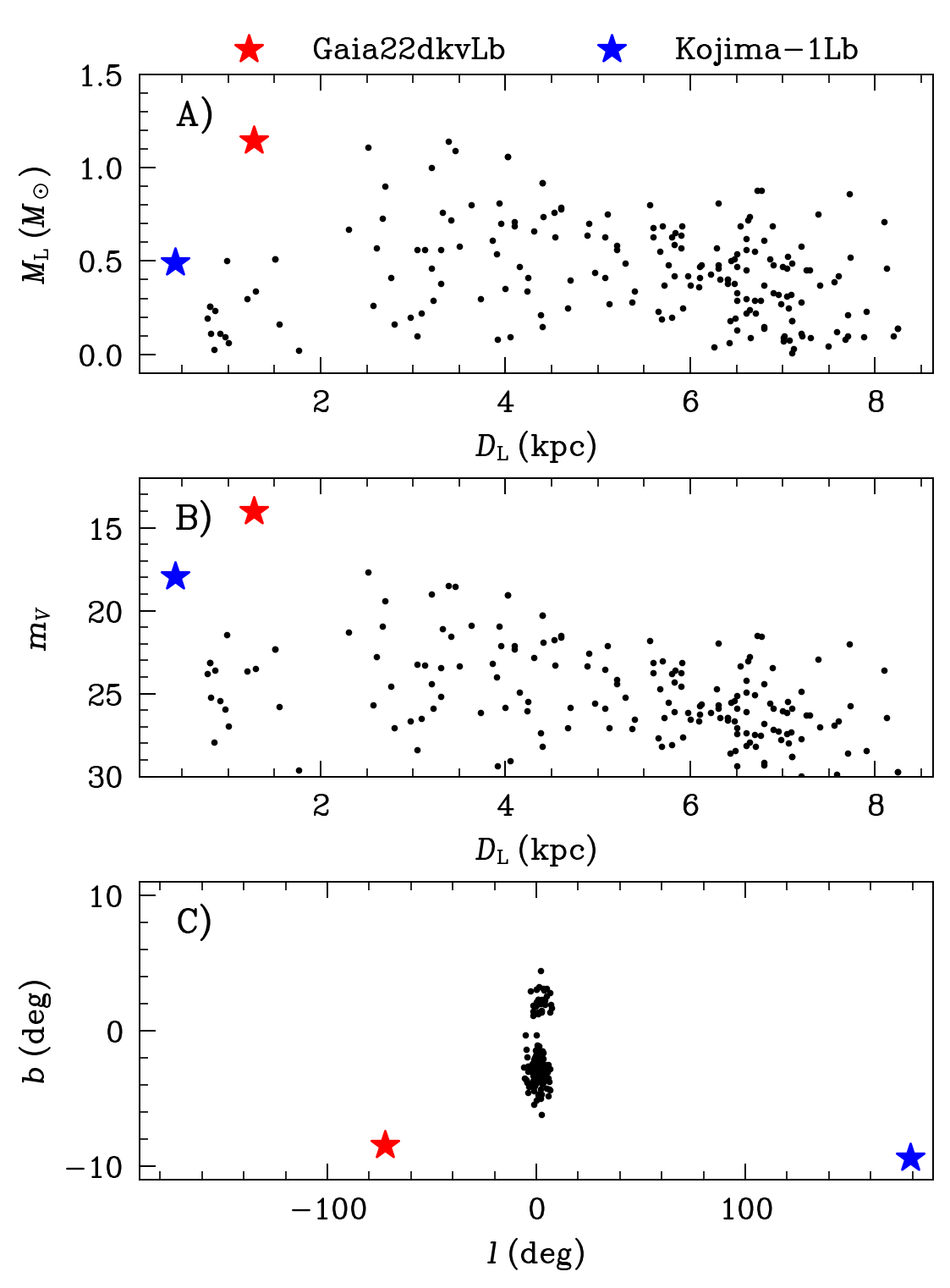}
	\caption{Host properties of microlensing planets. We use the parameters of the default models (modeldef=1) in the microlensing planet catalog of the NASA Exoplanet Archive for all discoveries in the bulge field (black dots). The host properties of Kojima-1Lb (blue star) are from \citet{Zang2020}, and those of Gaia22dkvLb are based on the globally best fit UPS$-$3.3$(u_0+)$ in this work. Panel A): Host  mass ($M_\L$) vs. distance ($D_\L$). Panel B):  Host brightness in $V$ band ($m_V$) vs. distance ($D_\L$). For the systems in the bulge field, we make crude $V$-band brightness estimates by assuming the lens star is a main-sequence star with stellar mass $M_\L$ at $D_\L$. We use the isochrone at a stellar age of $5$\,Gyr from the Dartmouth Stellar Evolution Database  \citep{Dartmouth} to estimate the stellar luminosities. Then we apply the extinction corrections using the near-infrared extinction maps by \citet{Gonzalez2012A} and convert it to $A_V$ with the relationship in \citet{Cardelli1989}. Panel C): The Galactic longitude ($l$) vs. latitude ($b$) distribution.
		\label{fig:hosts}}
\end{figure}

\begin{figure}[htbp]
	\epsscale{1.15}
	\plotone{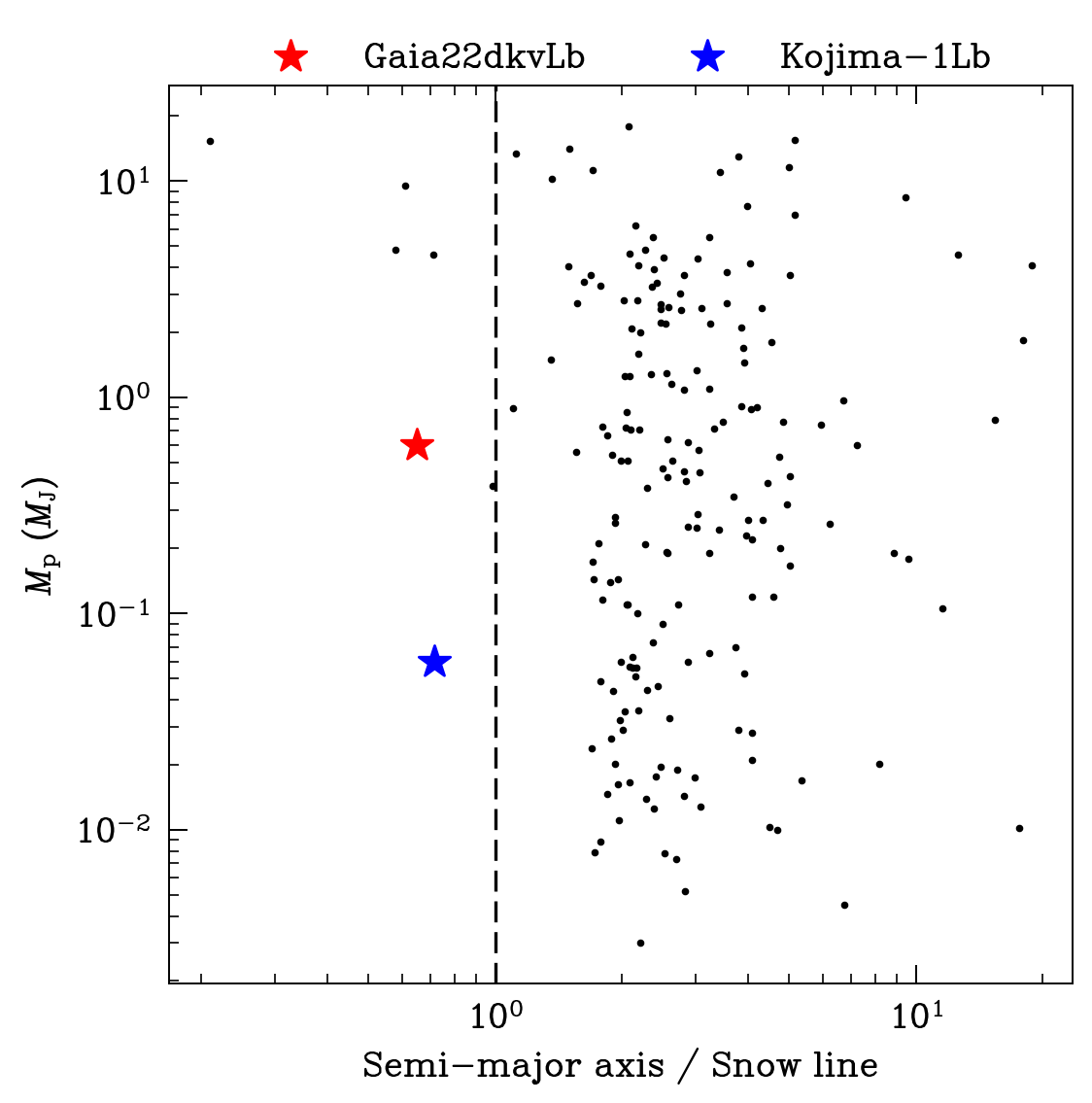}
	\caption{Semi-major axis (normalized to snow line) vs. planet mass distribution of microlensing planets (black dots: bulge field, red star: Gaia22dkvLb, blue star: Kojima1-Lb). We scale the projected separation by a factor of $\sqrt{3/2}$ (the averaged value assuming random orbital orientation) to estimate the  semi-major axis. The parameters are from the same sources as in Figure~\ref{fig:hosts}.
		\label{fig:snow line}}
\end{figure}

\subsection{Gaia astrometric lensing}
\label{sec:astrometric}
Our best-fit planetary solution lacks the $\theta_\E$ measurement due to the absence of significant finite-source effect. Astrometric microlensing is an alternative channel to measure $\theta_\E$
\citep{1995A&A...294..287H,1995AJ....110.1427M, 1995ApJ...453...37W,2000ApJ...534..213D}.
We perform a simplified PSPL analysis to evaluate the prospect of astrometric microlensing with the Gaia time-series data that will ultimately be released.

We simulate Gaia astrometric observations based on the PSPL parameters from our
best-fit planetary solution (UPS$-$3.3$(u_0+)$) along with the physical parameters estimated by assuming the blend to be the lens as evaluated in Table~\ref{tab:phy_params}. The epochs of observation from the start of Gaia mission to future predictions are based on output from the Gaia Observation Forecast Tool (GOST\footnote{\url{https://gaia.esac.esa.int/gost/}}).
We adopt the along-scan (AL) astrometric precision of $0.08\,\mathrm{mas}$ for each epoch based on the estimate from \citet{2018MNRAS.476.2013R} for a target with $G\sim 13$. { The across-scan (AC) error bars are two orders of magnitude larger than the AL errors, and we do not include AC errors in modeling.} See the black dots with red error bars in Figure~\ref{fig:astro_tracks} for the simulated data. Then we fit the simulated Gaia data with an astrometric microlensing model with 12 parameters, including 4 parameters $(t_0,\,u_0,\,t_\E,\,f_\B/f_\Sc)$ with priors derived from the fit to the light curve. We do not impose light-curve priors on the other 8 free parameters $(\mathrm{{RA_{base}}},\, \mathrm{{Dec_{base}}},\,\theta_\E,\, \pi_\mathrm{E,N},\, \pi_\mathrm{E,E},\, \mu_\mathrm{N,base},\,\mu_\mathrm{E,base},\, \pi_\mathrm{base})$, where $(\mathrm{{RA_{base}}}, \mathrm{{Dec_{base}}})$, $\bmu_\mathrm{base} = (\mu_\mathrm{N,base},\,\mu_\mathrm{E,base})$ are the coordinates and proper motion of the baseline object when un-lensed at the same reference epoch of the Gaia DR3 catalog. In practice, we fit $(\mathrm{\Delta{RA_\mathrm{base}}}, \mathrm{\Delta{Dec_\mathrm{base}}})$, which are differences from the input Gaia DR3 coordinates.

We sample the posteriors using \texttt{dynesty}
\citep{2020MNRAS.493.3132S,sergey_koposov_2023_7995596}, which implements Nested Sampling on the high-dimension parameter space. The posteriors of the 8 free parameters of the astrometric model are displayed in Figure~\ref{fig:astro_chi2}, and the centroid's trajectory from the best-fit model is shown as the solid black line in Figure~\ref{fig:astro_tracks}. The expected $1\sigma$ error of $\theta_\E$ from our MCMC posterior is $\sim 0.1\,\mathrm{mas}$, and the corresponding uncertainty of lens mass is $\sim 15\%$. Interestingly, the astrometric data alone can also constrain $\bpi_\E$ to a high precision ($\sigma_{\pi_{\rm E,N}}\approx0.02$ and $\sigma_{\pi_{\rm E,E}}\approx0.03$). We fit the simulated data to $u_0-$ models and we find that the best-fit $u_0-$ model has a $\Delta \chi^2 = {115}$ compared to $u_0+$. Therefore, the Gaia astrometric microlensing data can completely break the $u_0+/-$ degeneracy and definitively determine the microlens parallax and thus the lens mass and distance.

\section{Conclusion}\label{sec:discussion}
{ Gaia22dkvLb} is an extrasolar planet discovered from a microlensing event alerted by Gaia, and its location is far from the traditional bulge microlensing fields (see the bottom panel of Figure~\ref{fig:hosts}). The only other non-bulge microlensing planet Kojima-1Lb was found serendipitously, and it is difficult to quantify the detection efficiency and make statistical inferences. { Both Gaia22dkvLb and Kojima-1Lb are brighter than $V=12$ at the peak, and according to \citet{Han2008}, the all-sky rate for such bright events is very rare, only $\sim 0.1\,{\rm yr}^{-1}$. Therefore it is intriguing that exoplanets are (serendipitously) found in both events, probably implying a large planet frequency for near-field microlensing events based on the small number statistics of 2 events.} Systematic all-sky surveys such as Gaia can open up the largely untapped discovery potential of near-field microlensing. Our primary interpretation reveals the lens as to be nearby ($\sim 1.3$\,kpc) turn-off star, making it the most brightest microlensing planet host (see the middle panels of Figure~\ref{fig:hosts}). Thanks to the host's close distance and its high effective temperature, the Einstein radius of Gaia22dkv is well within the snow line, unlike most microlensing planets. As a Jovian planet orbiting a turn-off star, Gaia22dkvLb is one of the few microlensing planets detected inside the snow line (see Figure~\ref{fig:snow line}).  Such near-field events can provide far richer opportunities for follow-up studies than planetary systems found in the bulge field. Gaia astrometric and host spectroscopic observations can yield detailed characterizations on the host's physical and kinematic properties and chemical abundance, which can offer significant clues regarding planet formation and dynamical evolutions \citep{ZhuDong2021}. The combination of high planet mass, relatively small orbital separation and the exceptionally bright host makes Gaia22dkvLb the most promising microlensing planet for RV follow ups. Besides determining Gaia22dkvLb's orbital period and eccentricity, precision RV data can also be used to search for much shorter-period planets and probe the planet architecture, which is generally impossible for microlensing by itself. While microlensing does occasionally detect two planets, { their projected locations are typically near the Einstein ring at $\sim1$--4 AU \citep[see, e.g.][]{Kuang2023}, so it can hardly probe multiple planets with large semi-major-axis ratios.} {\it Kepler} and RV observations have revealed tantalizing evidence for a strong ``inner-outer'' correlation: $\sim 90\%$ of cold Jovian planets are found to be accompanied with close-in super Earths in the same systems \citep{Zhu2018, Bryan2019}. { Such a correlation may have implications in understanding the physical processes of planet formation \citep[see, e.g.,][]{Schlecker2021, Guo2023}.} Future RV follow ups can examine whether the cold Jovian planet of Gaia22dkvLb has an associated inner planet as would be likely according to the inner-out correlation.

\section*{Acknowledgment}
We thank Johanna Teske for stimulating discussion and Wei Zhu's team for attempting to observe the event. This work is supported by the National Key R\&D Program of China No. 2019YFA0405100, the National Natural Science Foundation of China (Grant No. 12133005) and the science research grants from the China Manned Space Project with No. CMS-CSST-2021-B12. SD acknowledges the New Cornerstone Science Foundation through the XPLORER PRIZE. RAS, EB gratefully acknowledge funding from NASA award 80NSSC19K029.
RFJ acknowledges support for this project provided by ANID's Millennium Science Initiative through grant ICN12\textunderscore 009, awarded to the Millennium Institute of Astrophysics (MAS), and by ANID's Basal project FB210003.
YT acknowledges the support of DFG priority program SPP 1992 ``Exploring the Diversity of Extrasolar Planets'' (TS 356/3-1). This research uses data obtained through the Telescope Access Program (TAP), which has been funded by the TAP member institutes. This research has made use of the NASA Exoplanet Archive, which is operated by the California Institute of Technology, under contract with the National Aeronautics and Space Administration under the Exoplanet Exploration Program. These observations are associated with program 16871. This work has made use of data from the European Space Agency (ESA) mission {\it Gaia} (\url{https://www.cosmos.esa.int/gaia}), processed by the {\it Gaia} Data Processing and Analysis Consortium (DPAC,
\url{https://www.cosmos.esa.int/web/gaia/dpac/consortium}). Funding for the DPAC has been provided by national institutions, in particular the institutions participating in the {\it Gaia} Multilateral Agreement. This research has made use of the VizieR catalogue access tool, CDS, Strasbourg, France. This work uses observations made at the Observatorio do Pico dos Dias/LNA (Brazil).
Some of the observations reported in this paper were obtained with the Southern African Large Telescope (SALT). Polish participation in SALT is funded by grant No. MEiN nr 2021/WK/01. This project has received funding from the EU Horizon 2020 research and innovation programme under grant agreement No. 101004719 (OPTICON-RadioNet Pilot).
This research was supported by the Polish National Science Centre grant number 2017/25/B/ST9/02805.
We acknowledge ESA Gaia, DPAC and the Photometric Science Alerts Team (http://gsaweb.ast.cam.ac.uk/alerts).

\facility {Gaia, LCOGT, Magellan:Clay (MIKE), Exoplanet Archive, BHTOM}
\software {PmPyeasy, DoPHOT, VBBinaryLensing, EMCEE, MIST, PyAstronomy, dynesty}

\bibliography{ref}{}

\begin{thebibliography}{}
\expandafter\ifx\csname natexlab\endcsname\relax\def\natexlab#1{#1}\fi
\providecommand{\url}[1]{\href{#1}{#1}}
\providecommand{\dodoi}[1]{doi:~\href{http://doi.org/#1}{\nolinkurl{#1}}}
\providecommand{\doeprint}[1]{\href{http://ascl.net/#1}{\nolinkurl{http://ascl.net/#1}}}
\providecommand{\doarXiv}[1]{\href{https://arxiv.org/abs/#1}{\nolinkurl{https://arxiv.org/abs/#1}}}

\bibitem[{{Allard} {et~al.}(2012){Allard}, {Homeier}, \&
  {Freytag}}]{Allard2012}
{Allard}, F., {Homeier}, D., \& {Freytag}, B. 2012, Philosophical Transactions
  of the Royal Society of London Series A, 370, 2765,
  \dodoi{10.1098/rsta.2011.0269}

\bibitem[{{An}(2005)}]{An2005}
{An}, J.~H. 2005, \mnras, 356, 1409, \dodoi{10.1111/j.1365-2966.2004.08581.x}

\bibitem[{{An} {et~al.}(2002){An}, {Albrow}, {Beaulieu}, {Caldwell}, {DePoy},
  {Dominik}, {Gaudi}, {Gould}, {Greenhill}, {Hill}, {Kane}, {Martin},
  {Menzies}, {Pogge}, {Pollard}, {Sackett}, {Sahu}, {Vermaak}, {Watson}, \&
  {Williams}}]{An2002}
{An}, J.~H., {Albrow}, M.~D., {Beaulieu}, J.~P., {et~al.} 2002, \apj, 572, 521,
  \dodoi{10.1086/340191}

\bibitem[{{Bachelet} {et~al.}(2022){Bachelet}, {Tsapras}, {Gould}, {Street},
  {Bennett}, {Hundertmark}, {Bozza}, {Bramich}, {Cassan}, {Dominik}, {Horne},
  {Mao}, {Saha}, {Wambsganss}, {Zang}, {ROME/REA Collaboration}, {Abe},
  {Barry}, {Bennett}, {Bhattacharya}, {Bond}, {Fukui}, {Fujii}, {Hirao},
  {Itow}, {Kirikawa}, {Kondo}, {Koshimoto}, {Matsubara}, {Matsumoto},
  {Miyazaki}, {Muraki}, {Olmschenk}, {Ranc}, {Okamura}, {Rattenbury}, {Satoh},
  {Sumi}, {Suzuki}, {Silva}, {Toda}, {Tristram}, {Vandorou}, {Yama}, {MOA
  Collaboration}, {Albrow}, {Chung}, {Han}, {Hwang}, {Jung}, {Ryu}, {Shin},
  {Shvartzvald}, {Yee}, {Cha}, {Kim}, {Kim}, {Lee}, {Lee}, {Lee}, {Park},
  {Pogge}, {KMTNet Collaboration}, {Udalski}, {Mr{\'o}z}, {Poleski}, {Skowron},
  {Szyma{\'n}ski}, {Soszy{\'n}ski}, {Pietrukowicz}, {Koz{\l}owski}, {Ulaczyk},
  {Rybicki}, {Iwanek}, {Wrona}, {Gromadzki}, \& {OGLE
  Collaboration}}]{2022AJ....164...75B}
{Bachelet}, E., {Tsapras}, Y., {Gould}, A., {et~al.} 2022, \aj, 164, 75,
  \dodoi{10.3847/1538-3881/ac78ed}

\bibitem[{{Bernstein} {et~al.}(2003){Bernstein}, {Shectman}, {Gunnels},
  {Mochnacki}, \& {Athey}}]{MIKE2003}
{Bernstein}, R., {Shectman}, S.~A., {Gunnels}, S.~M., {Mochnacki}, S., \&
  {Athey}, A.~E. 2003, in Society of Photo-Optical Instrumentation Engineers
  (SPIE) Conference Series, Vol. 4841, Instrument Design and Performance for
  Optical/Infrared Ground-based Telescopes, ed. M.~{Iye} \& A.~F.~M.
  {Moorwood}, 1694--1704, \dodoi{10.1117/12.461502}

\bibitem[{{Blanco-Cuaresma}(2019)}]{BlancoCuaresma2019}
{Blanco-Cuaresma}, S. 2019, \mnras, 486, 2075, \dodoi{10.1093/mnras/stz549}

\bibitem[{{Blanco-Cuaresma} {et~al.}(2014){Blanco-Cuaresma}, {Soubiran},
  {Heiter}, \& {Jofr{\'e}}}]{BlancoCuaresma2014}
{Blanco-Cuaresma}, S., {Soubiran}, C., {Heiter}, U., \& {Jofr{\'e}}, P. 2014,
  \aap, 569, A111, \dodoi{10.1051/0004-6361/201423945}

\bibitem[{{Boyajian} {et~al.}(2013){Boyajian}, {von Braun}, {van Belle},
  {Farrington}, {Schaefer}, {Jones}, {White}, {McAlister}, {ten Brummelaar},
  {Ridgway}, {Gies}, {Sturmann}, {Sturmann}, {Turner}, {Goldfinger}, \&
  {Vargas}}]{2013ApJ...771...40B}
{Boyajian}, T.~S., {von Braun}, K., {van Belle}, G., {et~al.} 2013, \apj, 771,
  40, \dodoi{10.1088/0004-637X/771/1/40}

\bibitem[{{Bozza}(2010)}]{2010MNRAS.408.2188B}
{Bozza}, V. 2010, \mnras, 408, 2188, \dodoi{10.1111/j.1365-2966.2010.17265.x}

\bibitem[{{Bozza} {et~al.}(2018){Bozza}, {Bachelet}, {Bartoli{\'c}}, {Heintz},
  {Hoag}, \& {Hundertmark}}]{2018MNRAS.479.5157B}
{Bozza}, V., {Bachelet}, E., {Bartoli{\'c}}, F., {et~al.} 2018, \mnras, 479,
  5157, \dodoi{10.1093/mnras/sty1791}

\bibitem[{{Bozza} {et~al.}(2021){Bozza}, {Khalouei}, \&
  {Bachelet}}]{VBBLastrometry}
{Bozza}, V., {Khalouei}, E., \& {Bachelet}, E. 2021, \mnras, 505, 126,
  \dodoi{10.1093/mnras/stab1376}

\bibitem[{{Brown} {et~al.}(2013){Brown}, {Baliber}, {Bianco}, {Bowman},
  {Burleson}, {Conway}, {Crellin}, {Depagne}, {De Vera}, {Dilday}, {Dragomir},
  {Dubberley}, {Eastman}, {Elphick}, {Falarski}, {Foale}, {Ford}, {Fulton},
  {Garza}, {Gomez}, {Graham}, {Greene}, {Haldeman}, {Hawkins}, {Haworth},
  {Haynes}, {Hidas}, {Hjelstrom}, {Howell}, {Hygelund}, {Lister}, {Lobdill},
  {Martinez}, {Mullins}, {Norbury}, {Parrent}, {Paulson}, {Petry}, {Pickles},
  {Posner}, {Rosing}, {Ross}, {Sand}, {Saunders}, {Shobbrook}, {Shporer},
  {Street}, {Thomas}, {Tsapras}, {Tufts}, {Valenti}, {Vander Horst}, {Walker},
  {White}, \& {Willis}}]{2013PASP..125.1031B}
{Brown}, T.~M., {Baliber}, N., {Bianco}, F.~B., {et~al.} 2013, \pasp, 125,
  1031, \dodoi{10.1086/673168}

\bibitem[{{Bryan} {et~al.}(2019){Bryan}, {Knutson}, {Lee}, {Fulton}, {Batygin},
  {Ngo}, \& {Meshkat}}]{Bryan2019}
{Bryan}, M.~L., {Knutson}, H.~A., {Lee}, E.~J., {et~al.} 2019, \aj, 157, 52,
  \dodoi{10.3847/1538-3881/aaf57f}

\bibitem[{{Buckley} {et~al.}(2006){Buckley}, {Swart}, \&
  {Meiring}}]{2006SPIE.6267E..0ZB}
{Buckley}, D. A.~H., {Swart}, G.~P., \& {Meiring}, J.~G. 2006, in Society of
  Photo-Optical Instrumentation Engineers (SPIE) Conference Series, Vol. 6267,
  Society of Photo-Optical Instrumentation Engineers (SPIE) Conference Series,
  ed. L.~M. {Stepp}, 62670Z, \dodoi{10.1117/12.673750}

\bibitem[{{Cardelli} {et~al.}(1989){Cardelli}, {Clayton}, \&
  {Mathis}}]{Cardelli1989}
{Cardelli}, J.~A., {Clayton}, G.~C., \& {Mathis}, J.~S. 1989, \apj, 345, 245,
  \dodoi{10.1086/167900}

\bibitem[{{Chen} {et~al.}(2022){Chen}, {Dong}, {Kochanek}, {Stanek}, {Post},
  {Stritzinger}, {Prieto}, {Filippenko}, {Kollmeier}, {Elias-Rosa}, {Katz},
  {Tomasella}, {Bose}, {Ashall}, {Benetti}, {Bersier}, {Brimacombe}, {Brink},
  {Brown}, {Buckley}, {Cappellaro}, {Christie}, {Fraser}, {Gromadzki},
  {Holoien}, {Hu}, {Kankare}, {Koff}, {Lundqvist}, {Mattila}, {Milne},
  {Morrell}, {Mu{\~n}oz}, {Mutel}, {Natusch}, {Nicolas}, {Pastorello},
  {Prentice}, {Roth}, {Shappee}, {Stone}, {Thompson}, {Villanueva}, \&
  {Zheng}}]{Chen2022}
{Chen}, P., {Dong}, S., {Kochanek}, C.~S., {et~al.} 2022, \apjs, 259, 53,
  \dodoi{10.3847/1538-4365/ac50b7}

\bibitem[{{Choi} {et~al.}(2016){Choi}, {Dotter}, {Conroy}, {Cantiello},
  {Paxton}, \& {Johnson}}]{2016ApJ...823..102C}
{Choi}, J., {Dotter}, A., {Conroy}, C., {et~al.} 2016, \apj, 823, 102,
  \dodoi{10.3847/0004-637X/823/2/102}

\bibitem[{{Crause} {et~al.}(2014){Crause}, {Sharples}, {Bramall}, {Schmoll},
  {Clark}, {Younger}, {Tyas}, {Ryan}, {Brink}, {Strydom}, {Buckley},
  {Wilkinson}, {Crawford}, \& {Depagne}}]{2014SPIE.9147E..6TC}
{Crause}, L.~A., {Sharples}, R.~M., {Bramall}, D.~G., {et~al.} 2014, in Society
  of Photo-Optical Instrumentation Engineers (SPIE) Conference Series, Vol.
  9147, Ground-based and Airborne Instrumentation for Astronomy V, ed. S.~K.
  {Ramsay}, I.~S. {McLean}, \& H.~{Takami}, 91476T, \dodoi{10.1117/12.2055635}

\bibitem[{{Dominik}(1999)}]{Dominik1999}
{Dominik}, M. 1999, \aap, 349, 108, \dodoi{10.48550/arXiv.astro-ph/9903014}

\bibitem[{{Dominik} \& {Sahu}(2000)}]{2000ApJ...534..213D}
{Dominik}, M., \& {Sahu}, K.~C. 2000, \apj, 534, 213, \dodoi{10.1086/308716}

\bibitem[{{Dong} {et~al.}(2009){Dong}, {Gould}, {Udalski}, {Anderson},
  {Christie}, {Gaudi}, {OGLE Collaboration}, {Jaroszy{\'n}ski}, {Kubiak},
  {Szyma{\'n}ski}, {Pietrzy{\'n}ski}, {Soszy{\'n}ski}, {Szewczyk}, {Ulaczyk},
  {Wyrzykowski}, {{\ensuremath{\mu}}FUN Collaboration}, {DePoy}, {Fox},
  {Gal-Yam}, {Han}, {L{\'e}pine}, {McCormick}, {Ofek}, {Park}, {Pogge}, {MOA
  Collaboration}, {Abe}, {Bennett}, {Bond}, {Britton}, {Gilmore}, {Hearnshaw},
  {Itow}, {Kamiya}, {Kilmartin}, {Korpela}, {Masuda}, {Matsubara}, {Motomura},
  {Muraki}, {Nakamura}, {Ohnishi}, {Okada}, {Rattenbury}, {Saito}, {Sako},
  {Sasaki}, {Sullivan}, {Sumi}, {Tristram}, {Yanagisawa}, {Yock}, {Yoshoika},
  {PLANET/RoboNet Collaborations}, {Albrow}, {Beaulieu}, {Brillant}, {Calitz},
  {Cassan}, {Cook}, {Coutures}, {Dieters}, {Dominis Prester}, {Donatowicz},
  {Fouqu{\'e}}, {Greenhill}, {Hill}, {Hoffman}, {Horne}, {J{\o}rgensen},
  {Kane}, {Kubas}, {Marquette}, {Martin}, {Meintjes}, {Menzies}, {Pollard},
  {Sahu}, {Vinter}, {Wambsganss}, {Williams}, {Bode}, {Bramich}, {Burgdorf},
  {Snodgrass}, {Steele}, {Doublier}, \& {Foellmi}}]{Dong2009ob071}
{Dong}, S., {Gould}, A., {Udalski}, A., {et~al.} 2009, \apj, 695, 970,
  \dodoi{10.1088/0004-637X/695/2/970}

\bibitem[{{Dong} {et~al.}(2019){Dong}, {M{\'e}rand}, {Delplancke-Str{\"o}bele},
  {Gould}, {Chen}, {Post}, {Kochanek}, {Stanek}, {Christie}, {Mutel},
  {Natusch}, {Holoien}, {Prieto}, {Shappee}, \& {Thompson}}]{Dong2019}
{Dong}, S., {M{\'e}rand}, A., {Delplancke-Str{\"o}bele}, F., {et~al.} 2019,
  \apj, 871, 70, \dodoi{10.3847/1538-4357/aaeffb}

\bibitem[{{Dotter} {et~al.}(2008){Dotter}, {Chaboyer}, {Jevremovi{\'c}},
  {Kostov}, {Baron}, \& {Ferguson}}]{Dartmouth}
{Dotter}, A., {Chaboyer}, B., {Jevremovi{\'c}}, D., {et~al.} 2008, \apjs, 178,
  89, \dodoi{10.1086/589654}

\bibitem[{{Foreman-Mackey} {et~al.}(2013){Foreman-Mackey}, {Hogg}, {Lang}, \&
  {Goodman}}]{emcee}
{Foreman-Mackey}, D., {Hogg}, D.~W., {Lang}, D., \& {Goodman}, J. 2013, PASP,
  125, 306, \dodoi{10.1086/670067}

\bibitem[{{Fukui} {et~al.}(2019){Fukui}, {Suzuki}, {Koshimoto}, {Bachelet},
  {Vanmunster}, {Storey}, {Maehara}, {Yanagisawa}, {Yamada}, {Yonehara},
  {Hirano}, {Bennett}, {Bozza}, {Mawet}, {Penny}, {Awiphan}, {Oksanen},
  {Heintz}, {Oberst}, {B{\'e}jar}, {Casasayas-Barris}, {Chen}, {Crouzet},
  {Hidalgo}, {Klagyivik}, {Murgas}, {Narita}, {Palle}, {Parviainen},
  {Watanabe}, {Kusakabe}, {Mori}, {Terada}, {de Leon}, {Hernandez}, {Luque},
  {Monelli}, {Monta{\~n}es-Rodriguez}, {Prieto-Arranz}, {Murata}, {Shugarov},
  {Kubota}, {Otsuki}, {Shionoya}, {Nishiumi}, {Nishide}, {Fukagawa}, {Onodera},
  {Villanueva}, {Street}, {Tsapras}, {Hundertmark}, {Kuzuhara}, {Fujita},
  {Beichman}, {Beaulieu}, {Alonso}, {Reichart}, {Kawai}, \&
  {Tamura}}]{Fukui2019}
{Fukui}, A., {Suzuki}, D., {Koshimoto}, N., {et~al.} 2019, \aj, 158, 206,
  \dodoi{10.3847/1538-3881/ab487f}

\bibitem[{{Gaia Collaboration} {et~al.}(2016){Gaia Collaboration}, {Prusti},
  {de Bruijne}, {Brown}, {Vallenari}, {Babusiaux}, {Bailer-Jones}, {Bastian},
  {Biermann}, {Evans}, {Eyer}, {Jansen}, {Jordi}, {Klioner}, {Lammers},
  {Lindegren}, {Luri}, {Mignard}, {Milligan}, {Panem}, {Poinsignon},
  {Pourbaix}, {Randich}, {Sarri}, {Sartoretti}, {Siddiqui}, {Soubiran},
  {Valette}, {van Leeuwen}, {Walton}, {Aerts}, {Arenou}, {Cropper}, {Drimmel},
  {H{\o}g}, {Katz}, {Lattanzi}, {O'Mullane}, {Grebel}, {Holland}, {Huc},
  {Passot}, {Bramante}, {Cacciari}, {Casta{\~n}eda}, {Chaoul}, {Cheek}, {De
  Angeli}, {Fabricius}, {Guerra}, {Hern{\'a}ndez}, {Jean-Antoine-Piccolo},
  {Masana}, {Messineo}, {Mowlavi}, {Nienartowicz}, {Ord{\'o}{\~n}ez-Blanco},
  {Panuzzo}, {Portell}, {Richards}, {Riello}, {Seabroke}, {Tanga},
  {Th{\'e}venin}, {Torra}, {Els}, {Gracia-Abril}, {Comoretto},
  {Garcia-Reinaldos}, {Lock}, {Mercier}, {Altmann}, {Andrae}, {Astraatmadja},
  {Bellas-Velidis}, {Benson}, {Berthier}, {Blomme}, {Busso}, {Carry},
  {Cellino}, {Clementini}, {Cowell}, {Creevey}, {Cuypers}, {Davidson}, {De
  Ridder}, {de Torres}, {Delchambre}, {Dell'Oro}, {Ducourant}, {Fr{\'e}mat},
  {Garc{\'\i}a-Torres}, {Gosset}, {Halbwachs}, {Hambly}, {Harrison}, {Hauser},
  {Hestroffer}, {Hodgkin}, {Huckle}, {Hutton}, {Jasniewicz}, {Jordan},
  {Kontizas}, {Korn}, {Lanzafame}, {Manteiga}, {Moitinho}, {Muinonen},
  {Osinde}, {Pancino}, {Pauwels}, {Petit}, {Recio-Blanco}, {Robin}, {Sarro},
  {Siopis}, {Smith}, {Smith}, {Sozzetti}, {Thuillot}, {van Reeven}, {Viala},
  {Abbas}, {Abreu Aramburu}, {Accart}, {Aguado}, {Allan}, {Allasia},
  {Altavilla}, {{\'A}lvarez}, {Alves}, {Anderson}, {Andrei}, {Anglada Varela},
  {Antiche}, {Antoja}, {Ant{\'o}n}, {Arcay}, {Atzei}, {Ayache}, {Bach},
  {Baker}, {Balaguer-N{\'u}{\~n}ez}, {Barache}, {Barata}, {Barbier}, {Barblan},
  {Baroni}, {Barrado y Navascu{\'e}s}, {Barros}, {Barstow}, {Becciani},
  {Bellazzini}, {Bellei}, {Bello Garc{\'\i}a}, {Belokurov}, {Bendjoya},
  {Berihuete}, {Bianchi}, {Bienaym{\'e}}, {Billebaud}, {Blagorodnova},
  {Blanco-Cuaresma}, {Boch}, {Bombrun}, {Borrachero}, {Bouquillon}, {Bourda},
  {Bouy}, {Bragaglia}, {Breddels}, {Brouillet}, {Br{\"u}semeister},
  {Bucciarelli}, {Budnik}, {Burgess}, {Burgon}, {Burlacu}, {Busonero}, {Buzzi},
  {Caffau}, {Cambras}, {Campbell}, {Cancelliere}, {Cantat-Gaudin}, {Carlucci},
  {Carrasco}, {Castellani}, {Charlot}, {Charnas}, {Charvet}, {Chassat},
  {Chiavassa}, {Clotet}, {Cocozza}, {Collins}, {Collins}, {Costigan}, {Crifo},
  {Cross}, {Crosta}, {Crowley}, {Dafonte}, {Damerdji}, {Dapergolas}, {David},
  {David}, {De Cat}, {de Felice}, {de Laverny}, {De Luise}, {De March}, {de
  Martino}, {de Souza}, {Debosscher}, {del Pozo}, {Delbo}, {Delgado},
  {Delgado}, {di Marco}, {Di Matteo}, {Diakite}, {Distefano}, {Dolding}, {Dos
  Anjos}, {Drazinos}, {Dur{\'a}n}, {Dzigan}, {Ecale}, {Edvardsson}, {Enke},
  {Erdmann}, {Escolar}, {Espina}, {Evans}, {Eynard Bontemps}, {Fabre},
  {Fabrizio}, {Faigler}, {Falc{\~a}o}, {Farr{\`a}s Casas}, {Faye}, {Federici},
  {Fedorets}, {Fern{\'a}ndez-Hern{\'a}ndez}, {Fernique}, {Fienga}, {Figueras},
  {Filippi}, {Findeisen}, {Fonti}, {Fouesneau}, {Fraile}, {Fraser}, {Fuchs},
  {Furnell}, {Gai}, {Galleti}, {Galluccio}, {Garabato}, {Garc{\'\i}a-Sedano},
  {Gar{\'e}}, {Garofalo}, {Garralda}, {Gavras}, {Gerssen}, {Geyer}, {Gilmore},
  {Girona}, {Giuffrida}, {Gomes}, {Gonz{\'a}lez-Marcos},
  {Gonz{\'a}lez-N{\'u}{\~n}ez}, {Gonz{\'a}lez-Vidal}, {Granvik}, {Guerrier},
  {Guillout}, {Guiraud}, {G{\'u}rpide}, {Guti{\'e}rrez-S{\'a}nchez}, {Guy},
  {Haigron}, {Hatzidimitriou}, {Haywood}, {Heiter}, {Helmi}, {Hobbs},
  {Hofmann}, {Holl}, {Holland}, {Hunt}, {Hypki}, {Icardi}, {Irwin}, {Jevardat
  de Fombelle}, {Jofr{\'e}}, {Jonker}, {Jorissen}, {Julbe}, {Karampelas},
  {Kochoska}, {Kohley}, {Kolenberg}, {Kontizas}, {Koposov}, {Kordopatis},
  {Koubsky}, {Kowalczyk}, {Krone-Martins}, {Kudryashova}, {Kull}, {Bachchan},
  {Lacoste-Seris}, {Lanza}, {Lavigne}, {Le Poncin-Lafitte}, {Lebreton},
  {Lebzelter}, {Leccia}, {Leclerc}, {Lecoeur-Taibi}, {Lemaitre}, {Lenhardt},
  {Leroux}, {Liao}, {Licata}, {Lindstr{\o}m}, {Lister}, {Livanou}, {Lobel},
  {L{\"o}ffler}, {L{\'o}pez}, {Lopez-Lozano}, {Lorenz}, {Loureiro},
  {MacDonald}, {Magalh{\~a}es Fernandes}, {Managau}, {Mann}, {Mantelet},
  {Marchal}, {Marchant}, {Marconi}, {Marie}, {Marinoni}, {Marrese},
  {Marschalk{\'o}}, {Marshall}, {Mart{\'\i}n-Fleitas}, {Martino}, {Mary},
  {Matijevi{\v{c}}}, {Mazeh}, {McMillan}, {Messina}, {Mestre}, {Michalik},
  {Millar}, {Miranda}, {Molina}, {Molinaro}, {Molinaro}, {Moln{\'a}r},
  {Moniez}, {Montegriffo}, {Monteiro}, {Mor}, {Mora}, {Morbidelli}, {Morel},
  {Morgenthaler}, {Morley}, {Morris}, {Mulone}, {Muraveva}, {Musella},
  {Narbonne}, {Nelemans}, {Nicastro}, {Noval}, {Ord{\'e}novic},
  {Ordieres-Mer{\'e}}, {Osborne}, {Pagani}, {Pagano}, {Pailler}, {Palacin},
  {Palaversa}, {Parsons}, {Paulsen}, {Pecoraro}, {Pedrosa}, {Pentik{\"a}inen},
  {Pereira}, {Pichon}, {Piersimoni}, {Pineau}, {Plachy}, {Plum}, {Poujoulet},
  {Pr{\v{s}}a}, {Pulone}, {Ragaini}, {Rago}, {Rambaux}, {Ramos-Lerate},
  {Ranalli}, {Rauw}, {Read}, {Regibo}, {Renk}, {Reyl{\'e}}, {Ribeiro},
  {Rimoldini}, {Ripepi}, {Riva}, {Rixon}, {Roelens}, {Romero-G{\'o}mez},
  {Rowell}, {Royer}, {Rudolph}, {Ruiz-Dern}, {Sadowski}, {Sagrist{\`a}
  Sell{\'e}s}, {Sahlmann}, {Salgado}, {Salguero}, {Sarasso}, {Savietto},
  {Schnorhk}, {Schultheis}, {Sciacca}, {Segol}, {Segovia}, {Segransan},
  {Serpell}, {Shih}, {Smareglia}, {Smart}, {Smith}, {Solano}, {Solitro},
  {Sordo}, {Soria Nieto}, {Souchay}, {Spagna}, {Spoto}, {Stampa}, {Steele},
  {Steidelm{\"u}ller}, {Stephenson}, {Stoev}, {Suess}, {S{\"u}veges}, {Surdej},
  {Szabados}, {Szegedi-Elek}, {Tapiador}, {Taris}, {Tauran}, {Taylor},
  {Teixeira}, {Terrett}, {Tingley}, {Trager}, {Turon}, {Ulla}, {Utrilla},
  {Valentini}, {van Elteren}, {Van Hemelryck}, {van Leeuwen}, {Varadi},
  {Vecchiato}, {Veljanoski}, {Via}, {Vicente}, {Vogt}, {Voss}, {Votruba},
  {Voutsinas}, {Walmsley}, {Weiler}, {Weingrill}, {Werner}, {Wevers},
  {Whitehead}, {Wyrzykowski}, {Yoldas}, {{\v{Z}}erjal}, {Zucker}, {Zurbach},
  {Zwitter}, {Alecu}, {Allen}, {Allende Prieto}, {Amorim},
  {Anglada-Escud{\'e}}, {Arsenijevic}, {Azaz}, {Balm}, {Beck}, {Bernstein},
  {Bigot}, {Bijaoui}, {Blasco}, {Bonfigli}, {Bono}, {Boudreault}, {Bressan},
  {Brown}, {Brunet}, {Bunclark}, {Buonanno}, {Butkevich}, {Carret}, {Carrion},
  {Chemin}, {Ch{\'e}reau}, {Corcione}, {Darmigny}, {de Boer}, {de Teodoro}, {de
  Zeeuw}, {Delle Luche}, {Domingues}, {Dubath}, {Fodor}, {Fr{\'e}zouls},
  {Fries}, {Fustes}, {Fyfe}, {Gallardo}, {Gallegos}, {Gardiol}, {Gebran},
  {Gomboc}, {G{\'o}mez}, {Grux}, {Gueguen}, {Heyrovsky}, {Hoar}, {Iannicola},
  {Isasi Parache}, {Janotto}, {Joliet}, {Jonckheere}, {Keil}, {Kim},
  {Klagyivik}, {Klar}, {Knude}, {Kochukhov}, {Kolka}, {Kos}, {Kutka}, {Lainey},
  {LeBouquin}, {Liu}, {Loreggia}, {Makarov}, {Marseille}, {Martayan},
  {Martinez-Rubi}, {Massart}, {Meynadier}, {Mignot}, {Munari}, {Nguyen},
  {Nordlander}, {Ocvirk}, {O'Flaherty}, {Olias Sanz}, {Ortiz}, {Osorio},
  {Oszkiewicz}, {Ouzounis}, {Palmer}, {Park}, {Pasquato}, {Peltzer}, {Peralta},
  {P{\'e}turaud}, {Pieniluoma}, {Pigozzi}, {Poels}, {Prat}, {Prod'homme},
  {Raison}, {Rebordao}, {Risquez}, {Rocca-Volmerange}, {Rosen}, {Ruiz-Fuertes},
  {Russo}, {Sembay}, {Serraller Vizcaino}, {Short}, {Siebert}, {Silva},
  {Sinachopoulos}, {Slezak}, {Soffel}, {Sosnowska}, {Strai{\v{z}}ys}, {ter
  Linden}, {Terrell}, {Theil}, {Tiede}, {Troisi}, {Tsalmantza}, {Tur},
  {Vaccari}, {Vachier}, {Valles}, {Van Hamme}, {Veltz}, {Virtanen}, {Wallut},
  {Wichmann}, {Wilkinson}, {Ziaeepour}, \& {Zschocke}}]{Gaia2016}
{Gaia Collaboration}, {Prusti}, T., {de Bruijne}, J.~H.~J., {et~al.} 2016,
  \aap, 595, A1, \dodoi{10.1051/0004-6361/201629272}

\bibitem[{{Gaia Collaboration} {et~al.}(2023){Gaia Collaboration}, {Vallenari},
  {Brown}, {Prusti}, {de Bruijne}, {Arenou}, {Babusiaux}, {Biermann},
  {Creevey}, {Ducourant}, {Evans}, {Eyer}, {Guerra}, {Hutton}, {Jordi},
  {Klioner}, {Lammers}, {Lindegren}, {Luri}, {Mignard}, {Panem}, {Pourbaix},
  {Randich}, {Sartoretti}, {Soubiran}, {Tanga}, {Walton}, {Bailer-Jones},
  {Bastian}, {Drimmel}, {Jansen}, {Katz}, {Lattanzi}, {van Leeuwen}, {Bakker},
  {Cacciari}, {Casta{\~n}eda}, {De Angeli}, {Fabricius}, {Fouesneau},
  {Fr{\'e}mat}, {Galluccio}, {Guerrier}, {Heiter}, {Masana}, {Messineo},
  {Mowlavi}, {Nicolas}, {Nienartowicz}, {Pailler}, {Panuzzo}, {Riclet}, {Roux},
  {Seabroke}, {Sordo}, {Th{\'e}venin}, {Gracia-Abril}, {Portell}, {Teyssier},
  {Altmann}, {Andrae}, {Audard}, {Bellas-Velidis}, {Benson}, {Berthier},
  {Blomme}, {Burgess}, {Busonero}, {Busso}, {C{\'a}novas}, {Carry}, {Cellino},
  {Cheek}, {Clementini}, {Damerdji}, {Davidson}, {de Teodoro}, {Nu{\~n}ez
  Campos}, {Delchambre}, {Dell'Oro}, {Esquej}, {Fern{\'a}ndez-Hern{\'a}ndez},
  {Fraile}, {Garabato}, {Garc{\'\i}a-Lario}, {Gosset}, {Haigron}, {Halbwachs},
  {Hambly}, {Harrison}, {Hern{\'a}ndez}, {Hestroffer}, {Hodgkin}, {Holl},
  {Jan{\ss}en}, {Jevardat de Fombelle}, {Jordan}, {Krone-Martins}, {Lanzafame},
  {L{\"o}ffler}, {Marchal}, {Marrese}, {Moitinho}, {Muinonen}, {Osborne},
  {Pancino}, {Pauwels}, {Recio-Blanco}, {Reyl{\'e}}, {Riello}, {Rimoldini},
  {Roegiers}, {Rybizki}, {Sarro}, {Siopis}, {Smith}, {Sozzetti}, {Utrilla},
  {van Leeuwen}, {Abbas}, {{\'A}brah{\'a}m}, {Abreu Aramburu}, {Aerts},
  {Aguado}, {Ajaj}, {Aldea-Montero}, {Altavilla}, {{\'A}lvarez}, {Alves},
  {Anders}, {Anderson}, {Anglada Varela}, {Antoja}, {Baines}, {Baker},
  {Balaguer-N{\'u}{\~n}ez}, {Balbinot}, {Balog}, {Barache}, {Barbato},
  {Barros}, {Barstow}, {Bartolom{\'e}}, {Bassilana}, {Bauchet}, {Becciani},
  {Bellazzini}, {Berihuete}, {Bernet}, {Bertone}, {Bianchi}, {Binnenfeld},
  {Blanco-Cuaresma}, {Blazere}, {Boch}, {Bombrun}, {Bossini}, {Bouquillon},
  {Bragaglia}, {Bramante}, {Breedt}, {Bressan}, {Brouillet}, {Brugaletta},
  {Bucciarelli}, {Burlacu}, {Butkevich}, {Buzzi}, {Caffau}, {Cancelliere},
  {Cantat-Gaudin}, {Carballo}, {Carlucci}, {Carnerero}, {Carrasco},
  {Casamiquela}, {Castellani}, {Castro-Ginard}, {Chaoul}, {Charlot}, {Chemin},
  {Chiaramida}, {Chiavassa}, {Chornay}, {Comoretto}, {Contursi}, {Cooper},
  {Cornez}, {Cowell}, {Crifo}, {Cropper}, {Crosta}, {Crowley}, {Dafonte},
  {Dapergolas}, {David}, {David}, {de Laverny}, {De Luise}, {De March}, {De
  Ridder}, {de Souza}, {de Torres}, {del Peloso}, {del Pozo}, {Delbo},
  {Delgado}, {Delisle}, {Demouchy}, {Dharmawardena}, {Di Matteo}, {Diakite},
  {Diener}, {Distefano}, {Dolding}, {Edvardsson}, {Enke}, {Fabre}, {Fabrizio},
  {Faigler}, {Fedorets}, {Fernique}, {Fienga}, {Figueras}, {Fournier},
  {Fouron}, {Fragkoudi}, {Gai}, {Garcia-Gutierrez}, {Garcia-Reinaldos},
  {Garc{\'\i}a-Torres}, {Garofalo}, {Gavel}, {Gavras}, {Gerlach}, {Geyer},
  {Giacobbe}, {Gilmore}, {Girona}, {Giuffrida}, {Gomel}, {Gomez},
  {Gonz{\'a}lez-N{\'u}{\~n}ez}, {Gonz{\'a}lez-Santamar{\'\i}a},
  {Gonz{\'a}lez-Vidal}, {Granvik}, {Guillout}, {Guiraud},
  {Guti{\'e}rrez-S{\'a}nchez}, {Guy}, {Hatzidimitriou}, {Hauser}, {Haywood},
  {Helmer}, {Helmi}, {Sarmiento}, {Hidalgo}, {Hilger}, {H{\l}adczuk}, {Hobbs},
  {Holland}, {Huckle}, {Jardine}, {Jasniewicz}, {Jean-Antoine Piccolo},
  {Jim{\'e}nez-Arranz}, {Jorissen}, {Juaristi Campillo}, {Julbe}, {Karbevska},
  {Kervella}, {Khanna}, {Kontizas}, {Kordopatis}, {Korn}, {K{\'o}sp{\'a}l},
  {Kostrzewa-Rutkowska}, {Kruszy{\'n}ska}, {Kun}, {Laizeau}, {Lambert},
  {Lanza}, {Lasne}, {Le Campion}, {Lebreton}, {Lebzelter}, {Leccia}, {Leclerc},
  {Lecoeur-Taibi}, {Liao}, {Licata}, {Lindstr{\o}m}, {Lister}, {Livanou},
  {Lobel}, {Lorca}, {Loup}, {Madrero Pardo}, {Magdaleno Romeo}, {Managau},
  {Mann}, {Manteiga}, {Marchant}, {Marconi}, {Marcos}, {Marcos Santos},
  {Mar{\'\i}n Pina}, {Marinoni}, {Marocco}, {Marshall}, {Martin Polo},
  {Mart{\'\i}n-Fleitas}, {Marton}, {Mary}, {Masip}, {Massari},
  {Mastrobuono-Battisti}, {Mazeh}, {McMillan}, {Messina}, {Michalik}, {Millar},
  {Mints}, {Molina}, {Molinaro}, {Moln{\'a}r}, {Monari}, {Mongui{\'o}},
  {Montegriffo}, {Montero}, {Mor}, {Mora}, {Morbidelli}, {Morel}, {Morris},
  {Muraveva}, {Murphy}, {Musella}, {Nagy}, {Noval}, {Oca{\~n}a}, {Ogden},
  {Ordenovic}, {Osinde}, {Pagani}, {Pagano}, {Palaversa}, {Palicio},
  {Pallas-Quintela}, {Panahi}, {Payne-Wardenaar}, {Pe{\~n}alosa Esteller},
  {Penttil{\"a}}, {Pichon}, {Piersimoni}, {Pineau}, {Plachy}, {Plum}, {Poggio},
  {Pr{\v{s}}a}, {Pulone}, {Racero}, {Ragaini}, {Rainer}, {Raiteri}, {Rambaux},
  {Ramos}, {Ramos-Lerate}, {Re Fiorentin}, {Regibo}, {Richards}, {Rios Diaz},
  {Ripepi}, {Riva}, {Rix}, {Rixon}, {Robichon}, {Robin}, {Robin}, {Roelens},
  {Rogues}, {Rohrbasser}, {Romero-G{\'o}mez}, {Rowell}, {Royer}, {Ruz Mieres},
  {Rybicki}, {Sadowski}, {S{\'a}ez N{\'u}{\~n}ez}, {Sagrist{\`a} Sell{\'e}s},
  {Sahlmann}, {Salguero}, {Samaras}, {Sanchez Gimenez}, {Sanna},
  {Santove{\~n}a}, {Sarasso}, {Schultheis}, {Sciacca}, {Segol}, {Segovia},
  {S{\'e}gransan}, {Semeux}, {Shahaf}, {Siddiqui}, {Siebert}, {Siltala},
  {Silvelo}, {Slezak}, {Slezak}, {Smart}, {Snaith}, {Solano}, {Solitro},
  {Souami}, {Souchay}, {Spagna}, {Spina}, {Spoto}, {Steele},
  {Steidelm{\"u}ller}, {Stephenson}, {S{\"u}veges}, {Surdej}, {Szabados},
  {Szegedi-Elek}, {Taris}, {Taylor}, {Teixeira}, {Tolomei}, {Tonello}, {Torra},
  {Torra}, {Torralba Elipe}, {Trabucchi}, {Tsounis}, {Turon}, {Ulla}, {Unger},
  {Vaillant}, {van Dillen}, {van Reeven}, {Vanel}, {Vecchiato}, {Viala},
  {Vicente}, {Voutsinas}, {Weiler}, {Wevers}, {Wyrzykowski}, {Yoldas}, {Yvard},
  {Zhao}, {Zorec}, {Zucker}, \& {Zwitter}}]{GaiaDR3}
{Gaia Collaboration}, {Vallenari}, A., {Brown}, A.~G.~A., {et~al.} 2023, \aap,
  674, A1, \dodoi{10.1051/0004-6361/202243940}

\bibitem[{{Gaudi}(1998)}]{1998ApJ...506..533G}
{Gaudi}, B.~S. 1998, \apj, 506, 533, \dodoi{10.1086/306256}

\bibitem[{{Gonzalez} {et~al.}(2012){Gonzalez}, {Rejkuba}, {Zoccali}, {Valenti},
  {Minniti}, {Schultheis}, {Tobar}, \& {Chen}}]{Gonzalez2012A}
{Gonzalez}, O.~A., {Rejkuba}, M., {Zoccali}, M., {et~al.} 2012, \aap, 543, A13,
  \dodoi{10.1051/0004-6361/201219222}

\bibitem[{{Gould}(1992)}]{Gould1992}
{Gould}, A. 1992, \apj, 392, 442, \dodoi{10.1086/171443}

\bibitem[{{Gould}(2003)}]{Gould03}
---. 2003, arXiv e-prints, astro.
\newblock \doarXiv{astro-ph/0310577}

\bibitem[{{Gould}(2004)}]{Gould2004}
---. 2004, \apj, 606, 319, \dodoi{10.1086/382782}

\bibitem[{{Gould}(2022)}]{masada}
---. 2022, arXiv e-prints, arXiv:2209.12501, \dodoi{10.48550/arXiv.2209.12501}

\bibitem[{{Gould} \& {Loeb}(1992)}]{1992ApJ...396..104G}
{Gould}, A., \& {Loeb}, A. 1992, \apj, 396, 104, \dodoi{10.1086/171700}

\bibitem[{{GRAVITY Collaboration} {et~al.}(2017){GRAVITY Collaboration},
  {Abuter}, {Accardo}, {Amorim}, {Anugu}, {{\'A}vila}, {Azouaoui}, {Benisty},
  {Berger}, {Blind}, {Bonnet}, {Bourget}, {Brandner}, {Brast}, {Buron},
  {Burtscher}, {Cassaing}, {Chapron}, {Choquet}, {Cl{\'e}net}, {Collin},
  {Coud{\'e} Du Foresto}, {de Wit}, {de Zeeuw}, {Deen},
  {Delplancke-Str{\"o}bele}, {Dembet}, {Derie}, {Dexter}, {Duvert}, {Ebert},
  {Eckart}, {Eisenhauer}, {Esselborn}, {F{\'e}dou}, {Finger}, {Garcia}, {Garcia
  Dabo}, {Garcia Lopez}, {Gendron}, {Genzel}, {Gillessen}, {Gonte}, {Gordo},
  {Grould}, {Gr{\"o}zinger}, {Guieu}, {Haguenauer}, {Hans}, {Haubois}, {Haug},
  {Haussmann}, {Henning}, {Hippler}, {Horrobin}, {Huber}, {Hubert}, {Hubin},
  {Hummel}, {Jakob}, {Janssen}, {Jochum}, {Jocou}, {Kaufer}, {Kellner},
  {Kendrew}, {Kern}, {Kervella}, {Kiekebusch}, {Klein}, {Kok}, {Kolb}, {Kulas},
  {Lacour}, {Lapeyr{\`e}re}, {Lazareff}, {Le Bouquin}, {L{\`e}na}, {Lenzen},
  {L{\'e}v{\^e}que}, {Lippa}, {Magnard}, {Mehrgan}, {Mellein}, {M{\'e}rand},
  {Moreno-Ventas}, {Moulin}, {M{\"u}ller}, {M{\"u}ller}, {Neumann}, {Oberti},
  {Ott}, {Pallanca}, {Panduro}, {Pasquini}, {Paumard}, {Percheron}, {Perraut},
  {Perrin}, {Pfl{\"u}ger}, {Pfuhl}, {Phan Duc}, {Plewa}, {Popovic}, {Rabien},
  {Ram{\'\i}rez}, {Ramos}, {Rau}, {Riquelme}, {Rohloff}, {Rousset},
  {Sanchez-Bermudez}, {Scheithauer}, {Sch{\"o}ller}, {Schuhler}, {Spyromilio},
  {Straubmeier}, {Sturm}, {Suarez}, {Tristram}, {Ventura}, {Vincent},
  {Waisberg}, {Wank}, {Weber}, {Wieprecht}, {Wiest}, {Wiezorrek}, {Wittkowski},
  {Woillez}, {Wolff}, {Yazici}, {Ziegler}, \& {Zins}}]{grav17}
{GRAVITY Collaboration}, {Abuter}, R., {Accardo}, M., {et~al.} 2017, \aap, 602,
  A94, \dodoi{10.1051/0004-6361/201730838}

\bibitem[{{Grevesse} {et~al.}(2007){Grevesse}, {Asplund}, \&
  {Sauval}}]{Grevesse2007}
{Grevesse}, N., {Asplund}, M., \& {Sauval}, A.~J. 2007, \ssr, 130, 105,
  \dodoi{10.1007/s11214-007-9173-7}

\bibitem[{{Griest} \& {Hu}(1992)}]{xallarap}
{Griest}, K., \& {Hu}, W. 1992, \apj, 397, 362, \dodoi{10.1086/171793}

\bibitem[{{Griest} \& {Safizadeh}(1998)}]{GriestSafizadeh1998}
{Griest}, K., \& {Safizadeh}, N. 1998, \apj, 500, 37, \dodoi{10.1086/305729}

\bibitem[{{Guo} {et~al.}(2021){Guo}, {Chen}, {Yuan}, {Huang}, {Liu}, {Yang},
  {Li}, {Sun}, \& {Liu}}]{guo2021}
{Guo}, H.~L., {Chen}, B.~Q., {Yuan}, H.~B., {et~al.} 2021, \apj, 906, 47,
  \dodoi{10.3847/1538-4357/abc68a}

\bibitem[{{Guo} \& {Kokubo}(2023)}]{Guo2023}
{Guo}, K., \& {Kokubo}, E. 2023, \apj, 955, 109,
  \dodoi{10.3847/1538-4357/acf31d}

\bibitem[{{Gustafsson} {et~al.}(2008){Gustafsson}, {Edvardsson}, {Eriksson},
  {J{\o}rgensen}, {Nordlund}, \& {Plez}}]{Gustafsson2008}
{Gustafsson}, B., {Edvardsson}, B., {Eriksson}, K., {et~al.} 2008, \aap, 486,
  951, \dodoi{10.1051/0004-6361:200809724}

\bibitem[{{Han}(2008)}]{Han2008}
{Han}, C. 2008, \apj, 681, 806, \dodoi{10.1086/588083}

\bibitem[{{Han} \& {Gould}(1997)}]{HanGould1997}
{Han}, C., \& {Gould}, A. 1997, \apj, 480, 196, \dodoi{10.1086/303944}

\bibitem[{{Han} {et~al.}(2019){Han}, {Yee}, {Udalski}, {Bond}, {Bozza},
  {Cassan}, {Hirao}, {Dong}, {Kollmeier}, {Morrell}, {Boutsia}, {authors},
  {Albrow}, {Chung}, {Gould}, {Hwang}, {Lee}, {Ryu}, {Shin}, {Shvartzvald},
  {Jung}, {Kim}, {Kim}, {Cha}, {Kim}, {Kim}, {Hong}, {Kim}, {Lee}, {Lee},
  {Park}, {Pogge}, {Zang}, {KMTNet Collaboration}, {Mr{\'o}z}, {Szyma{\'n}ski},
  {Skowron}, {Poleski}, {Soszy{\'n}ski}, {Pietrukowicz}, {Koz{\l}owski},
  {Ulaczyk}, {Rybicki}, {Iwanek}, {Wrona}, {OGLE Collaboration}, {Abe},
  {Barry}, {Bennett}, {Bhattacharya}, {Donachie}, {Fukui}, {Itow}, {Kawasaki},
  {Kondo}, {Koshimoto}, {Li}, {Matsubara}, {Muraki}, {Miyazaki}, {Nagakane},
  {Ranc}, {Rattenbury}, {Suematsu}, {Sullivan}, {Sumi}, {Suzuki}, {Tristram},
  {Yonehara}, \& {MOA Collaboration}}]{Han2019}
{Han}, C., {Yee}, J.~C., {Udalski}, A., {et~al.} 2019, \aj, 158, 102,
  \dodoi{10.3847/1538-3881/ab2df4}

\bibitem[{{Han} {et~al.}(2022){Han}, {Bond}, {Yee}, {Zang}, {Albrow}, {Chung},
  {Gould}, {Hwang}, {Jung}, {Kim}, {Lee}, {Ryu}, {Shin}, {Shvartzvald}, {Cha},
  {Kim}, {Kim}, {Lee}, {Lee}, {Park}, {Pogge}, {Abe}, {Barry}, {Bennett},
  {Bhattacharya}, {Hirao}, {Fujii}, {Fukui}, {Itow}, {Kirikawa}, {Kondo},
  {Koshimoto}, {Matsubara}, {Matsumoto}, {Muraki}, {Miyazaki}, {Ranc},
  {Okamura}, {Rattenbury}, {Satoh}, {Sumi}, {Suzuki}, {Ishitani Silva}, {Toda},
  {Tristram}, {Yama}, {Yonehara}, {Cooper}, {Dimitrov}, {Dong}, {Drummond},
  {Green}, {Hennerley}, {Liu}, {Mao}, {Maoz}, {Penny}, \& {Yang}}]{Han2022}
{Han}, C., {Bond}, I.~A., {Yee}, J.~C., {et~al.} 2022, \aap, 658, A94,
  \dodoi{10.1051/0004-6361/202142495}

\bibitem[{{Hodgkin} {et~al.}(2021){Hodgkin}, {Harrison}, {Breedt}, {Wevers},
  {Rixon}, {Delgado}, {Yoldas}, {Kostrzewa-Rutkowska}, {Wyrzykowski}, {van
  Leeuwen}, {Blagorodnova}, {Campbell}, {Eappachen}, {Fraser}, {Ihanec},
  {Koposov}, {Kruszy{\'n}ska}, {Marton}, {Rybicki}, {Brown}, {Burgess},
  {Busso}, {Cowell}, {De Angeli}, {Diener}, {Evans}, {Gilmore}, {Holland},
  {Jonker}, {van Leeuwen}, {Mignard}, {Osborne}, {Portell}, {Prusti},
  {Richards}, {Riello}, {Seabroke}, {Walton}, {{\'A}brah{\'a}m}, {Altavilla},
  {Baker}, {Bastian}, {O'Brien}, {de Bruijne}, {Butterley}, {Carrasco},
  {Casta{\~n}eda}, {Clark}, {Clementini}, {Copperwheat}, {Cropper},
  {Damljanovic}, {Davidson}, {Davis}, {Dennefeld}, {Dhillon}, {Dolding},
  {Dominik}, {Esquej}, {Eyer}, {Fabricius}, {Fridman}, {Froebrich}, {Garralda},
  {Gomboc}, {Gonz{\'a}lez-Vidal}, {Guerra}, {Hambly}, {Hardy}, {Holl},
  {Hourihane}, {Japelj}, {Kann}, {Kiss}, {Knigge}, {Kolb}, {Komossa},
  {K{\'o}sp{\'a}l}, {Kov{\'a}cs}, {Kun}, {Leto}, {Lewis}, {Littlefair},
  {Mahabal}, {Mundell}, {Nagy}, {Padeletti}, {Palaversa}, {Pigulski},
  {Pretorius}, {van Reeven}, {Ribeiro}, {Roelens}, {Rowell}, {Schartel},
  {Scholz}, {Schwope}, {Sip{\H{o}}cz}, {Smartt}, {Smith}, {Serraller},
  {Steeghs}, {Sullivan}, {Szabados}, {Szegedi-Elek}, {Tisserand}, {Tomasella},
  {van Velzen}, {Whitelock}, {Wilson}, \& {Young}}]{GSC}
{Hodgkin}, S.~T., {Harrison}, D.~L., {Breedt}, E., {et~al.} 2021, \aap, 652,
  A76, \dodoi{10.1051/0004-6361/202140735}

\bibitem[{{Hog} {et~al.}(1995){Hog}, {Novikov}, \&
  {Polnarev}}]{1995A&A...294..287H}
{Hog}, E., {Novikov}, I.~D., \& {Polnarev}, A.~G. 1995, \aap, 294, 287

\bibitem[{{Hwang} {et~al.}(2013){Hwang}, {Choi}, {Bond}, {Sumi}, {Han},
  {Gaudi}, {Gould}, {Bozza}, {Beaulieu}, {Tsapras}, {Abe}, {Bennett},
  {Botzler}, {Chote}, {Freeman}, {Fukui}, {Fukunaga}, {Harris}, {Itow},
  {Koshimoto}, {Ling}, {Masuda}, {Matsubara}, {Muraki}, {Namba}, {Ohnishi},
  {Rattenbury}, {Saito}, {Sullivan}, {Sweatman}, {Suzuki}, {Tristram}, {Wada},
  {Yamai}, {Yock}, {Yonehara}, {MOA Collaboration}, {de Almeida}, {DePoy},
  {Dong}, {Jablonski}, {Jung}, {Kavka}, {Lee}, {Park}, {Pogge}, {Shin}, {Yee},
  {{\ensuremath{\mu}}FUN Collaboration}, {Albrow}, {Bachelet}, {Batista},
  {Brillant}, {Caldwell}, {Cassan}, {Cole}, {Corrales}, {Coutures}, {Dieters},
  {Dominis Prester}, {Donatowicz}, {Fouqu{\'e}}, {Greenhill}, {J{\o}rgensen},
  {Kane}, {Kubas}, {Marquette}, {Martin}, {Meintjes}, {Menzies}, {Pollard},
  {Williams}, {Wouters}, {PLANET Collaboration}, {Bramich}, {Dominik}, {Horne},
  {Browne}, {Hundertmark}, {Ipatov}, {Kains}, {Snodgrass}, {Steele}, {Street},
  \& {RoboNet Collaboration}}]{2013ApJ...778...55H}
{Hwang}, K.~H., {Choi}, J.~Y., {Bond}, I.~A., {et~al.} 2013, \apj, 778, 55,
  \dodoi{10.1088/0004-637X/778/1/55}

\bibitem[{{Juri{\'c}} {et~al.}(2008){Juri{\'c}}, {Ivezi{\'c}}, {Brooks},
  {Lupton}, {Schlegel}, {Finkbeiner}, {Padmanabhan}, {Bond}, {Sesar},
  {Rockosi}, {Knapp}, {Gunn}, {Sumi}, {Schneider}, {Barentine}, {Brewington},
  {Brinkmann}, {Fukugita}, {Harvanek}, {Kleinman}, {Krzesinski}, {Long},
  {Neilsen}, {Nitta}, {Snedden}, \& {York}}]{Juric2008}
{Juri{\'c}}, M., {Ivezi{\'c}}, {\v{Z}}., {Brooks}, A., {et~al.} 2008, \apj,
  673, 864, \dodoi{10.1086/523619}

\bibitem[{{Kelson}(2003)}]{Kelson2003}
{Kelson}, D.~D. 2003, \pasp, 115, 688, \dodoi{10.1086/375502}

\bibitem[{{Kelson} {et~al.}(2000){Kelson}, {Illingworth}, {van Dokkum}, \&
  {Franx}}]{Kelson2000}
{Kelson}, D.~D., {Illingworth}, G.~D., {van Dokkum}, P.~G., \& {Franx}, M.
  2000, \apj, 531, 159, \dodoi{10.1086/308445}

\bibitem[{{Kniazev} {et~al.}(2016){Kniazev}, {Gvaramadze}, \&
  {Berdnikov}}]{2016MNRAS.459.3068K}
{Kniazev}, A.~Y., {Gvaramadze}, V.~V., \& {Berdnikov}, L.~N. 2016, \mnras, 459,
  3068, \dodoi{10.1093/mnras/stw889}

\bibitem[{{Kniazev} {et~al.}(2017){Kniazev}, {Gvaramadze}, \&
  {Berdnikov}}]{2017ASPC..510..480K}
{Kniazev}, A.~Y., {Gvaramadze}, V.~V., \& {Berdnikov}, L.~N. 2017, in
  Astronomical Society of the Pacific Conference Series, Vol. 510, Stars: From
  Collapse to Collapse, ed. Y.~Y. {Balega}, D.~O. {Kudryavtsev}, I.~I.
  {Romanyuk}, \& I.~A. {Yakunin}, 480, \dodoi{10.48550/arXiv.1612.00292}

\bibitem[{{Kochanek} {et~al.}(2017){Kochanek}, {Shappee}, {Stanek}, {Holoien},
  {Thompson}, {Prieto}, {Dong}, {Shields}, {Will}, {Britt}, {Perzanowski}, \&
  {Pojma{\'n}ski}}]{2017PASP..129j4502K}
{Kochanek}, C.~S., {Shappee}, B.~J., {Stanek}, K.~Z., {et~al.} 2017, \pasp,
  129, 104502, \dodoi{10.1088/1538-3873/aa80d9}

\bibitem[{Koposov {et~al.}(2023)Koposov, Speagle, Barbary, Ashton, Bennett,
  Buchner, Scheffler, Cook, Talbot, Guillochon, Cubillos, Ramos, Johnson, Lang,
  Ilya, Dartiailh, Nitz, McCluskey, Archibald, Deil, Foreman-Mackey, Goldstein,
  Tollerud, Leja, Kirk, Pitkin, Sheehan, Cargile, Patel, \&
  Angus}]{sergey_koposov_2023_7995596}
Koposov, S., Speagle, J., Barbary, K., {et~al.} 2023, joshspeagle/dynesty:
  v2.1.2, v2.1.2,  Zenodo, \dodoi{10.5281/zenodo.7995596}

\bibitem[{{Kuang} {et~al.}(2023){Kuang}, {Zang}, {Mao}, {Zhang}, \&
  {Jiang}}]{Kuang2023}
{Kuang}, R., {Zang}, W., {Mao}, S., {Zhang}, J., \& {Jiang}, H. 2023, \mnras,
  520, 4540, \dodoi{10.1093/mnras/stad461}

\bibitem[{{Lindegren} {et~al.}(2021){Lindegren}, {Klioner}, {Hern{\'a}ndez},
  {Bombrun}, {Ramos-Lerate}, {Steidelm{\"u}ller}, {Bastian}, {Biermann}, {de
  Torres}, {Gerlach}, {Geyer}, {Hilger}, {Hobbs}, {Lammers}, {McMillan},
  {Stephenson}, {Casta{\~n}eda}, {Davidson}, {Fabricius}, {Gracia-Abril},
  {Portell}, {Rowell}, {Teyssier}, {Torra}, {Bartolom{\'e}}, {Clotet},
  {Garralda}, {Gonz{\'a}lez-Vidal}, {Torra}, {Abbas}, {Altmann}, {Anglada
  Varela}, {Balaguer-N{\'u}{\~n}ez}, {Balog}, {Barache}, {Becciani}, {Bernet},
  {Bertone}, {Bianchi}, {Bouquillon}, {Brown}, {Bucciarelli}, {Busonero},
  {Butkevich}, {Buzzi}, {Cancelliere}, {Carlucci}, {Charlot}, {Cioni},
  {Crosta}, {Crowley}, {del Peloso}, {del Pozo}, {Drimmel}, {Esquej}, {Fienga},
  {Fraile}, {Gai}, {Garcia-Reinaldos}, {Guerra}, {Hambly}, {Hauser},
  {Jan{\ss}en}, {Jordan}, {Kostrzewa-Rutkowska}, {Lattanzi}, {Liao}, {Licata},
  {Lister}, {L{\"o}ffler}, {Marchant}, {Masip}, {Mignard}, {Mints}, {Molina},
  {Mora}, {Morbidelli}, {Murphy}, {Pagani}, {Panuzzo}, {Pe{\~n}alosa Esteller},
  {Poggio}, {Re Fiorentin}, {Riva}, {Sagrist{\`a} Sell{\'e}s}, {Sanchez
  Gimenez}, {Sarasso}, {Sciacca}, {Siddiqui}, {Smart}, {Souami}, {Spagna},
  {Steele}, {Taris}, {Utrilla}, {van Reeven}, \& {Vecchiato}}]{Gaiaastrometry}
{Lindegren}, L., {Klioner}, S.~A., {Hern{\'a}ndez}, J., {et~al.} 2021, \aap,
  649, A2, \dodoi{10.1051/0004-6361/202039709}

\bibitem[{{Miyamoto} \& {Yoshii}(1995)}]{1995AJ....110.1427M}
{Miyamoto}, M., \& {Yoshii}, Y. 1995, \aj, 110, 1427, \dodoi{10.1086/117616}

\bibitem[{{Nucita} {et~al.}(2018){Nucita}, {Licchelli}, {De Paolis},
  {Ingrosso}, {Strafella}, {Katysheva}, \& {Shugarov}}]{Nucita2018}
{Nucita}, A.~A., {Licchelli}, D., {De Paolis}, F., {et~al.} 2018, \mnras, 476,
  2962, \dodoi{10.1093/mnras/sty448}

\bibitem[{{Paczynski}(1986)}]{1986ApJ...304....1P}
{Paczynski}, B. 1986, \apj, 304, 1, \dodoi{10.1086/164140}

\bibitem[{{Pepe} {et~al.}(2021){Pepe}, {Cristiani}, {Rebolo}, {Santos},
  {Dekker}, {Cabral}, {Di Marcantonio}, {Figueira}, {Lo Curto}, {Lovis},
  {Mayor}, {M{\'e}gevand}, {Molaro}, {Riva}, {Zapatero Osorio}, {Amate},
  {Manescau}, {Pasquini}, {Zerbi}, {Adibekyan}, {Abreu}, {Affolter}, {Alibert},
  {Aliverti}, {Allart}, {Allende Prieto}, {{\'A}lvarez}, {Alves}, {Avila},
  {Baldini}, {Bandy}, {Barros}, {Benz}, {Bianco}, {Borsa}, {Bourrier},
  {Bouchy}, {Broeg}, {Calderone}, {Cirami}, {Coelho}, {Conconi}, {Coretti},
  {Cumani}, {Cupani}, {D'Odorico}, {Damasso}, {Deiries}, {Delabre},
  {Demangeon}, {Dumusque}, {Ehrenreich}, {Faria}, {Fragoso}, {Genolet},
  {Genoni}, {G{\'e}nova Santos}, {Gonz{\'a}lez Hern{\'a}ndez}, {Hughes},
  {Iwert}, {Kerber}, {Knudstrup}, {Landoni}, {Lavie}, {Lillo-Box}, {Lizon},
  {Maire}, {Martins}, {Mehner}, {Micela}, {Modigliani}, {Monteiro}, {Monteiro},
  {Moschetti}, {Murphy}, {Nunes}, {Oggioni}, {Oliveira}, {Oshagh}, {Pall{\'e}},
  {Pariani}, {Poretti}, {Rasilla}, {Rebord{\~a}o}, {Redaelli}, {Santana
  Tschudi}, {Santin}, {Santos}, {S{\'e}gransan}, {Schmidt}, {Segovia},
  {Sosnowska}, {Sozzetti}, {Sousa}, {Span{\`o}}, {Su{\'a}rez Mascare{\~n}o},
  {Tabernero}, {Tenegi}, {Udry}, \& {Zanutta}}]{espresso}
{Pepe}, F., {Cristiani}, S., {Rebolo}, R., {et~al.} 2021, \aap, 645, A96,
  \dodoi{10.1051/0004-6361/202038306}

\bibitem[{{Poindexter} {et~al.}(2005){Poindexter}, {Afonso}, {Bennett},
  {Glicenstein}, {Gould}, {Szyma{\'n}ski}, \& {Udalski}}]{2005ApJ...633..914P}
{Poindexter}, S., {Afonso}, C., {Bennett}, D.~P., {et~al.} 2005, \apj, 633,
  914, \dodoi{10.1086/468182}

\bibitem[{{Rucinski}(1999)}]{Rucinski1999}
{Rucinski}, S. 1999, in Astronomical Society of the Pacific Conference Series,
  Vol. 185, IAU Colloq. 170: Precise Stellar Radial Velocities, ed. J.~B.
  {Hearnshaw} \& C.~D. {Scarfe}, 82, \dodoi{10.48550/arXiv.astro-ph/9807327}

\bibitem[{{Rybicki} {et~al.}(2018){Rybicki}, {Wyrzykowski}, {Klencki}, {de
  Bruijne}, {Belczy{\'n}ski}, \& {Chru{\'s}li{\'n}ska}}]{2018MNRAS.476.2013R}
{Rybicki}, K.~A., {Wyrzykowski}, {\L}., {Klencki}, J., {et~al.} 2018, \mnras,
  476, 2013, \dodoi{10.1093/mnras/sty356}

\bibitem[{{Schechter} {et~al.}(1993){Schechter}, {Mateo}, \&
  {Saha}}]{1993PASP..105.1342S}
{Schechter}, P.~L., {Mateo}, M., \& {Saha}, A. 1993, \pasp, 105, 1342,
  \dodoi{10.1086/133316}

\bibitem[{{Schlecker} {et~al.}(2021){Schlecker}, {Mordasini}, {Emsenhuber},
  {Klahr}, {Henning}, {Burn}, {Alibert}, \& {Benz}}]{Schlecker2021}
{Schlecker}, M., {Mordasini}, C., {Emsenhuber}, A., {et~al.} 2021, \aap, 656,
  A71, \dodoi{10.1051/0004-6361/202038554}

\bibitem[{{Sch{\"o}nrich} {et~al.}(2010){Sch{\"o}nrich}, {Binney}, \&
  {Dehnen}}]{Schonrich2010}
{Sch{\"o}nrich}, R., {Binney}, J., \& {Dehnen}, W. 2010, \mnras, 403, 1829,
  \dodoi{10.1111/j.1365-2966.2010.16253.x}

\bibitem[{{Shappee} {et~al.}(2014){Shappee}, {Prieto}, {Grupe}, {Kochanek},
  {Stanek}, {De Rosa}, {Mathur}, {Zu}, {Peterson}, {Pogge}, {Komossa}, {Im},
  {Jencson}, {Holoien}, {Basu}, {Beacom}, {Szczygie{\l}}, {Brimacombe},
  {Adams}, {Campillay}, {Choi}, {Contreras}, {Dietrich}, {Dubberley},
  {Elphick}, {Foale}, {Giustini}, {Gonzalez}, {Hawkins}, {Howell}, {Hsiao},
  {Koss}, {Leighly}, {Morrell}, {Mudd}, {Mullins}, {Nugent}, {Parrent},
  {Phillips}, {Pojmanski}, {Rosing}, {Ross}, {Sand}, {Terndrup}, {Valenti},
  {Walker}, \& {Yoon}}]{Shappee2014}
{Shappee}, B.~J., {Prieto}, J.~L., {Grupe}, D., {et~al.} 2014, \apj, 788, 48,
  \dodoi{10.1088/0004-637X/788/1/48}

\bibitem[{{Shen} \& {Turner}(2008)}]{Shen2008}
{Shen}, Y., \& {Turner}, E.~L. 2008, \apj, 685, 553, \dodoi{10.1086/590548}

\bibitem[{{Skowron} \& {Gould}(2012)}]{SkowronGould12}
{Skowron}, J., \& {Gould}, A. 2012, arXiv e-prints, arXiv:1203.1034,
  \dodoi{10.48550/arXiv.1203.1034}

\bibitem[{{Skowron} {et~al.}(2011){Skowron}, {Udalski}, {Gould}, {Dong},
  {Monard}, {Han}, {Nelson}, {McCormick}, {Moorhouse}, {Thornley}, {Maury},
  {Bramich}, {Greenhill}, {Koz{\l}owski}, {Bond}, {Poleski}, {Wyrzykowski},
  {Ulaczyk}, {Kubiak}, {Szyma{\'n}ski}, {Pietrzy{\'n}ski}, {Soszy{\'n}ski},
  {OGLE Collaboration}, {Gaudi}, {Yee}, {Hung}, {Pogge}, {DePoy}, {Lee},
  {Park}, {Allen}, {Mallia}, {Drummond}, {Bolt}, {{\ensuremath{\mu}}FUN
  Collaboration}, {Allan}, {Browne}, {Clay}, {Dominik}, {Fraser}, {Horne},
  {Kains}, {Mottram}, {Snodgrass}, {Steele}, {Street}, {Tsapras}, {RoboNet
  Collaboration}, {Abe}, {Bennett}, {Botzler}, {Douchin}, {Freeman}, {Fukui},
  {Furusawa}, {Hayashi}, {Hearnshaw}, {Hosaka}, {Itow}, {Kamiya}, {Kilmartin},
  {Korpela}, {Lin}, {Ling}, {Makita}, {Masuda}, {Matsubara}, {Muraki},
  {Nagayama}, {Miyake}, {Nishimoto}, {Ohnishi}, {Perrott}, {Rattenbury},
  {Saito}, {Skuljan}, {Sullivan}, {Sumi}, {Suzuki}, {Sweatman}, {Tristram},
  {Wada}, {Yock}, {MOA Collaboration}, {Beaulieu}, {Fouqu{\'e}}, {Albrow},
  {Batista}, {Brillant}, {Caldwell}, {Cassan}, {Cole}, {Cook}, {Coutures},
  {Dieters}, {Dominis Prester}, {Donatowicz}, {Kane}, {Kubas}, {Marquette},
  {Martin}, {Menzies}, {Sahu}, {Wambsganss}, {Williams}, {Zub}, \& {PLANET
  Collaboration}}]{Skowron2011}
{Skowron}, J., {Udalski}, A., {Gould}, A., {et~al.} 2011, \apj, 738, 87,
  \dodoi{10.1088/0004-637X/738/1/87}

\bibitem[{{Smith} {et~al.}(2003){Smith}, {Mao}, \& {Paczy{\'n}ski}}]{Smith03}
{Smith}, M.~C., {Mao}, S., \& {Paczy{\'n}ski}, B. 2003, \mnras, 339, 925,
  \dodoi{10.1046/j.1365-8711.2003.06183.x}

\bibitem[{{Smith} {et~al.}(2002){Smith}, {Mao}, {Wo{\'z}niak}, {Udalski},
  {Szyma{\'n}ski}, {Kubiak}, {Pietrzy{\'n}ski}, {Soszy{\'n}ski}, \&
  {{\.Z}ebru{\'n}}}]{Smith2002}
{Smith}, M.~C., {Mao}, S., {Wo{\'z}niak}, P., {et~al.} 2002, \mnras, 336, 670,
  \dodoi{10.1046/j.1365-8711.2002.05811.x}

\bibitem[{{Speagle}(2020)}]{2020MNRAS.493.3132S}
{Speagle}, J.~S. 2020, \mnras, 493, 3132, \dodoi{10.1093/mnras/staa278}

\bibitem[{{Tonry} {et~al.}(2012){Tonry}, {Stubbs}, {Lykke}, {Doherty},
  {Shivvers}, {Burgett}, {Chambers}, {Hodapp}, {Kaiser}, {Kudritzki},
  {Magnier}, {Morgan}, {Price}, \& {Wainscoat}}]{PS1}
{Tonry}, J.~L., {Stubbs}, C.~W., {Lykke}, K.~R., {et~al.} 2012, \apj, 750, 99,
  \dodoi{10.1088/0004-637X/750/2/99}

\bibitem[{{Tonry} {et~al.}(2018){Tonry}, {Denneau}, {Flewelling}, {Heinze},
  {Onken}, {Smartt}, {Stalder}, {Weiland}, \& {Wolf}}]{Refcat2}
{Tonry}, J.~L., {Denneau}, L., {Flewelling}, H., {et~al.} 2018, \apj, 867, 105,
  \dodoi{10.3847/1538-4357/aae386}

\bibitem[{{Volgenau} {et~al.}(2022){Volgenau}, {Harbeck}, {Lindstrom},
  {Collom}, {Street}, \& {Johnson}}]{TOM}
{Volgenau}, N., {Harbeck}, D., {Lindstrom}, W., {et~al.} 2022, in Society of
  Photo-Optical Instrumentation Engineers (SPIE) Conference Series, Vol. 12186,
  Observatory Operations: Strategies, Processes, and Systems IX, ed. D.~S.
  {Adler}, R.~L. {Seaman}, \& C.~R. {Benn}, 121860W, \dodoi{10.1117/12.2628704}

\bibitem[{{Walker}(1995)}]{1995ApJ...453...37W}
{Walker}, M.~A. 1995, \apj, 453, 37, \dodoi{10.1086/176367}

\bibitem[{Winn \& Fabrycky(2015)}]{WinnFabrycky2014}
Winn, J.~N., \& Fabrycky, D.~C. 2015, Annual Review of Astronomy and
  Astrophysics, 53, 409, \dodoi{10.1146/annurev-astro-082214-122246}

\bibitem[{{Wyrzykowski} {et~al.}(2023){Wyrzykowski}, {Kruszy{\'n}ska},
  {Rybicki}, {Holl}, {Lec{\oe}ur-Ta{\"\i}bi}, {Mowlavi}, {Nienartowicz},
  {Jevardat de Fombelle}, {Rimoldini}, {Audard}, {Garcia-Lario}, {Gavras},
  {Evans}, {Hodgkin}, \& {Eyer}}]{Wyrzykowski2023}
{Wyrzykowski}, {\L}., {Kruszy{\'n}ska}, K., {Rybicki}, K.~A., {et~al.} 2023,
  \aap, 674, A23, \dodoi{10.1051/0004-6361/202243756}

\bibitem[{{Yee} {et~al.}(2012){Yee}, {Shvartzvald}, {Gal-Yam}, {Bond},
  {Udalski}, {Koz{\l}owski}, {Han}, {Gould}, {Skowron}, {Suzuki}, {Abe},
  {Bennett}, {Botzler}, {Chote}, {Freeman}, {Fukui}, {Furusawa}, {Itow},
  {Kobara}, {Ling}, {Masuda}, {Matsubara}, {Miyake}, {Muraki}, {Ohmori},
  {Ohnishi}, {Rattenbury}, {Saito}, {Sullivan}, {Sumi}, {Suzuki}, {Sweatman},
  {Takino}, {Tristram}, {Wada}, {MOA Collaboration}, {Szyma{\'n}ski}, {Kubiak},
  {Pietrzy{\'n}ski}, {Soszy{\'n}ski}, {Poleski}, {Ulaczyk}, {Wyrzykowski},
  {Pietrukowicz}, {OGLE Collaboration}, {Allen}, {Almeida}, {Batista}, {Bos},
  {Christie}, {DePoy}, {Dong}, {Drummond}, {Finkelman}, {Gaudi}, {Gorbikov},
  {Henderson}, {Higgins}, {Jablonski}, {Kaspi}, {Manulis}, {Maoz}, {McCormick},
  {McGregor}, {Monard}, {Moorhouse}, {Mu{\~n}oz}, {Natusch}, {Ngan}, {Ofek},
  {Pogge}, {Santallo}, {Tan}, {Thornley}, {Shin}, {Choi}, {Park}, {Lee}, {Koo},
  \& {{\ensuremath{\mu}}FUN Collaboration}}]{2012ApJ...755..102Y}
{Yee}, J.~C., {Shvartzvald}, Y., {Gal-Yam}, A., {et~al.} 2012, \apj, 755, 102,
  \dodoi{10.1088/0004-637X/755/2/102}

\bibitem[{{Yi} {et~al.}(2022){Yi}, {Gu}, {Zhang}, {Zheng}, {Sun}, {Wang},
  {Bai}, {Wang}, {Wu}, {Bai}, {Wang}, {Zhang}, {Dong}, {Shao}, {Li}, {Zhang},
  {Huang}, {Yang}, {Yu}, {Mu}, {Fu}, {Qi}, {Guo}, {Fang}, {Zheng}, {Li}, {Shi},
  {Chen}, \& {Liu}}]{Yi2022}
{Yi}, T., {Gu}, W.-M., {Zhang}, Z.-X., {et~al.} 2022, Nature Astronomy, 6,
  1203, \dodoi{10.1038/s41550-022-01766-0}

\bibitem[{{Yoo} {et~al.}(2004){Yoo}, {DePoy}, {Gal-Yam}, {Gaudi}, {Gould},
  {Han}, {Lipkin}, {Maoz}, {Ofek}, {Park}, {Pogge}, {Mu-Fun Collaboration},
  {Udalski}, {Soszy{\'n}ski}, {Wyrzykowski}, {Kubiak}, {Szyma{\'n}ski},
  {Pietrzy{\'n}ski}, {Szewczyk}, {{\.Z}ebru{\'n}}, \& {OGLE
  Collaboration}}]{Yoo2004}
{Yoo}, J., {DePoy}, D.~L., {Gal-Yam}, A., {et~al.} 2004, \apj, 603, 139,
  \dodoi{10.1086/381241}

\bibitem[{{Zakamska} {et~al.}(2011){Zakamska}, {Pan}, \& {Ford}}]{Zakamska2011}
{Zakamska}, N.~L., {Pan}, M., \& {Ford}, E.~B. 2011, \mnras, 410, 1895,
  \dodoi{10.1111/j.1365-2966.2010.17570.x}

\bibitem[{{Zang} {et~al.}(2020){Zang}, {Dong}, {Gould}, {Calchi Novati},
  {Chen}, {Yang}, {Li}, {Mao}, {Alton}, {Brimacombe}, {Carey}, {Christie},
  {Delplancke-Str{\"o}bele}, {Feliz}, {Gaudi}, {Green}, {Hu}, {Jayasinghe},
  {Koff}, {Kurtenkov}, {M{\'e}rand}, {Minev}, {Mutel}, {Natusch}, {Roth},
  {Shvartzvald}, {Sun}, {Vanmunster}, \& {Zhu}}]{Zang2020}
{Zang}, W., {Dong}, S., {Gould}, A., {et~al.} 2020, \apj, 897, 180,
  \dodoi{10.3847/1538-4357/ab9749}

\bibitem[{{Zhang} \& {Yuan}(2023)}]{2023ApJS..264...14Z}
{Zhang}, R., \& {Yuan}, H. 2023, \apjs, 264, 14,
  \dodoi{10.3847/1538-4365/ac9dfa}

\bibitem[{{Zhu} \& {Dong}(2021)}]{ZhuDong2021}
{Zhu}, W., \& {Dong}, S. 2021, \araa, 59, 291,
  \dodoi{10.1146/annurev-astro-112420-020055}

\bibitem[{{Zhu} \& {Wu}(2018)}]{Zhu2018}
{Zhu}, W., \& {Wu}, Y. 2018, \aj, 156, 92, \dodoi{10.3847/1538-3881/aad22a}

\bibitem[{{Zieli{\'n}ski} {et~al.}(2019){Zieli{\'n}ski}, {Wyrzykowski},
  {Rybicki}, {Ko{\l}aczkowski}, {Bru{\'s}}, \& {Miko{\l}ajczyk}}]{CPCS}
{Zieli{\'n}ski}, P., {Wyrzykowski}, {\L}., {Rybicki}, K., {et~al.} 2019,
  Contributions of the Astronomical Observatory Skalnate Pleso, 49, 125

\end{thebibliography}
\bibliographystyle{aasjournal}
\listofchanges
\end{document}